\documentclass[prx,twocolumn,nofootinbib,citeautoscript,longbibliography,notitlepage,superscriptaddress]{revtex4-1}

\usepackage{graphicx}
\usepackage{dcolumn}
\usepackage{bm}
\usepackage{color}
\usepackage{amsmath}
\usepackage{tabularx,graphicx}
\usepackage{epstopdf}
\usepackage{latexsym}
\usepackage{amssymb}
\usepackage{amsmath}
\usepackage{color, colortbl}
\usepackage{psfrag}
\usepackage{bbm}
\usepackage{bm}
\usepackage{titlesec}
\usepackage{dsfont}
\usepackage{feynmp}
\usepackage{slashed}
\usepackage{multirow}
\usepackage[tight]{subfigure}
\usepackage{comment}
\usepackage{float}
\usepackage[papersize={8.5in,11in}]{geometry}

\usepackage{color}
\definecolor{darkblue}{rgb}{0.,0.,0.4}
\definecolor{darkred}{rgb}{0.5,0.,0.}
\definecolor{BlueViolet}{RGB}{138,43,226}
\definecolor{SkyBlue}{RGB}{30,144,255}
\definecolor{DarkGreen}{RGB}{0,100,0}

\usepackage[pdftex,colorlinks=true,linkcolor=darkblue,citecolor=blue,urlcolor=darkred]{hyperref}

\renewcommand{\epsilon}{\varepsilon}

\usepackage{cases}

\allowdisplaybreaks[4]

\UseRawInputEncoding

\begin{document}



\title{Fixed points and critical temperature near quantum critical points in $d$-wave cuprate superconductors}

\date{\today}

\author{Qi-Qi Yue}
\altaffiliation{These authors contributed equally to this work.}
\affiliation{Department of Physics, Tianjin University, Tianjin 300072, P.R. China}
\affiliation{Department of Advanced Optical and Microelectronic Equipment, Shanghai Institute of Optics and Fine Mechanics (SIOM), Chinese Academy of Sciences (CAS), Shanghai 201800, P.R. China}
\affiliation{Center of Materials Science and Optoelectronics Engineering, University of Chinese Academy of Sciences,
Beijing 100049, P.R. China}

\author{Yi-Sheng Fu}
\altaffiliation{These authors contributed equally to this work.}
\affiliation{Department of Physics, Tianjin University, Tianjin 300072, P.R. China}
\affiliation{Institute of High Energy Physics, Chinese Academy of Sciences, Beijing 100049, P.R. China}
\affiliation{School of Physical Sciences, University of Chinese Academy of Sciences, Beijing 100049, P. R. China}

\author{Jing Wang}
\altaffiliation{Corresponding author: jing$\textunderscore$wang@tju.edu.cn}
\affiliation{Department of Physics, Tianjin University, Tianjin 300072, P.R. China}
\affiliation{Tianjin Key Laboratory of Low Dimensional Materials Physics and
Preparing Technology, Tianjin University, Tianjin 300072, P.R. China}

\begin{abstract}
We study the critical behavior driven by potential quantum critical points (QCPs) termed
as $\tau_{0,x,y,z}$-Type QCPs beneath the superconducting dome of the $d$-wave cuprate superconductors.
To comprehensively capture the distinct degrees of freedom in the vicinity of these QCPs, we construct
a phenomenological effective theory based on the Landau-Ginzburg-Wilson framework and then employ
the renormalization group approach to derive the coupled flow equations of all interaction
parameters, incorporating all relevant one-loop corrections. Decoding these flow equations
yields a series of unique properties arising from strong quantum fluctuations around QCPs.
On one hand, the interaction parameters flow toward several fixed points (FPs) at
certain critical energy scales. We identify two different types of FPs designated at
the clean limit. FP-I is characterized by the divergence of the quadratic parameter
and exhibits robustness against variations in interaction parameters.
In contrast, FP-II is dominated by the cubic and quartic interaction parameters,
and it is sensitive to initial conditions, leading to five subclasses: FP-IIA, FP-IIB, FP-IIC, FP-IID,
and FP-IIE. In addition, we find that disorder scattering can influence fermion velocities and critical energy scales,
and even destabilize certain FPs around the $\tau_{x,z}$-QCPs, driving the system toward a preempted
disorder-induced FP. On the other hand, we find that quantum fluctuations play a critical role in
shaping the critical temperature ($T_c$) as the system approaches these QCPs. Near the $\tau_x$-QCP,
$T_c$ is considerably suppressed for both FP-I and FP-II. In contrast, near the $\tau_0$-QCP,
$T_c$ undergoes a substantial decrease for FP-I but only a slight decrease for FP-II. Conversely,
$T_c$ exhibits an increasing trend near the $\tau_y$-QCP, with a pronounced peak at $v_{\Delta 0}/v_{F0} \sim 0.25$.
However, numerical analysis suggests that the $\tau_z$-QCP is unlikely to be physically realizable.
Additionally, we realize that $T_c$ can also be modified by the emergence of disorder-induced FPs
in the vicinity of the $\tau_x$-QCP. These findings would provide valuable insights into the critical
low-energy properties of $d$-wave cuprate superconductors and related materials.

\end{abstract}


\maketitle


\section{Introduction}\label{Sec_intro}

The study of $d$-wave cuprate superconductors has garnered significant theoretical and experimental interest over the past three decades~\cite{Lee2006RMP,Vojta2000PRL,
Vojta2000PRB,Vojta2000IJMPB,Sachdev2000Science,Sachdev2003RMP,Sachdev2008PRB,Sachdev2011PT,
Wang2011PRB,Fradkin2012NPhys,Kivelson2014PNAS,Fradkin2015RMP,
Dagotto1994RMP,Dagotto2005Science,Kivelson1995Nature,Kivelson1998Nature,
Kivelson2003RMP_DFS,Sigrist1991RMP,Sigrist1995RMP,
Tinkham1996Book,Anderson1997Book,Phillips2020NPhys,
Kim-Kivelson2008PRB,She2010PRB,She2015PRB,Xu2008PRB,Larkin2005Book,Takagi1992PRL,Norman2011Science,Shekhter2013Nature,
Bozovic2016Nature,Phillips2022Science,Ramakrishnan2025}.
These materials are renowned for their unconventional pairing mechanisms, the anomalous behavior of their normal,
and the coexistence of multiple phases below the critical temperature~\cite{Lee2006RMP,Sachdev2003RMP,Kivelson2014PNAS,Fradkin2015RMP,Kivelson1998Nature,
Norman2011Science,Shekhter2013Nature,Phillips2022Science,Ramakrishnan2025}.
Notably, they exhibit a distinct $d_{x^2-y^2}$ superconducting gap~\cite{Lee2006RMP,Ding1996Nature,Loeser1996Science,
Valla1999Science,Orenstein2000Science,Yoshida2003PRL,Dagotto1994RMP}, which vanishes at four points ($\pm\pi/4,\pm\pi/4$)
in the first Brillouin zone known as the nodal points~\cite{Lee2006RMP,Fradkin2015RMP,Dagotto1994RMP}.
At these nodal points, gapless quasiparticles (QPs) are excited even at the lowest-energy limit under
the superconducting dome~\cite{Orenstein2000Science,Sachdev2000Science,Sachdev2003RMP,Lee2006RMP,Sachdev2011PT}.
In addition, the coexistence of different phases is usually accompanied by some quantum phase transition (QPT)~\cite{Vojta2003RPP,Sachdev2011Book,Coleman2005Nature}, during which certain critical bosonic
modes can emerge.

The QPs are typically considered nearly noninteracting, but they can couple to
critical bosonic modes near a QPT due to strong quantum fluctuations~\cite{Vojta2000PRB,Vojta2000PRL,Vojta2000IJMPB,Paaske2001PRL,Kim-Kivelson2008PRB,
Sachdev2008PRB,Xu2008PRB,Sachdev2009PRB,
Liu2012PRB,Liu2013NJP,She2015PRB}. These quantum critical degrees of freedom are assumed to play
a pivotal role in the anomalous behavior observed in $d$-wave superconductors ~\cite{Orenstein2000Science,Coleman2005Nature,Lee2006RMP,Fradkin2012NPhys,
Kivelson2014PNAS,Fradkin2015RMP,Sachdev2011Book,Vojta2003RPP,
Moon2010PRB,Moon2012PRB,Moon2016PRB,Moon2016SRep,Wang-EM2014PRB,
Yoshida2003PRL,Paaske2001PRL,Sachdev2009PRB,Liu2012PRB,Liu2013NJP}.
Several theoretical frameworks have been proposed to study the interplay between quantum criticality
and the QPs in these compounds~\cite{Lee1993PRL,
Vojta2000PRL,Vojta2000PRB,Coleman2005Nature,Dagotto2005Science,
Sachdev2000Science,Sachdev2011PT,Castellani1997ZPB,She2011PRL}. In particular, Vojta \emph{et al.}~\cite{Vojta2000PRL,Vojta2000PRB,Vojta2000IJMPB} suggest the existence of a quantum critical
point (QCP) within the superconducting dome~\cite{Sachdev2011Book,Vojta2003RPP}.
This QCP marks a transition from the $d_{x^2-y^2}$ superconducting state to a new state denoted by $d_{x^2-y^2}+X$ as illustrated in Fig.~\ref{fig1}, where the $X$ state is associated with certain a symmetry breaking. The group-theory analysis~\cite{Vojta2000PRL,Vojta2000PRB,Vojta2000IJMPB} identifies several potential candidates,
which can be further categorized into four distinct types, namely, $\tau_{0}$-Type, $\tau_{x}$-Type, $\tau_{y}$-Type, and $\tau_{z}$-Type QCPs~\cite{Vojta2000PRL,Vojta2000PRB,Wang2013PRB}.

In recent years, significant efforts have been devoted to studying these
QCPs in $d$-wave superconductors~\cite{Sachdev2008PRB,Kim-Kivelson2008PRB,Xu2008PRB,Vojta2009AP,
Sachdev2002PRB,Kivelson2009PRB,Moon2010PRB,Fradkin2010ARCMP,
Kivelson1998Nature,Kivelson2001PRB,Vojta2000PRB,Vojta2000PRL,
Sachdev2009PRB,Kim2010PRB,Wang2011PRB}.
Among these, the $\tau_x$-Type QCP attracted considerable attention, which is related to a nematic QPT
with breaking $C_4$ symmetry down to $C_2$ symmetry~\cite{Keimer2008Science,Kim2010Nature,Sachdev2011Book,Kivelson1998Nature,
Metzner2000PRL,Kivelson2001PRB}. In particular, Huh and Sachdev~\cite{Sachdev2008PRB} carefully examine the tendencies of
fermion velocities ($v_{F,\Delta}$) nearby the $\tau_x$-Type QCP.
Then, the fates of $v_{F,\Delta}$ as approaching the $\tau_y$-Type and $\tau_z$-Type QCPs have also
been investigated~\cite{Wang2013PRB,Wang2015PLA,Wang2013NJP} and distinct kinds of fixed points for fermion velocities
in the lowest-energy regime have been identified.
These unusual behavior of fermion velocities are expected to play a crucial role in shaping
the low-energy physical quantities~\cite{Lee1993PRL,Durst2000PRB,Mesot1999PRL,Vojta2009AP,
Kim-Kivelson2008PRB,Xu2008PRB,Wang2011PRB,Wang2013PRB,Wang2015PLA,Wang2013NJP,She2015PRB,RZW2022NPB,Durst2000PRB}.

While important progress~\cite{Sachdev2008PRB,Wang2013PRB,
Kim-Kivelson2008PRB,Xu2008PRB,Wang2011PRB,Wang2015PLA,Wang2013NJP,Liu2012PRB,She2015PRB,RZW2022NPB} has been made in understanding the critical behavior near these QPTs, several intriguing issues remain to be addressed to fully characterize the quantum criticality in the vicinity of these QCPs. On one hand, to simplify the analysis,
the underlying quantum fluctuations of SC order have usually been
neglected in previous studies. However, as the system approaches the QCP, quantum criticality can coax this very
ingredient to mutually interact with other degrees of freedom. As studied in Refs.~\cite{Kleinert2003NPB,Wang2014PRD,Wang2017PRB},
the amplitude and phase fluctuations of a complex scalar field can ferociously compete with
other kinds of fields, leading to nontrivial critical effects. In principle, the SC order parameter
in $d$-wave superconductors can be described by a complex scalar field,
which yields both the amplitude and phase fluctuations nearby the QCP.
These fluctuations are expected to couple with those of the $X$-state order parameter,
and their mutual interactions can influence all other interaction parameters.
As a consequence, this coupling can be expected to significantly modify the low-energy behavior of the QCPs.
On the other hand, most previous studies have focused on the $\tau_x$-type QCP, leaving the critical properties of other candidate QCPs ($\tau_y$, $\tau_z$, and $\tau_0$ types) insufficiently explored. In addition, it is particularly important to determine whether these candidate QCPs can emerge under physically realistic conditions, despite being theoretically possible from the perspective of group theory. Furthermore, as the system approaches physically realizable QCPs, it is crucial to investigate how the critical temperature is affected by the interplay between quantum fluctuations of the $X$-state order parameter and those of the superconducting order parameter.

These interesting issues motivate us to systematically investigate the critical
consequences and distinctions among various types of QCPs on the physics of
related quantum critical regions. This requires us to consider the combination of strong quantum
fluctuations of all order parameters and their interplay with other degrees of freedom.
In order to take into account these ingredients unbiasedly, we employ the momentum-shell renormalization group (RG) approach~\cite{Shankar1994RMP,Wilson1975RMP,Polchinski1992}, which equally treats all relevant critical degrees of freedom
near the putative QPT from a $d$-wave superconducting state to a $d_{x^2-y^2}+X$ state,
as illustrated in Fig.~\ref{fig1}. By incorporating all one-loop corrections, we derive a set of coupled
RG equations for all the interaction parameters. These equations provide detailed insights into the quantum
critical behavior near the candidate QCPs.

We commence with performing a numerical analysis of these flow equations. The results indicate that the overall tendencies of
the interaction parameters are dictated by several fixed points (FPs) that are designated in Sec.~\ref{Sec_FP}
at certain critical energy scale $l = l_c$ depending on the initial conditions. These fixed points are closely
linked to the low-energy critical behavior of the system. In the clean limit, the system evolves toward two distinct
types of fixed points: FP-I and FP-II. FP-I is characterized by the divergence of the quadratic parameter,
which remains largely independent of other interaction parameters. In contrast, FP-II is dominated by the
cubic and quartic interaction parameters, and it is sensitive to alterations in interaction parameters, yielding
five subclasses, i.e., FP-IIA, FP-IIB, FP-IIC, FP-IID,
and FP-IIE, due to the strong competition among order parameters. In addition, we examine the stabilities of these
FPs in the presence of three types of disorder scattering: random mass, random gauge potential, and random chemical potential.
Numerical analysis demonstrates that both random mass and random gauge potential provide negligible impacts on FPs.
However, the random chemical potential can significantly influence both the fermion velocities and the critical energy scales.
This may lead to the emergence of a preempted disorder-induced fixed point, as illustrated in Fig.~\ref{Fig_schematic_l_c}.

To proceed, we systematically investigate the behavior of critical
temperature ($T_c$) as we access FPs for all QCPs in Fig.~\ref{fig1}.
A detailed analysis demonstrates that $T_c$ around $\tau_x$-QCP
is significantly suppressed for both FP-I and FP-II. In contrast, near the $\tau_0$-QCP,
it undergoes a substantial decrease for FP-I but only a slight decrease for FP-II.
Interestingly, $T_c$ exhibits an increasing trend near the $\tau_y$-QCP, with a pronounced
peak at $v_{\Delta 0}/v_{F0}\sim 0.25$. This indicates that the initial anisotropy of the fermion
velocities ($v_{\Delta0} / v_{F0}$) can also quantitatively influence $T_c$. Additionally, we notice
that $T_c$ is insensitive to both random mass and random gauge potential but can
be considerably altered by the random chemical potential in the vicinity of the $\tau_x$-QCP.
Regarding the $\tau_z$-QCP, we find that the enhancement of $T_c$ lies well beyond the range
of critical temperatures $T_c$ typically observed in $d$-wave superconductors~\cite{Lee2006RMP,Fradkin2015RMP,Keimer2008Science,Orenstein2000Science,
Bozovic2016Nature,Phillips2022Science,Ramakrishnan2025}.
This suggests that such a QCP is unlikely to be physically realizable.

The remainder of this paper is organized as follows. In Sec.~\ref{Sec_eff-theory}, we establish
a low-energy effective field theory to describe the physical properties surrounding the QCP in Fig.~\ref{fig1}.
By adopting the standard procedure of RG framework, we within Sec.~\ref{Sec_RG}
derive the coupled RG equations of all interacting parameters. Section~\ref{Sec_FP}
presents a systematic analysis of the low-energy behavior of these interaction
parameters and identifies distinct fixed points as the system approaches four candidate QCPs.
In Sec.~\ref{Sec_T_c}, we carefully examine the effects of quantum fluctuations and competing
orders on the critical temperature of $d$-wave superconductor near the QCPs. Finally, we provide
a concise summary of our results in Sec.~\ref{Sec_summary}.

\section{Effective theory}\label{Sec_eff-theory}

\begin{figure}
\hspace{-0.68cm}
\includegraphics[width=3.5in]{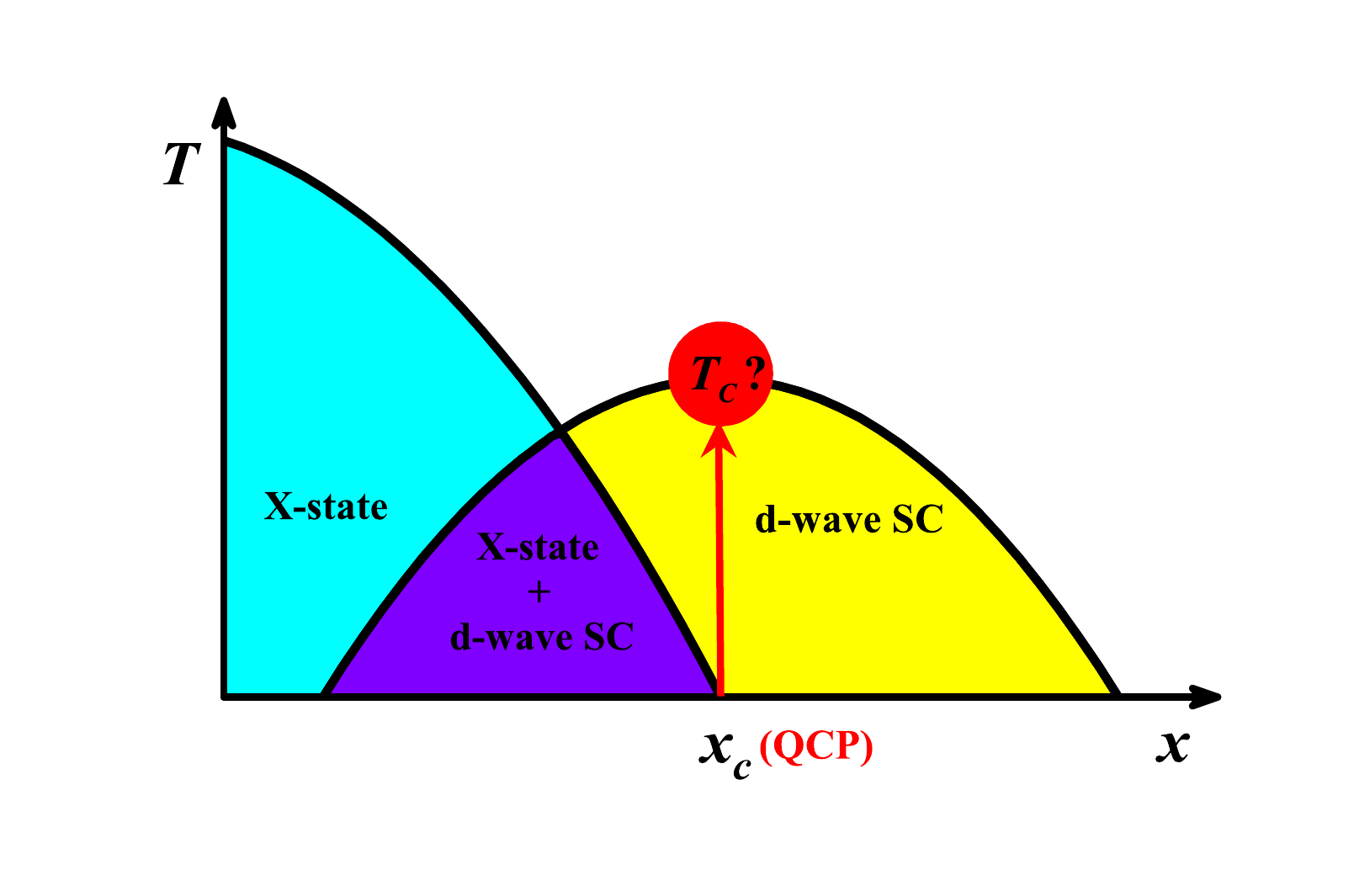}
\vspace{-0.9cm}
\caption{(Color online) Schematic illustration for an underlying quantum phase
transition (QPT) between the $d$-wave superconducting (SC) state and
an $X$ state with certain symmetry breaking beneath the SC dome via tuning the nonthermal doping variable~\cite{Vojta2000PRL,Vojta2000PRB,Vojta2000IJMPB,Sachdev2011PT}. Here, $T_c$ represents the critical temperature
that is dependent upon the doping, while the point $x_c$ denotes the quantum critical point (QCP) near the
optimal doping, which separates the ordered $X$ state plus $d-$wave SC
and the pure $d$-wave SC state. Within the quantum critical region above the QCP, the critical behavior
would be expected owing to the strong quantum fluctuations induced by the QPT, which will
be presented in Sec.~\ref{Sec_T_c}.}
\label{fig1}
\end{figure}

We within this work put our focus on the critical behavior around the quantum critical point (QCP) in the
superconducting dome of $d$-wave superconductor.  As schematically illustrated in Fig.~\ref{fig1}, there exist distinct
kinds of degrees of freedom that interact with each other in the vicinity of the QCP, including the gapless fermionic quasiparticles
(QPs) and the fluctuations of order parameters as well as the disorder scatterings.

In order to capture all these ingredients,
we arrive at the following phenomenological effective model,
\begin{eqnarray}
S^\star_{\mathrm{eff}}=S_{\mathrm{eff}}+S_{\mathrm{dis}}.\label{Eq_S-phenomen}
\end{eqnarray}
where $S_{\mathrm{eff}}$ collects the contributions from fermionic QPs as well as $X$-state and superconducting order parameter,
and $S_{\mathrm{dis}}$ represents the disorder scatterings. In what follows, we explicitly present
each of these physical components in detail.

\subsection{Fermionic quasiparticles}\label{subsec_phi_SC}

To be specific, the fermionic sector takes the form of~\cite{Sachdev2008PRB},
\begin{eqnarray}
S_{\Psi}\!\!&=&\!\!\int\!\!\frac{d^{2}\mathbf{k}}{(2\pi)^{2}}\frac{d\omega}{2\pi}\Psi^{\dag}
_{1a}(-i\omega\!+\!v_{F}k_{x}\tau_{z}\!+\!v_{\Delta}k_{y}\tau_{x})\Psi_{1a}\nonumber\\
&&\!\!\!\!\!\!+\!\!\int\!\!\frac{d^{2}\mathbf{k}}{(2\pi)^{2}}\frac{d\omega}{2\pi}\Psi^{\dag}
_{2a}(-i\omega\!+\!v_{F}k_{y}\tau_{z}\!\!+\!v_{\Delta}k_{x}\tau_{x})\Psi_{2a},\label{Eq_Psi}
\end{eqnarray}
with $\tau_{x,y,z}$ being Pauli matrices. Here, $\Psi^\dagger_{1a}$ denotes two gapless QPs excited from the nodes at $(\frac{\pi}{4}$,$\frac{\pi}{4})$ and $(-\frac{\pi}{4}$,$-\frac{\pi}{4})$, while $\Psi^\dagger_{2a}$ describes the
QPs from the other two nodes~\cite{Vojta2000PRL,Vojta2000PRB,Vojta2000IJMPB,Sachdev2008PRB}.
In addittion, $k_{x,y}$ represent two mutually perpendicular momenta,
and $v_{F,\Delta}$ specify the Fermi velocity
and the gap velocity, respectively.
To proceed, we are going to present the remaining parts
in Eq.~(\ref{Eq_S-phenomen}) one by one in the following sections.

\subsection{Fluctuations of order parameters}\label{subsec_phi_SC}

As depicted in Fig.~\ref{fig1}, the superconducting dome is separated into two distinct regions owing to the QPT.
The right-hand side corresponds to the $d$-wave superconducting (SC) state. In comparison, the $X$-state order parameter emerges
on the left-hand side accompanied by certain symmetry breaking. In the vicinity of the QCP, the fluctuations of order parameters
can bring important contributions to the critical behavior.

At first, let us consider the fluctuation of the $X$ state denoted by $\phi$.
In principle, there exist four distinct kinds of $X$ states due to different symmetry breakings~\cite{Vojta2000PRL,Vojta2000PRB,Vojta2000IJMPB}, which are helpful
to cluster the four sorts of potential QCPs at $x_c$, namely $\tau_{x,y,z}$-QCP and $\tau_{0}$-QCP with $\tau_0$ denoting the $2 \times 2$ identity matrix. These fluctuations can interact with the fermionic QPs near QCPs~\cite{Vojta2000PRL,Vojta2000PRB,Vojta2000IJMPB,RZW2022NPB},
\begin{eqnarray}
S_{\Psi\phi}=
\begin{cases}
\int d^{2}\mathbf{x}d\tau[\lambda_{0}\phi(\Psi^{\dag}_{1}\mathcal{M}_{1}\Psi_{1}+\Psi^{\dag}_{2}\mathcal{M}_{2}\Psi_{2})],\\
\int d^{2}\mathbf{x}d\tau[\lambda_{0}(\phi_{0A}\Psi^{\dag}_{1}\mathcal{M}_{1}\Psi_{1}+\phi_{0B}\Psi^{\dag}_{2}\mathcal{M}_{2}\Psi_{2})],
\end{cases}
\label{Eq_S_Psi_phi}
\end{eqnarray}
with $\lambda_0$ specifying the coupling strength. Here, $\phi$ together with $\phi_{0A,0B}$ denote the $X$-state fluctuations and the matrices $\mathcal{M}_{1,2}$ characterize unique features of related QCPs, namely $\mathcal{M}_{1,2}=\tau_{x},\tau_{y},\tau_{z}$ and $\mathcal{M}_{1,2}=\tau_0$ corresponding to the $\tau_{x,y,z}$-QCP and $\tau_{0}$-QCP that owns  two components $\phi_{0A}$ and $\phi_{0B}$, respectively~\cite{Vojta2000PRL,Vojta2000PRB,Vojta2000IJMPB}.

As to the free part of the $X$ state, it generally can be written as follows,
\begin{eqnarray}
S_{\phi}&=&\frac{1}{2}\int\frac{d^{3}q}{(2\pi)^{3}}(-2r+q^{2})\phi^{2},\label{Eq_S-phi0}
\end{eqnarray}
where $\phi^2$ are replaced by $(\phi_{0A}^2+\phi_{0B}^2)$ for the $\tau_0$-QCP and
the parameter $r$ measures the distance away from the QCP. It is worth highlighting
that one-loop corrections due to the interplay $S_{\Psi\phi}$ can qualitatively modify
such a free part. As shown in Refs.~\cite{Sachdev2008PRB,Wang2013PRB,Wang2013NJP},
one-loop polarization of the order parameter of the $X$ state can be
expressed as
\begin{eqnarray}
\Pi^{x}(\mathbf{q})&=&\frac{1}{16v_Fv_\Delta}\frac{\epsilon^2+v^2_F q_x^2}
{\sqrt{\epsilon^2+v^2_F q_x^2+v^2_\Delta q_y^2}}+(q_x\leftrightarrow q_y),\\
\Pi^{y}(\mathbf{q})&=&\frac{1}{16v_Fv_\Delta}\sqrt{\epsilon^2+v^2_F q_x^2+v^2_\Delta q_y^2}+(q_x\leftrightarrow q_y),\\
\Pi^{z}(\mathbf{q})&=&\frac{1}{16v_Fv_\Delta}\frac{\epsilon^2+v^2_\Delta q_y^2}
{\sqrt{\epsilon^2+v^2_F q_x^2+v^2_\Delta q_y^2}}+(q_x\leftrightarrow q_y),
\end{eqnarray}
and
\begin{eqnarray}
\Pi^{0}_A(q_x,q_y)&=&-\frac{1}{16v_Fv_\Delta}\sqrt{\epsilon^2+v^2_F q_x^2+v^2_\Delta q_y^2},\\
\Pi^{0}_B(q_y,q_x)&=&-\frac{1}{16v_Fv_\Delta}\sqrt{\epsilon^2+v^2_F q_y^2+v^2_\Delta q_x^2}.
\end{eqnarray}
It can be found that $\Pi^{i}(q)$ with $i=0,x,y,z$ is proportional to $q$ and hence it dominates over
the dynamic term with a quadratic $q$ in the low-energy region.
In addition, the parameter $r$ is negligible as we primarily focus on the physics near the QCP
shown in Fig.~\ref{fig1}. Accordingly, the low-energy ``free term" of the $X$ state can be recast into
\begin{eqnarray}
S_{\phi}&=&\frac{1}{2}\int\frac{d^{3}q}{(2\pi)^{3}}[\Pi^{i}(q)]\phi^{2},\label{Eq_S-phi}
\end{eqnarray}
which yields the renormalized propagator of the $X$-state order parameter,
\begin{eqnarray}
G_{\phi}(q)=\frac{1}{\Pi^{i}(q)}.\label{Eq_G_phi}
\end{eqnarray}
with $i=\tau_{x,y,z,0}$ for distinct kinds of QCPs.

Next, we move to consider the fluctuation of the SC order parameter. Generally, the competition between SC and
$X$-state order parameters takes the form of~\cite{Wang2013NJP,Wang2014PRD,Kleinert2003NPB}
\begin{eqnarray}
\mathcal{L}_{\mathrm{compe}}&=&\mathcal{L}_\psi+\mathcal{L}_{\phi}+\mathcal{L}_{\psi\phi},\label{Eq_L_compe}
\end{eqnarray}
with
\begin{eqnarray}
\mathcal{L}_\psi&=&\partial_\mu\psi^\dagger\partial_\mu\psi-\alpha|\psi|^2+\frac{\beta}{2}|\psi|^4,\\
\mathcal{L}_{\phi}&=&\frac{1}{2}(\partial_\mu\phi)^2-2r\phi^2+\frac{u}{2}\phi^4,\\
\mathcal{L}_{\psi\phi}&=&\lambda|\psi|^2\phi^2,\label{Eq_L_psi_phi}
\end{eqnarray}
where $\psi$ describes the SC order parameter and $\alpha,\beta,\gamma,u$ are the related parameters.
As the QCP shown in Fig.~\ref{fig1} is approached, it is of particular importance to point out that $\psi$
acquires a finite vacuum expectation value due to vacuum degeneracy~\cite{Kleinert2003NPB},
\begin{eqnarray}
\langle\psi\rangle\equiv V_0=\sqrt{\frac{\alpha}{\beta}}.
\end{eqnarray}
In order to capture the SC fluctuation around such an expectation value, we introduce two gapless field
$h$ and $\eta$ to specify the amplitude fluctuation and the phase fluctuation, respectively. Accordingly, the $\psi$ can be reexpressed as~\cite{Kleinert2003NPB}
\begin{eqnarray}
\psi=V_0+\frac{1}{\sqrt{2}}(h+i\eta).\label{Eq_psi_h_eta}
\end{eqnarray}
where $\langle h \rangle=\langle \eta \rangle=0$ and their free propagators can be obtained as
\begin{eqnarray}
G_h(k)&=&\frac{1}{k^2+2\alpha_h},\label{Eq_G_h}\\
G_\eta(k)&=&\frac{1}{k^2}\label{Eq_G_eta}.
\end{eqnarray}
Then, after inserting Eq.~(\ref{Eq_psi_h_eta}) into Eqs.~(\ref{Eq_L_compe})-(\ref{Eq_L_psi_phi}), as well
as combining Eq.~(\ref{Eq_Psi}) and Eq.~(\ref{Eq_S_Psi_phi}), and carrying out several
calculations and discarding the unimportant constant terms, we are finally left with the following~\cite{Wang2013NJP,Wang2014PRD,Kleinert2003NPB}
effective action,
\begin{eqnarray}
S_{\mathrm{eff}}&=&S_\Psi+S_\phi+S_h+S_\eta+S_{\Psi\phi}+S_{\eta ^2h}\nonumber\\
&&+S_{\phi^2h}+S_{h^2\eta^2}+S_{h^2\phi^2}+S_{\eta^2\phi^2}.\label{Eq-S-eff}
\end{eqnarray}
Here, $S_{\Psi}$, $S_{\phi}$, and $S_{h,\eta}$ correspond to the contributions from fermionic QPs, $X$-state, and superconducting
fluctuations, respectively. In addition, the other terms collect the interactions among these distinct kinds of
degrees of freedom. Specifically, they are written as
\begin{eqnarray}
S_{\phi}&=&\int\frac{d^2\mathbf{q}d\epsilon}{(2\pi)^3}[2\alpha_\phi+\Pi^i(\mathbf{q},\epsilon)]\phi^2\nonumber\\
    &&+\frac{\beta_\phi}{2}\int\prod_{m=1}^4\frac{d^2\mathbf{q}_md\epsilon_m}{(2\pi)^3}\mathcal{X}(m)\phi^4,\label{Eq-S-phi}\\
S_h&=&\int\frac{d^2\mathbf{k}d\omega}{(2\pi)^3}(2\alpha_h+\mathbf{k}^2+\omega^2)h^2\nonumber\\
    &&+\gamma_h\int\prod_{m=1}^3\frac{d^2\mathbf{k}_md\omega_m}{(2\pi)^3}\mathcal{X}(m)h^3\nonumber\\
    &&+\frac{\beta_h}{2}\int\prod_{m=1}^4\frac{d^2\mathbf{k}_md\omega_m}{(2\pi)^3}\mathcal{X}(m)h^4,\label{Eq-S-h}\\
S_\eta&=&\int\frac{d^2\mathbf{k}d\omega}{(2\pi)^3}(\mathbf{k}^2+\omega^2)\eta^2\nonumber\\
    &&+\frac{\beta_\eta}{2}\int\prod_{m=1}^4\frac{d^2\mathbf{k}_md\omega_m}{(2\pi)^3}\mathcal{X}(m)\eta^4,\label{Eq-S-eta}\\
S_{\eta^2h}&=&\gamma_{\eta h}\int\prod_{i=1,2}\frac{d^2\mathbf{k}_id\omega_i}{(2\pi)^3}\frac{d^2\mathbf{k}'d\omega'}{(2\pi)^3}\nonumber\\
	&&\times\eta(\mathbf{k}_i,\omega_i)h(\mathbf{k}',\omega')\mathcal{Y}^{(\mathbf{k},\mathbf{k}')}_{(\omega,\omega')},\label{Eq-S-etah}\\
S_{\phi^2h}&=&\gamma_{\phi h}\int\prod_{i=1,2}\frac{d^2\mathbf{q}_id\epsilon_i}{(2\pi)^3}\frac{d^2\mathbf{k}d\omega}{(2\pi)^3}\nonumber\\
	&&\times\phi(\mathbf{q}_i,\epsilon_i)h(\mathbf{k},\omega)\mathcal{Y}^{(\mathbf{q},\mathbf{k})}_{(\epsilon,\omega)},\label{Eq-S-phih}\\
S_{h^2\eta^2}&=&\lambda_{\eta h}\int\prod_{i=1,2}\frac{d^2\mathbf{k}_id\omega_id^2\mathbf{k}_i'd\omega_i'}{(2\pi)^6}\nonumber\\
    &&\times\eta(\mathbf{k}_i,\omega_i)h(\mathbf{k}_i',\omega_i')\mathcal{Z}^{(\mathbf{k},\mathbf{k}')}_{(\omega,\omega')},\label{Eq-S-heta}\\
S_{h^2\phi^2}&=&\lambda_{h\phi}\int\prod_{i=1,2}\frac{d^2\mathbf{k}_id\omega_id^2\mathbf{q}_id\epsilon_i}{(2\pi)^6}\nonumber\\
    &&\times h(\mathbf{k}_i,\omega_i)\phi(\mathbf{q}_i,\epsilon_i)\mathcal{Z}^{(\mathbf{k},\mathbf{q})}_{(\omega,\epsilon)},\label{Eq-S-hphi}\\
S_{\eta^2\phi^2}&=&\lambda_{\eta\phi}\int\prod_{i=1,2}\frac{d^2\mathbf{k}_id\omega_id^2\mathbf{q}_id\epsilon_i}{(2\pi)^6}\nonumber\\
    &&\times\eta(\mathbf{k}_i,\omega_i)\phi(\mathbf{q}_i,\epsilon_i)\mathcal{Z}^{(\mathbf{k},\mathbf{q})}_{(\omega,\epsilon)},\label{Eq-S-eta-phi}
\end{eqnarray}
where the related coefficients are designated as
\begin{eqnarray}
\alpha_h&=&\alpha,\quad \alpha_\phi=\lambda\alpha/\beta+r,\\[2mm]
\beta_h&=&\beta_\eta=\beta/4,\quad\beta_\phi=u,\\[2mm]
\gamma_h&=&\gamma_{\eta h}=\sqrt{2\alpha\beta}/2,\quad \gamma_{\phi h}=\lambda\sqrt{2\alpha/\beta},\\[2mm]
\lambda_{\eta h}&=&\beta/4,\quad \lambda_{h\phi}=\lambda_{\eta \phi}=\lambda/2,\\[2mm]
\mathcal{X}(m)&\equiv&\delta^2(\Sigma\mathbf{k}_m)\delta(\Sigma\omega_m),\\[2mm]
\mathcal{Y}^{(\mathbf{a},\mathbf{b})}_{(p,q)}&\equiv&\delta^2(\mathbf{a}_1\!+\!\mathbf{a}_2\!+\!\mathbf{b})
\delta(p_1\!+\!p_2\!+\!q),\\[2mm]
\mathcal{Z}^{(\mathbf{a},\mathbf{b})}_{(p,q)}&\equiv&\delta^2(\mathbf{a}_1\!+\!\mathbf{a}_2\!+\!\mathbf{b}_1
\!+\!\mathbf{b}_2)\delta(p_1\!+\!p_2\!+\!q_1\!+\!q_2).
\end{eqnarray}

\subsection{Disorder scatterings and total effective action}\label{subsec_dis_eff}

In real systems, disorder scatterings may also be of particular importance~\cite{Lee1985RMP,Mirlin2008RMP,Nersesyan1995NPB,Fiete2016PRB,Efremov11}. Without loss of generality, we
consider the disorder $A(\mathbf{x})$ a quneched, Gaussian white noise potential, which is defined by the following
correction functions~\cite{Nersesyan1995NPB},
\begin{eqnarray}
\langle A(\mathbf{x})\rangle&=&0,\\
\langle A(\mathbf{x_1})A(\mathbf{x_2})\rangle&=&g\delta^2(\mathbf{x_1}-\mathbf{x_2}),\label{Eq-dis-distribution}
\end{eqnarray}
where the parameter $g$ represents the concentration of impurity. It therefore yields the interplay between
fermionic QPs and disorder as~\cite{Nersesyan1995NPB,Stauber2005PRB}
\begin{eqnarray}
S_{\mathrm{dis}}&=&v_\Gamma\int d^2\mathbf{x}\Psi^\dagger(\mathbf{x})\Gamma\Psi(\mathbf{x})A(\mathbf{x}),\label{Eq-S-dis}
\end{eqnarray}
where $v_\Gamma$ measures the strength of fermion-disorder coupling and the matrix $\Gamma$ characterizes the feature of disorder.
Specifically, $\Gamma=\tau_0$, $\Gamma=\tau_y$, and $\Gamma=(\tau_x,\tau_z)$ correspond to a random chemical potential, a random mass,
and a random gauge potential, respectively~\cite{Nersesyan1995NPB,Stauber2005PRB,Wang2011PRB}.

With these in hand, we are going to start from the effective action~(\ref{Eq_S-phenomen})
and carry out the RG analysis armed with the
one-loop corrections as well as reveal the potential critical behavior induced by the QCP below
the SC dome in the forthcoming sections.

\section{RG analysis}\label{Sec_RG}

On the basis of the effective action~(\ref{Eq_S-phenomen}), we employ the Wilsonian momentum-shell RG method~\cite{Wilson1975RMP,Polchinski1992,Shankar1994RMP} to
establish the energy-dependent flows of all interaction
parameters as approaching the QCP. These evolutions carry the low-energy information
and are closely associated with the critical behavior.

Following the RG formalism~\cite{Wilson1975RMP,Polchinski1992,Shankar1994RMP},
we need to integrate out the high-energy modes (fast modes) of fields in the effective
theory~(\ref{Eq_S-phenomen}) to obtain the one-loop corrections arising from all the interactions.
Such fast modes are confined to a momentum shell $b\Lambda<k<\Lambda$
where $\Lambda$ represents an ultraviolet cutoff and $b$ is designated as $b=e^{-l}$ with $l$ being a running
scale that goes toward infinity at
the lowest energy~\cite{Shankar1994RMP,Kim-Kivelson2008PRB,Sachdev2008PRB,
She2010PRB,She2015PRB,Wang2011PRB,Wang2013PRB,Wang2014PRD,Wang2015PRB,Wang2017PRB,Wang2017QBCP,
Vafek2012PRB,Vafek2014PRB,Wang2022SST}. For convenience, we rescale the momenta and energy
by $\Lambda$, i.e., $k\rightarrow k/\Lambda$
and $\omega\rightarrow\omega=\omega/\Lambda$. Then, after lengthy but straightforward algebraic calculations,
we obtain the one-loop corrections, which are presented in the Appendix~\ref{Appendix_one-loop-corrections}.

In order to perform the RG analysis, we subsequently consider the RG rescaling transformations of
momenta, energy, and fields that connect successive steps of the RG procedure~\cite{Wilson1975RMP,Polchinski1992,Shankar1994RMP}.
For the fermionic and the $X$-state parts, we assume that
the $-i\omega$ term in $S_{\Psi}$~(\ref{Eq_Psi}) and the Yukawa coupling $\lambda_0$
in $S_{\Psi\phi}$~(\ref{Eq_S_Psi_phi}) remain invariant under the RG transformations~\cite{Shankar1994RMP,Sachdev2008PRB,Sachdev2011Book}.
In addition, the disorder distribution~(\ref{Eq-dis-distribution}) is also invariant~\cite{Wang2011PRB,Nersesyan1995NPB,Stauber2005PRB}. These together yield the following RG transformations,
\begin{eqnarray}
k&\rightarrow&k'e^{-l},\label{Eq_scaling_1}\\
\omega&\rightarrow&\omega'e^{-l},\\
\Psi_{1,2}(k,\omega)&\rightarrow&\Psi'_{1,2}(k',\omega')e^{\frac{1}{2}\int ^l_0 (4-\zeta_f)d l},\\
\phi(k,\omega)&\rightarrow&\phi'(k',\omega')e^{\frac{1}{2}\int ^l_0 (4-\zeta_b)d l},\\
A(k)&\rightarrow&A'(k')e^{l}.
\end{eqnarray}
Here, $\zeta_f$ and $\zeta_b$ serve as the anomalous dimensions that
are determined by one-loop corrections presented in the Appendix~\ref{Appendix_one-loop-corrections} and are expressed as,
\begin{eqnarray}
\zeta_f&=&(C_g-\lambda_0^2\mathcal{A}^i_1),
\end{eqnarray}
in tandem with
\begin{numcases}{\zeta_b=}
1+2\lambda_0^2(\mathcal{A}^i_1+\mathcal{R}^i),\!\! &
$\Gamma=\tau_0$\label{RG_zeta_0}\\
1+2\lambda_0^2(\mathcal{A}^i_1+\mathcal{R}^i)-4C_g,\!\! &
$\Gamma=\tau_y$\label{RG_zeta_y}\\
1+2\lambda_0^2(\mathcal{A}^i_1+\mathcal{R}^i)-2C_g,\!\! &
$\Gamma=\tau_x$\label{RG_zeta_x}\\
1+2\lambda_0^2(\mathcal{A}^i_1+\mathcal{R}^i)-2C_g,\!\! &
$\Gamma=\tau_z$\label{RG_zeta_z}
\end{numcases}
where $\Gamma=\tau_{0,y,x,z}$ corresponds to different kinds of disorders and
$C_g$ is defined as
\begin{eqnarray}
C_g&\equiv&\frac{gv_\Gamma^2}{2\pi v_Fv_\Delta}.
\end{eqnarray}
Here, the relevant coefficients $\mathcal{A}^i_1$ and $\mathcal{R}^i$ are
designated in the Appendix~\ref{Appendix_one-loop-corrections}, with the superscript $i=\tau_{x,y,z,0}$
denoting the $\tau_{x,y,z,0}-$QCP discussed in Sec.~\ref{subsec_phi_SC}.
Under these transformations, the field
$h$ and $\eta$ that characterize the SC part should transform accordingly so that the
free parts of actions $S_{h}$ and $S_{\eta}$ remain
unchanged. As a consequence, $h$ and $\eta$ can be rescaled as~\cite{Wang2014PRD,Wang2022SST,Wang2017PRB,Kleinert2003NPB}
\begin{eqnarray}
h(\mathbf{k},\omega) &\rightarrow& h'(\mathbf{k'},\omega')e^{\frac{5}{2}l},\\
\eta(\mathbf{k},\omega) &\rightarrow& \eta'(\mathbf{k'},\omega')e^{\frac{5}{2}l}.\label{Eq_scaling_2}
\end{eqnarray}

To proceed, we are able to derive the coupled RG flows of all interaction parameters that dictate the critical behaviors
around the QCP~\cite{Shankar1994RMP,Kim-Kivelson2008PRB,Sachdev2008PRB,
She2010PRB,She2015PRB,Wang2011PRB} by combining the above RG transformation scalings~(\ref{Eq_scaling_1})-(\ref{Eq_scaling_2}) with
all one-loop corrections to interaction parameters detailed in the Appendix~\ref{Appendix_one-loop-corrections}. For conciseness, we list the coupled RG equations for all interaction
parameters near the type $i$-QCP, with the superscripts $i=\tau_{x,y,z,0}$ respectively
denoting the $\tau_{x,y,z,0}-$QCP~\cite{Shankar1994RMP,Kim-Kivelson2008PRB,Sachdev2008PRB,
She2010PRB,She2015PRB,Wang2011PRB,Wang2013PRB,Wang2014PRD,Wang2015PRB,Wang2017PRB,Wang2017QBCP,
Vafek2012PRB,Vafek2014PRB,Wang2022SST},
\begin{widetext}
\begin{eqnarray}
\frac{d v_F}{dl}&=&[\lambda_0^2(\mathcal{A}^i_1-\mathcal{A}^i_2)-C_g]v_F,\label{RG_v_F}\\
 \frac{d v_\Delta}{dl}&=&[\lambda_0^2(\mathcal{A}^i_1-\mathcal{A}^i_3)-C_g]v_\Delta,\label{RG_v_Delta}\\
\frac{d \frac{v_\Delta}{v_F}}{d l}
&=&\lambda_0^2(\mathcal{A}^i_2-\mathcal{A}^i_3)(v_\Delta/v_F),\label{RG_v_Delta/v_F}\\
\frac{d\alpha_h}{dl}
&=&2\alpha_h+
\frac{1}{4\pi^2}\left[ -3\gamma_{h}^2-\gamma_{\eta h}^2
-\frac{64}{\pi}\gamma_{\phi h}^2v_F^2v_\Delta^2
\mathcal{F}^i_{2}(\theta,\varphi) +\frac{3}{4}\beta_h(1-2\alpha_h)+\frac{1}{2}\lambda_{\eta h}
+\frac{2v_Fv_\Delta}{\pi}\lambda_{h\phi}\mathcal{F}^i_1(\theta,\varphi) \right],\label{RG_alpha_h}\\
\frac{d\gamma_h}{dl}
&=&\frac{3}{2}\gamma_h+\frac{1}{4\pi^2}\left[8\gamma_{h}^3(1-6\alpha_h)+\gamma_h\beta_h(1-4\alpha_h)
+\lambda_{\eta h}\gamma_{\eta h}+\frac{\lambda_{h\phi}\gamma_{\phi h}}
{2\pi}(16v_Fv_\Delta)^2\mathcal{F}^i_{2}(\theta,\varphi)\right],\label{RG_gamma_h}\\
\frac{d\gamma_{\eta h}}{dl}
&=&\frac{3}{2}\gamma_{\eta h}+\frac{1}{4\pi^2}
\left[2\gamma_{\eta h}^3(1-2\alpha_h)
+\beta_{\eta}\gamma_{\eta h}+\frac{2\lambda_{\eta h}\gamma_{h}(1-4\alpha_h)
+\lambda_{\eta\phi}\gamma_{\phi h}}
{2\pi}(16v_Fv_\Delta)^2\mathcal{F}^i_{2}(\theta,\varphi)\right],\label{RG_gamma_etah}\\
\frac{d\gamma_{\phi h}}{dl}
&=&\frac{(1-2\zeta_b)}{2}\gamma_{\phi h}+\frac{1}{2\pi^2}\bigg[\frac{1}{4\pi}\gamma_{\phi h}^3(1-2\alpha_h)(16v_Fv_\Delta)^2
\mathcal{F}^i_{2}(\theta,\varphi)+
\lambda_{h\phi}\gamma_{h}(1-4\alpha_h)+\lambda_{\eta\phi}
\gamma_{\eta h}\nonumber\\
&&+\frac{\beta_{\phi}\gamma_{\phi h}}
{8\pi}(16v_Fv_\Delta)^2\mathcal{F}^i_{2}(\theta,\varphi)\bigg],\label{RG_gamma_phih}\\
\frac{d\beta_h}{dl}
&=&\beta_h+\frac{1}{2\pi^2}
\left[4\beta_h\gamma_{h}^2(1-6\alpha_h)+\frac{\beta^2_h}{2}(1-4\alpha_h)
+2\lambda_{\eta h}^2+\frac{\lambda_{h\phi}^2}
{2\pi}(16v_Fv_\Delta)^2\mathcal{F}^i_{2}(\theta,\varphi)\right],\label{RG_beta_h}\\
\frac{d\beta_\eta}{dl}
&=&\beta_\eta+\frac{1}{2\pi^2}\left[\beta_\eta\gamma_{\eta h}^2(1-2\alpha_h)+\frac{\beta_\eta^2}{2}
+2\lambda_{\eta h}^2(1-4\alpha_h)+\frac{\lambda_{h\phi}^2}
{2\pi}(16v_Fv_\Delta)^2\mathcal{F}^i_{2}(\theta,\varphi)\right],\label{RG_beta_eta}\\
\frac{d\beta_\phi}{dl}
&=&-(1+2\zeta_b)\beta_\phi+\frac{1}{2\pi^2}\bigg[
\frac{1}{4\pi}\beta_\phi\gamma_{\phi h}^2(1-2\alpha_h)(16v_Fv_\Delta)^2
\mathcal{F}^i_{2}(\theta,\varphi)+2\lambda_{\phi h}^2(1-4\alpha_h)\nonumber\\
&&+2\lambda_{\eta\phi}^2
+\frac{\beta^2_\phi}{8\pi}(16v_Fv_\Delta)^2\mathcal{F}^i_{2}(\theta,\varphi)
\bigg],\label{RG_beta_phi}\\
\frac{d\lambda_{\eta h}}{dl}
&=&\lambda_{\eta h}+\frac{1}{2\pi^2}\bigg[
4\lambda_{\eta h}\gamma_{h}^2(1-6\alpha_h)+
\lambda_{\eta h}\gamma_{\eta h}^2(1-2\alpha_h)
+\frac{1}{2}\lambda_{\eta h}\beta_{h}(1-4\alpha_h)+\frac{1}{2}\lambda_{\eta h}\beta_\eta
\nonumber\\
&&+\frac{1}{2}\lambda^2_{\eta h}(1-2\alpha_h)
+\frac{\lambda_{\eta\phi}\lambda_{h\phi}}
{4\pi}(16v_Fv_\Delta)^2\mathcal{F}^i_{2}(\theta,\varphi)\bigg],\label{RG_lambda_etah}\\
\frac{d\lambda_{h\phi}}{dl}
&=&-\zeta_b\lambda_{h\phi}+\frac{1}{2\pi^2}\bigg[
4\lambda_{h\phi}\gamma_{h}^2(1-6\alpha_h)
+\frac{1}{4\pi}\lambda_{h\phi}\gamma_{\phi h}^2(1-2\alpha_h)
(16v_Fv_\Delta)^2\mathcal{F}^i_2(\theta,\varphi)
+\lambda_{\eta h}\lambda_{\eta\phi}
+\frac{1}{2}\lambda_{h\phi}\beta_{h}(1-4\alpha_h)\nonumber\\
&&+\frac{\beta_\phi\lambda_{h\phi}}{8\pi}(16v_Fv_\Delta)^2\mathcal{F}^i_{2}(\theta,\varphi)
+\frac{1}{8\pi}\lambda^2_{h\phi}(1-2\alpha_h)
(16v_Fv_\Delta)\mathcal{F}^i_1(\theta,\varphi)\bigg],\label{RG_lambda_hphi}\\
\frac{d\lambda_{\eta\phi}}{dl}
&=&-\zeta_b\lambda_{\eta\phi}+\frac{1}{4\pi^2}\bigg[
2\lambda_{\eta\phi}\gamma_{\eta h}^2(1-2\alpha_h)
+\frac{1}{2\pi}\lambda_{\eta\phi}\gamma_{\phi h}^2(1-2\alpha_h)(16v_Fv_\Delta)^2
\mathcal{F}^i_{2}(\theta,\varphi)-2\lambda_{\eta\phi}\lambda_{h\phi}(1-4\alpha_h)
+\beta_{\eta}\lambda_{\eta\phi}\nonumber\\
&&+\frac{\beta_\phi\lambda_{\eta\phi}}{4\pi}(16v_Fv_\Delta)^2\mathcal{F}^i_{2}(\theta,\varphi)
+\frac{1}{4\pi}\lambda^2_{\eta\phi}(16v_Fv_\Delta)\mathcal{F}^i_1(\theta,\varphi)\bigg]\label{RG_lambda_etaphi},
\end{eqnarray}
\end{widetext}
in tandem with the flows of disorder couplings
\begin{numcases}{}
\frac{d v_\Gamma}{dl}=0,\!\!&
$\Gamma=\tau_0$\label{RG_Gamma_0}\\
\frac{dv_\Gamma}{dl}=(\lambda^2_0\mathcal{A}^i_3-\lambda^2_0\mathcal{A}^i_2-2C_g)v_\Gamma,\!\!&
$\Gamma=\tau_y$\label{RG_Gamma_y}\\
\frac{dv_\Gamma}{dl}=(\lambda_0^2\mathcal{A}^i_1-\lambda^2_0\mathcal{A}^i_3-C_g)v_\Gamma,\!\!&
$\Gamma=\tau_x$\label{RG_Gamma_x}\\
\frac{dv_\Gamma}{dl}=(\lambda_0^2\mathcal{A}^i_1-\lambda^2_0\mathcal{A}^i_2-C_g)v_\Gamma,\!\!&
$\Gamma=\tau_z$\label{RG_Gamma_z}
\end{numcases}
where $\Gamma=\tau_{0,y,x,z}$ again corresponds to distinct kinds of disorders,
and the related coefficients $\mathcal{A}^i_{1,2,3}$ and $\mathcal{F}^i_{1,2}$ with the superscripts $i=\tau_{x,y,z,0}$
denoting the $\tau_{x,y,z,0}-$QCP are provided in the Appendix~\ref{Appendix_one-loop-corrections}.
Here, it is necessary to stress that the Yukawa coupling $\lambda_0$ does not flow as it is chosen as
the free fixed point~\cite{Sachdev2008PRB}. In addition, $\alpha_\phi$ in Eq.~(\ref{Eq-S-phi}) is also considered to be a very small
constant in that the focus is put on the quantum critical regime
nearby the QCP~\cite{Vojta2000PRL,Vojta2000PRB,Vojta2000IJMPB,Sachdev2011PT,Vojta2003RPP}.

As the QCP is approached as illustrated in Fig.~\ref{fig1}, the interaction parameters couple intimately with
each other through these RG evolutions~(\ref{RG_v_F})-(\ref{RG_Gamma_z}).
On one hand, the basic tendencies of the interaction parameters can be dictated
by several fixed points as detailed in Sec.~\ref{Sec_FP} in the parameter space.
On the other hand, the critical behavior near these fixed points may give rise to significant
physical implications, such as the superfluid density and the critical temperature,
which are influenced by the strong quantum fluctuations induced by the QCP.
These will be addressed in the forthcoming Sec.~\ref{Sec_T_c}.

\begin{figure}
\hspace{-0.68cm}
\includegraphics[width=3.5in]{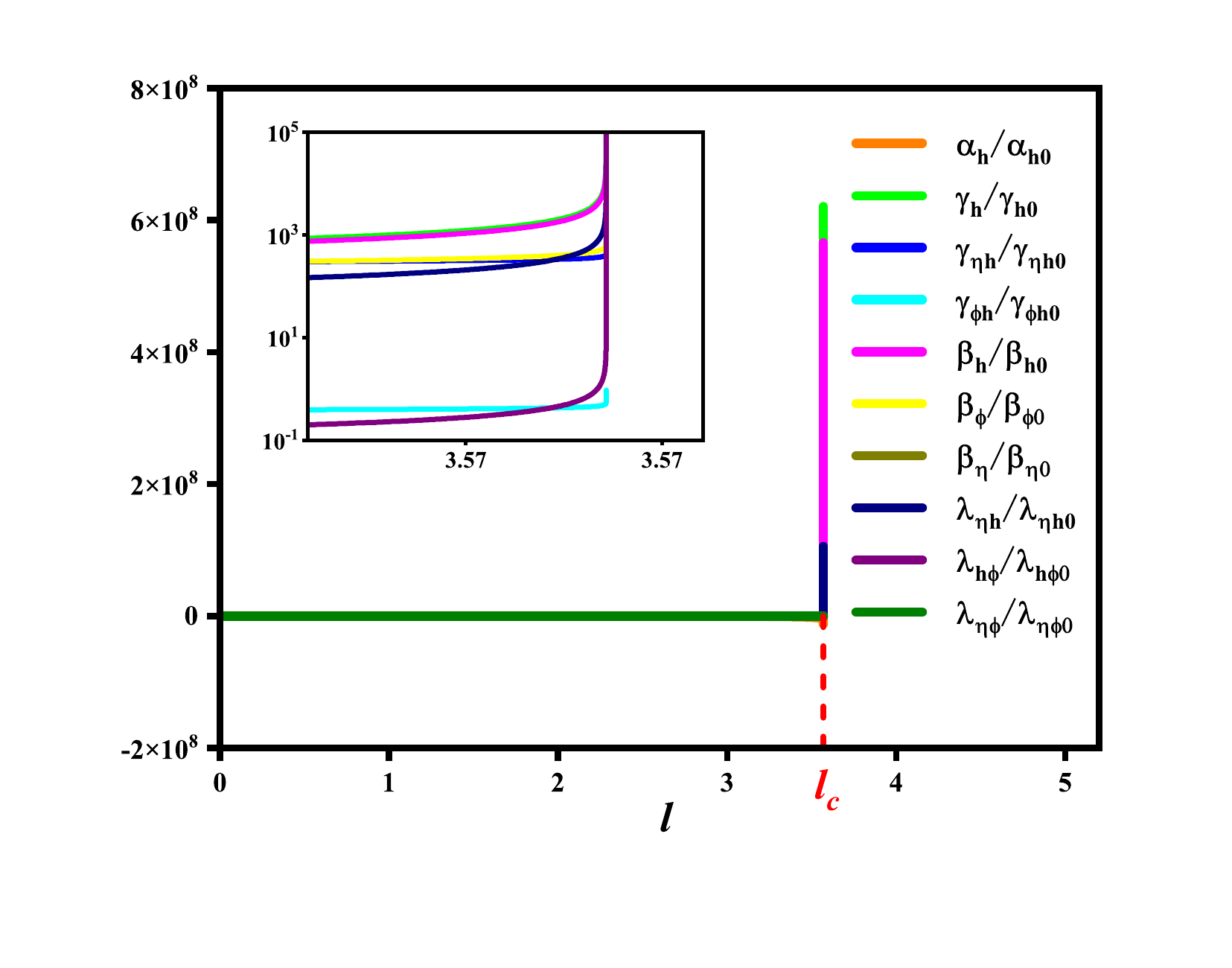}
\vspace{-0.9cm}
\caption{(Color online) Energy-dependent evolutions of interaction parameters for a representative initial condition around the
$\tau_x$-QCP (the basic results are analogous for other kinds of QCPs). Here, the critical energy scale is denoted by $l_c$.}
\label{fig_4_1}
\end{figure}

\begin{figure}
\centering
\subfigure[]{
\includegraphics[width=1.5in]{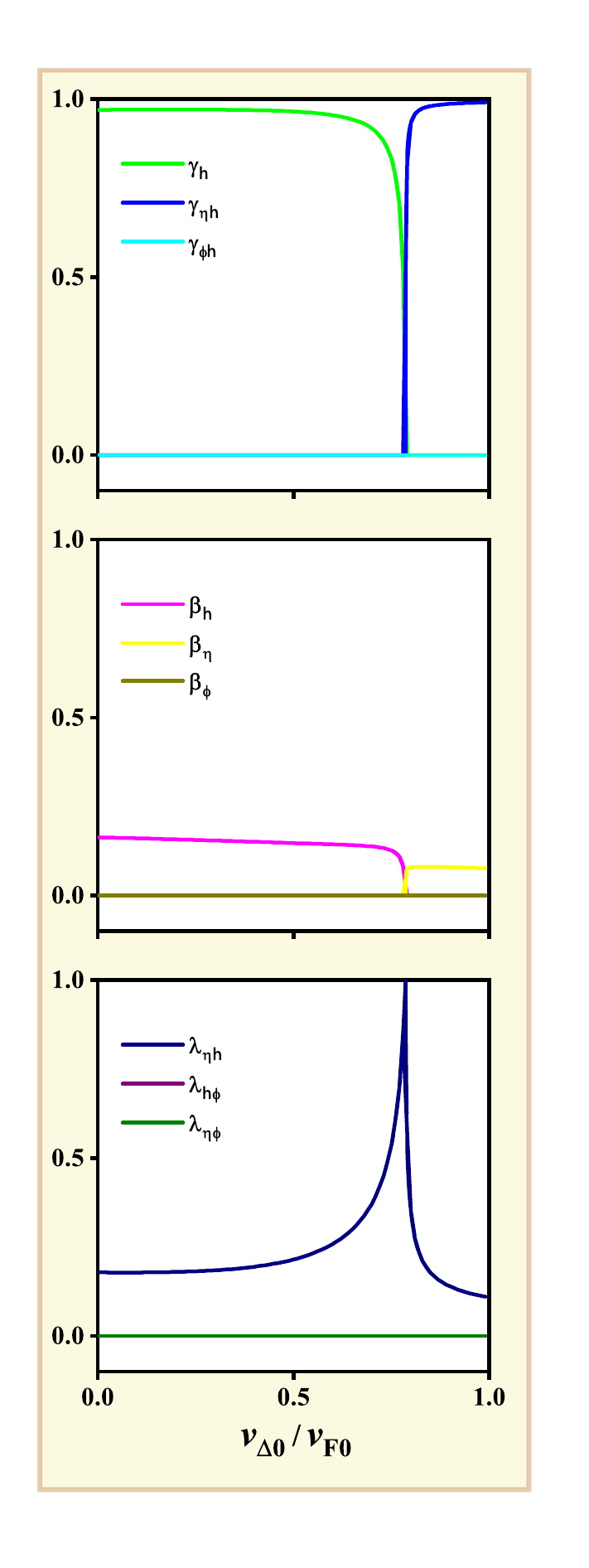}
}
\subfigure[]{
\includegraphics[width=1.5in]{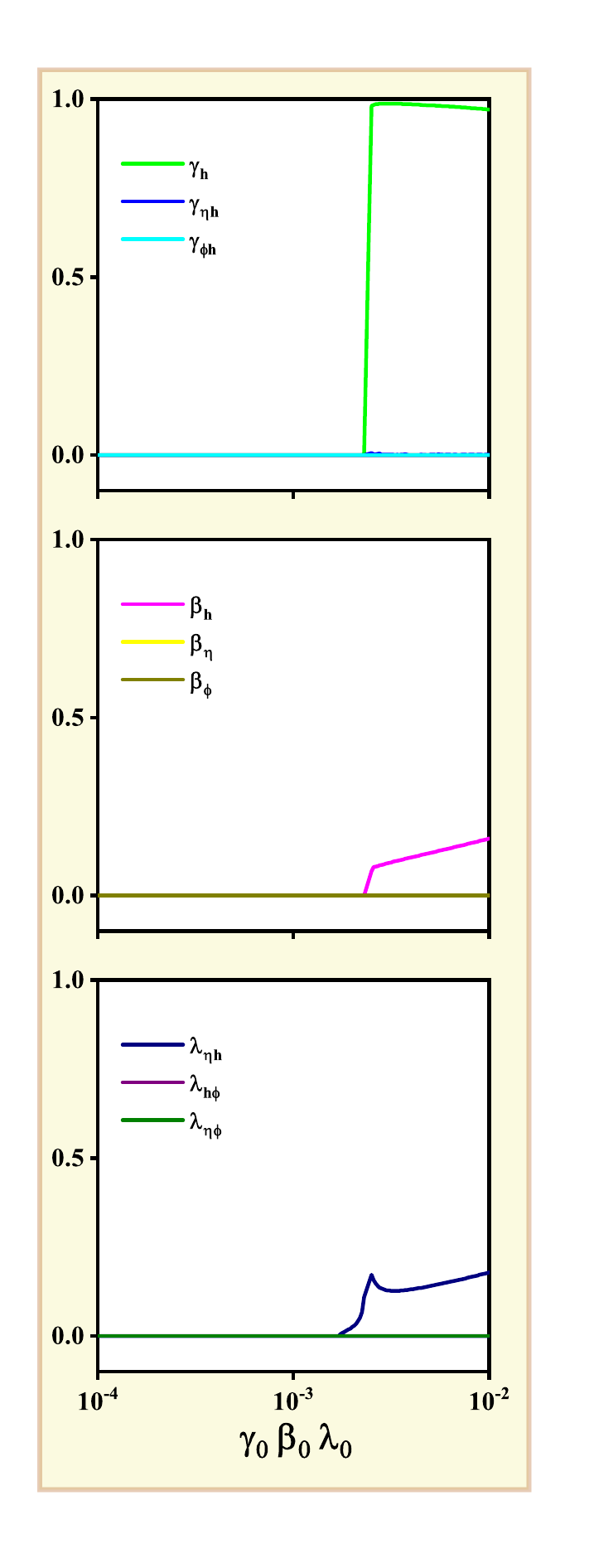}
}
\caption{(Color online) Evolutions of the concrete values of FPs with variations of initial conditions: (a) the fermion velocities
$v_{\Delta0} / v_{F0}$ and (b) the interaction parameters $\gamma_0 \beta_0 \lambda_0$.
The horizontal axis denotes the initial values of related parameters
and the vertical axis represent the values of interaction parameters at the FP with $l=l_c$.
With increasing $v_{\Delta0} / v_{F0}$, the FP-IIA can be driven to the FP-IID,  whereas decreasing $\gamma_0 \beta_0 \lambda_0$,
it would be replaced by FP-I. }\label{Fig_4_2_1}
\end{figure}

\section{Fixed points driven by the quantum critical points}\label{Sec_FP}

Starting from the coupled RG equations~(\ref{RG_v_F})-(\ref{RG_Gamma_z}) that encapsulate the ordering
competition among quantum fluctuations, we are now in a proper position to investigate the low-energy tendencies of all parameters
around the distinct kinds of QCPs.

\subsection{Fixed points and classification}\label{Subsec_eff_FP}

To investigate the energy-dependent tendencies of interaction parameters, we perform a numerical
analysis of the coupled RG flow equations~(\ref{RG_v_F})-(\ref{RG_Gamma_z}). For a representative initial condition for the
$\tau_x$-QCP, Fig.~\ref{fig_4_1} illustrates how the interaction parameters evolve with the running parameter $l$ and shows that some parameters diverge at a critical energy scale denoted by $l_c$. It is of particular importance to highlight that such divergent behavior of
interaction parameters is free of the initial conditions.

Principally, this implies that the effective theory remains valid only within the range $l\in[0,l_c)$, beyond
$l=l_c$ the RG equations are invalid and thus nonphysical behavior may appear. In this sense, the key point becomes how the interaction
parameters evolve as the system approaches such a point. Since the basic tendencies of the interaction parameters are dictated
by the point $l=l_c$, we can consider the evolutions in the interaction-parameter space and
designate the very point at $l=l_c$ as a fixed point (FP). The FPs generally characterize
the low-energy fates of interaction parameters and govern the low-energy critical behavior
~\cite{Maiti2010PRB,Vojta2003RPP,
Roy2018PRX,Wang2017QBCP,Vafek2012PRB,Vafek2014PRB,
Wang2020PRB,Chubukov2012ARCMP,Chubukov2016PRX,Nandkishore2012NP,Wang2020NPB}.

For the sake of completeness, we hereafter will revolve around the potential FPs and then uncover the distinct critical
properties in the vicinity of all potential FPs. After performing the detailed numerical calculations for all kinds of
QCPs, there exist two types of FPs depending upon the initial values of interaction parameters.
As to the Type-I FP, the system is dominated by the quadratic parameter $\alpha_h$. In comparison,
the Type-II FP is dictated by the cubic and quartic interaction parameters, namely
$\gamma_i$, $\beta_j$ and $\lambda_k$ with $i = h, \eta h, \phi h$, $j = h, \eta, \phi$
and $k = \eta h, \phi h, \eta \phi$. Under these circumstances, we introduce a concise
notation $(\alpha,\gamma,\beta,\lambda)$ to specify a FP, where $\alpha$, $\gamma$, $\beta$, and $\lambda$
are employed to denote the dominant quadratic, cubic, quartic self-interaction and
quartic competing parameters, respectively. In order to facilitate the analysis, one can
rescale all interaction parameters by certain dominant variable to
obtain a rescaled FP~\cite{Vafek2012PRB,Vafek2014PRB,Roy2018PRX,Wang2017QBCP,Maiti2010PRB,Chubukov2016PRX,
Nandkishore2012NP,Wang2020NPB,Wang2021NPB}, which consists of several relatively small coordinates.

Accordingly, the Type-I FP can be expressed as FP-I$(1,0,0,0)$ rescaled by the dominant quadratic parameter.
This rescaling renders it is insusceptible to the concrete initial values of interaction parameters. As for the Type-II FP, we can
rescale the $\gamma_i$, $\beta_j$ and $\lambda_k$ by $\kappa \equiv \sqrt{\sum_{ijk} (\gamma_i^2 + \beta_j^2 + \lambda_k^2)}$
and then make transformations $\gamma_i / \kappa \to \gamma_i$, $\beta_j / \kappa \to \beta_j$ and $\lambda_k / \kappa \to \lambda_k$.
Under this rescaling, the Type-II FP can be written as FP-II$(0,\gamma_i,\beta_j,\lambda_k)$.
It is of importance to point out that, unlike the Type-I FP, the Type-II FP is much more sensitive to the initial
conditions. The detailed numerical analysis suggests that there are five distinct situations, i.e.,
FP-IIA$(0,\gamma_h,\beta_h,\lambda_{\eta h})$, FP-IIB$(0,\gamma_h,\beta_h,\lambda_{\phi h})$, FP-IIC$(0,\gamma_h,\beta_\eta,\lambda_{\eta h})$, FP-IID$(0,\gamma_{\eta h},\beta_\eta,\lambda_{\eta h})$, and
FP-IIE$(0,\gamma_{\phi h},\beta_\phi,\lambda_{\phi h})$.

Compared to the Type-I FP, we would like to highlight that the concrete values for the five situations of type-II FP are relatively insensitive to the quadratic parameter but they are heavily
dependent upon the initial conditions. It is therefore necessary to
clarify this issue before studying the critical behavior.

\subsection{Influence of initial conditions on fixed points at clean limit}\label{Sec_FPs_clean}

In order to uncover the dependence of FPs on the initial conditions, it is essential to consider three key
sorts of parameters: the interaction couplings,
fermion velocities and disorder scatterings. For simplicity, we
categorize the starting points of interaction couplings into two cases, i.e., Case-Special in which
all nine interaction parameters are assigned the same initial value,
and Case-General in which the initial values of quadratic ($\gamma$), cubic ($\beta$), and quartic ($\lambda$)
interactions are independently specified.
In both cases, the initial values of fermion velocities $v_{F,\Delta}$ and disorder strength $v_\Gamma$
can also impose an effect on the FPs.

We put our focus on the clean limit in this subsection and
defer the disorder effects to the forthcoming subsection~\ref{Subsec_impurity}. At the clean limit,
we find that, as shown in Fig.~\ref{Fig_4_2_1} for the $\tau_x$-QCP (the basic results are analogous for other QCPs),
the critical values of interaction parameters at $l=l_c$ are
sensitive to the starting values of $v_{\Delta}/v_{F}$ and $\gamma,\beta,\lambda$.
In other words, tuning the initial conditions can drive one FP to transition into another.
Subsequently, we are going to provide a more detailed discussion of how FPs depend on initial conditions near all kinds of QCPs.

\subsubsection{Case-Special initial conditions}\label{Subsec_equal}

\begin{figure}
\centering
\subfigure[]{
\includegraphics[width=0.23\textwidth]{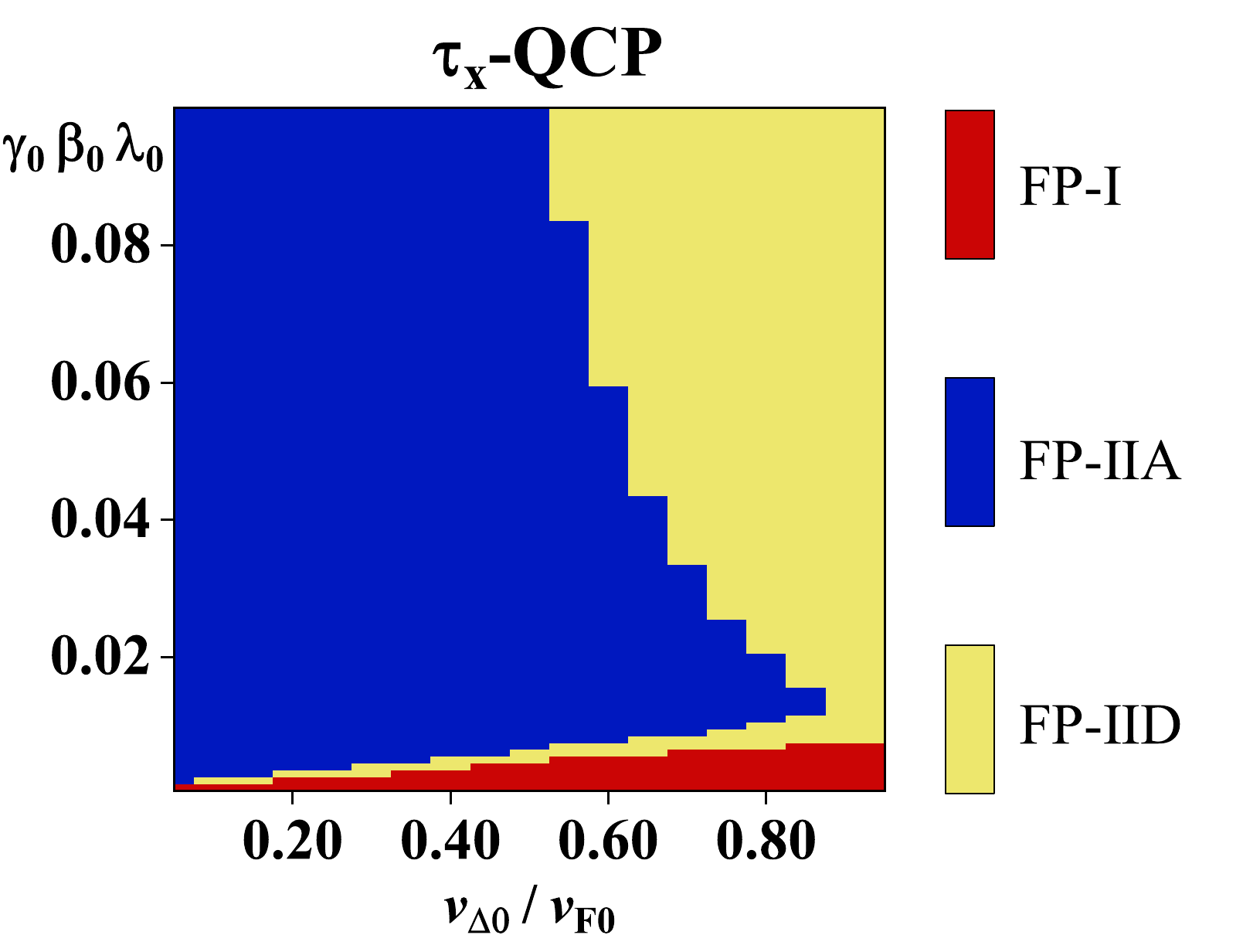}\label{Fig_4_2_2x}
}\quad\hspace{-5mm}
\subfigure[]{
\includegraphics[width=0.23\textwidth]{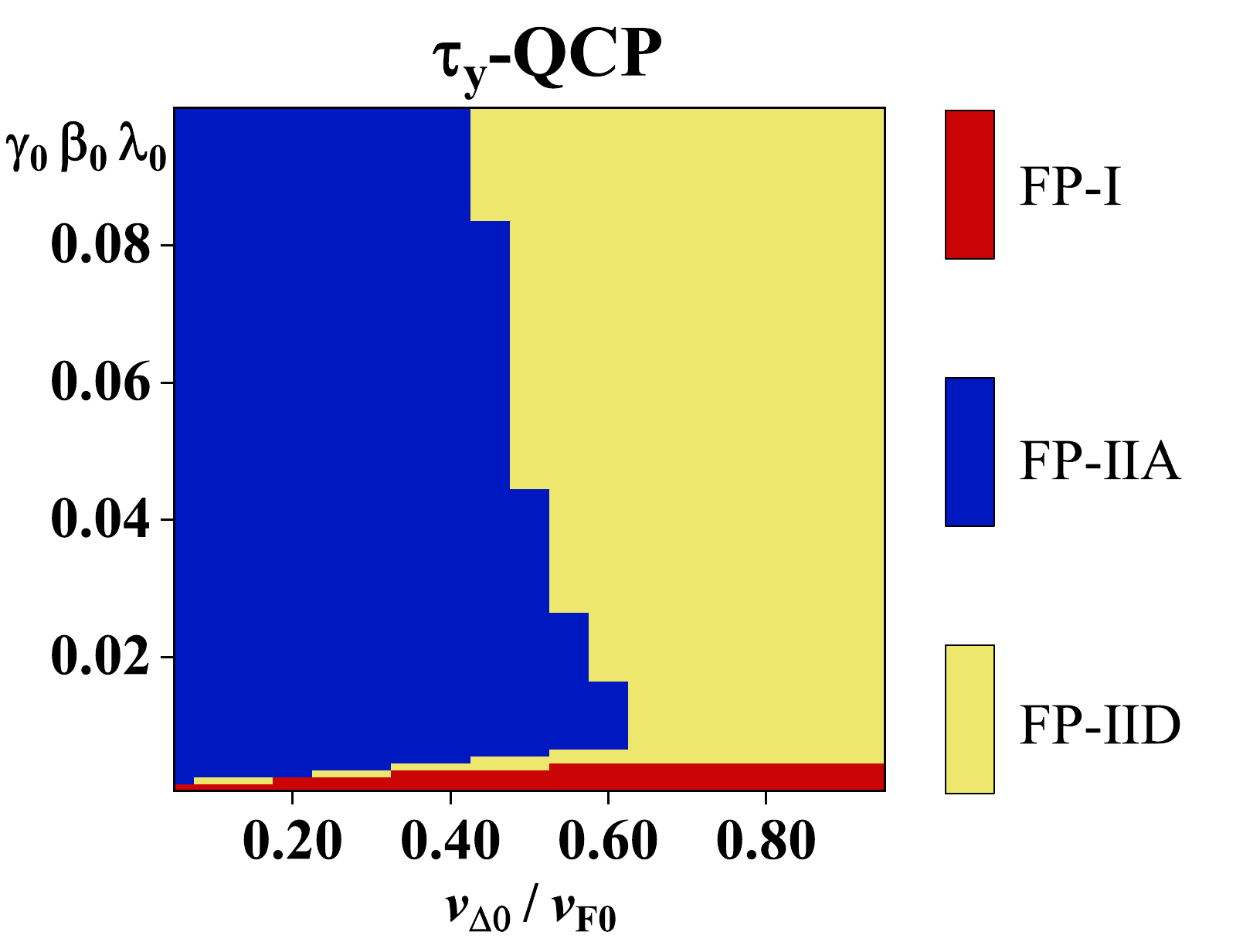}\label{Fig_4_2_2y}
}\quad
\subfigure[]{
\includegraphics[width=0.23\textwidth]{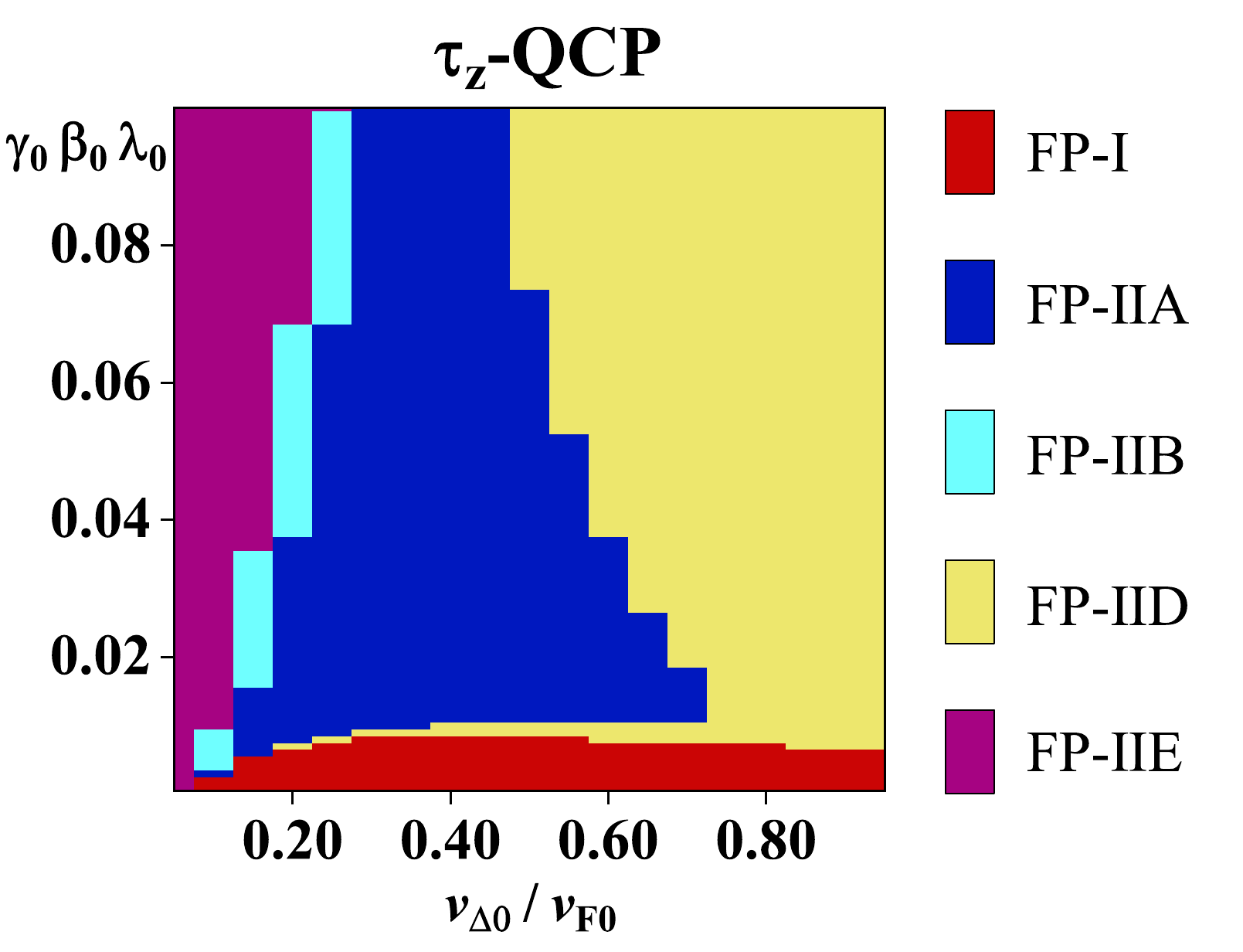}\label{Fig_4_2_2z}
}\quad\hspace{-5mm}
\subfigure[]{
\includegraphics[width=0.23\textwidth]{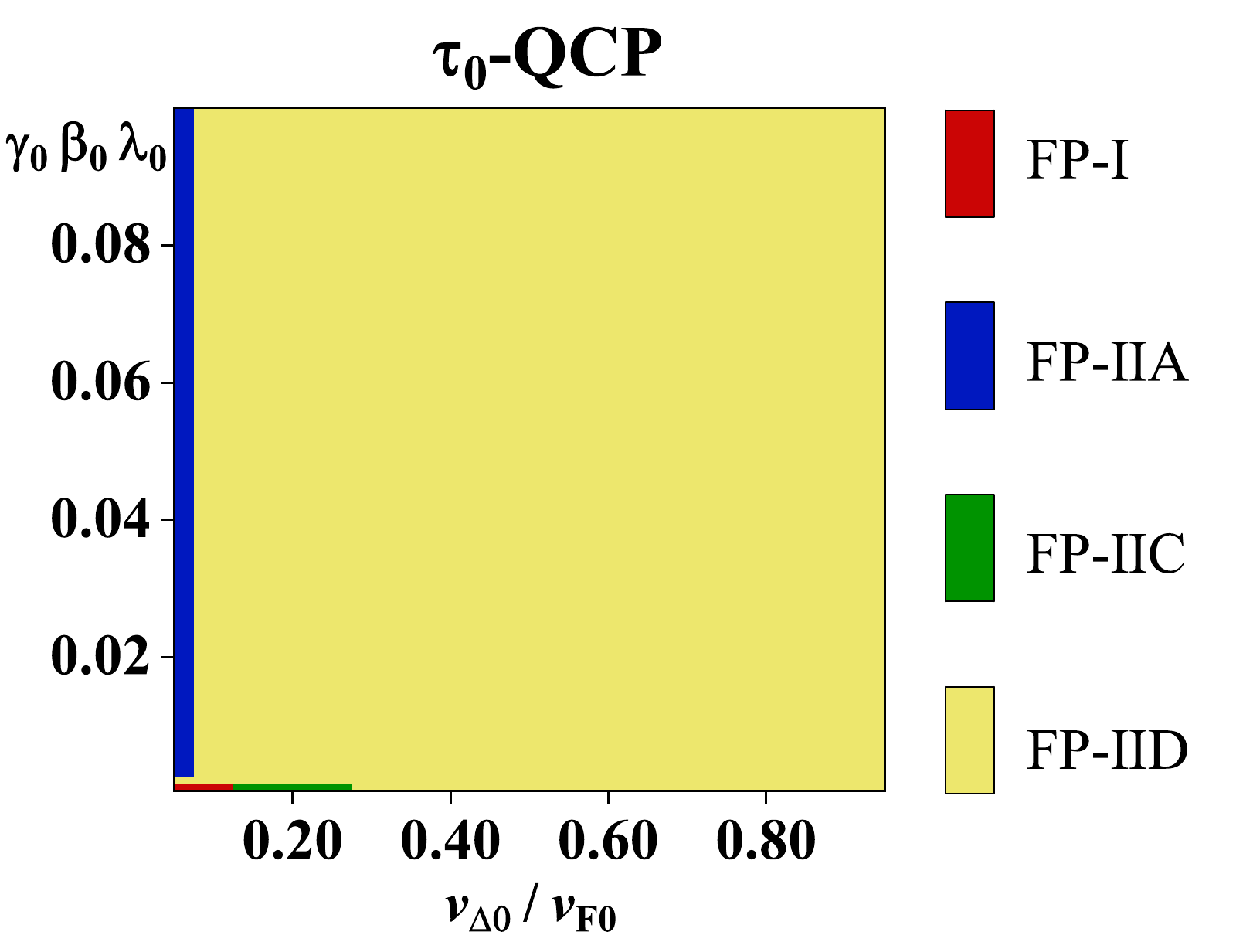}\label{Fig_4_2_20}
}\quad
\caption{(Color online) The dependence of FPs on the initial condition as the system approaches:
(a) $\tau_{x}$-QCP,  (b) $\tau_{y}$-QCP, (c) $\tau_{z}$-QCP, and  (d)$\tau_{0}$-QCP.
The horizontal axis and the vertical axis serve as the initial values of fermion velocities and
interaction parameters, respectively. }
\label{Fig_4_2_2_2D}
\end{figure}

\begin{figure}
\hspace{-0.68cm}
\includegraphics[width=3in]{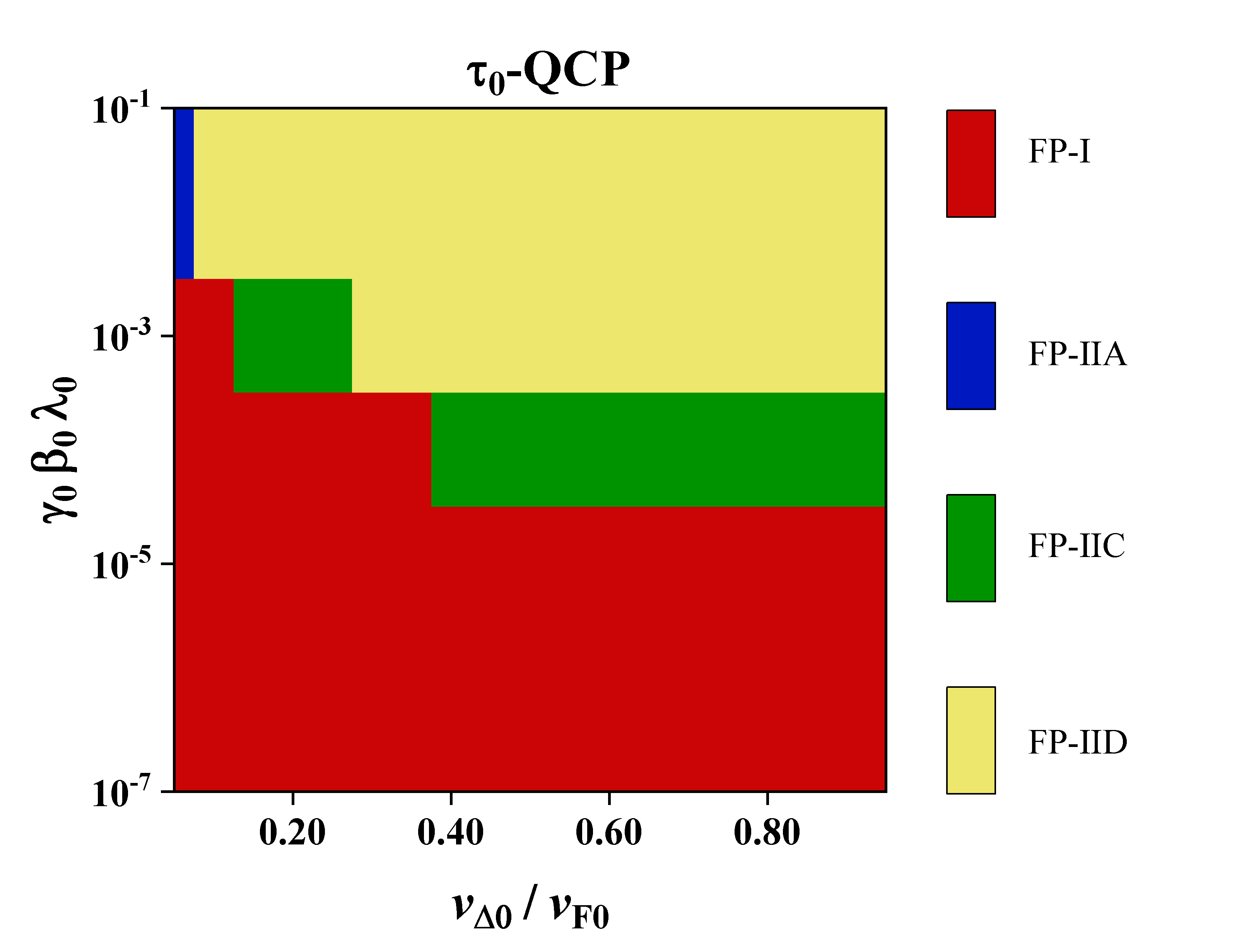}
\caption{(Color online) The dependence of FPs on the initial condition as the system approaches the $\tau_{0}$-QCP with
enlarging the small value regime of Fig.~\ref{Fig_4_2_2_2D}(d). The horizontal axis and the vertical
axis serve as the initial values of fermion velocities and
interaction parameters, respectively. }
\label{Fig_4_2_20log}
\end{figure}

At first, let us consider the Case-Special initial condition.
After performing the detailed numerical analysis, we notice that the initial values of $\gamma,\beta,\lambda$
($\gamma_0 \beta_0 \lambda_0$) are crucial to determine the fates of FPs.
When $\gamma_0 \beta_0 \lambda_0$ values are small enough, the system is always driven to the
Type-I FP (FP-I), where the quadratic parameter is dominant. The related results are  depicted in
Fig.~\ref{Fig_4_2_2_2D}(a)-(c) for $\gamma_0 \beta_0 \lambda_0 \textless 10^{-3}$
and Fig.~\ref{Fig_4_2_2_2D}(d)-Fig~\ref{Fig_4_2_20log} for $\gamma_0 \beta_0 \lambda_0 \textless 10^{-5}$
as approaching $\tau_{x,y,z}$-QCP and $\tau_{0}$-QCP, respectively.

However, with increasing $\gamma_0 \beta_0 \lambda_0$, both the interaction parameters and
fermion velocities play an important role and the system exhibits distinct FPs around different kinds of QCPs.
Considering the $\tau_{x,y}$-QCP, Fig.~\ref{Fig_4_2_2_2D}(a) and Fig.~\ref{Fig_4_2_2_2D}(b) indicate that
these two QCPs share analogous results. Specifically, in addition to the Type-I FP, the systems can flow towards
either FP-IIA or FP-IID. It can be found that the initial value of fermion velocity anisotropy $v_{\Delta0} / v_{F0}$
is essential to determine the final FP. Once $v_{\Delta0} / v_{F0}$ is small, FP-IIA dominates over FP-IID. Instead,
the system is always governed by FP-IID as long as $v_{\Delta0} / v_{F0}$ is appropriate.
For the $\tau_{z}$-QCP, as illustrated in Fig.~\ref{Fig_4_2_2z}, four candidate FPs, including
FP-IIA, FP-IIB, FP-IID, and FP-IIE, emerge with variations of $\gamma_0 \beta_0 \lambda_0$ and $v_{\Delta0} / v_{F0}$.
In particular, compared to the $\tau_{x,y}$-QCP, the strong fluctuation of order parameter around
$\tau_{z}$-QCP induces FP-IIB and FP-IIE in the small $v_{\Delta0} / v_{F0}$ region.
With respect to the $\tau_{0}$-QCP, it is clear from Fig.~\ref{Fig_4_2_2_2D}(d)
that FP-D predominates among all the FPs observed around the $\tau_{x,y,z}$ QCPs, with
FP-IIA appearing only at very small $v_{\Delta0} / v_{F0}$. After enlarging the region with the
small $\gamma_0 \beta_0 \lambda_0$ in Fig.~\ref{Fig_4_2_2_2D}(d),
it is of particular interest to stress that FP-IIC, which is absent in other QCPs,
develops within a confined region as depicted in Fig.~\ref{Fig_4_2_20log}.

\subsubsection{Case-General initial conditions}\label{Subsec_unequal}

\begin{figure*}
\centering
\subfigure[]{
\includegraphics[width=1.5in]{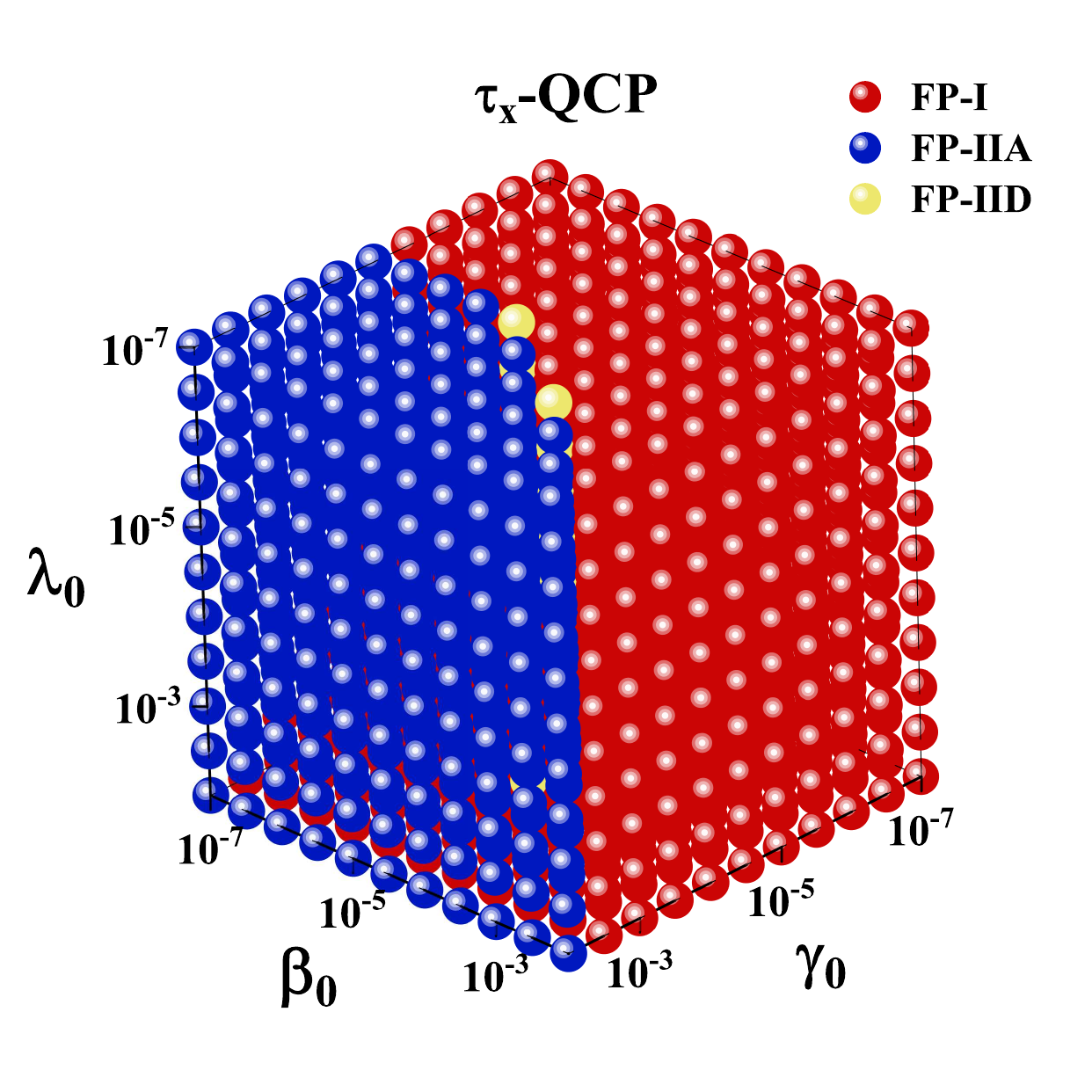}
}
\subfigure[]{
\includegraphics[width=1.5in]{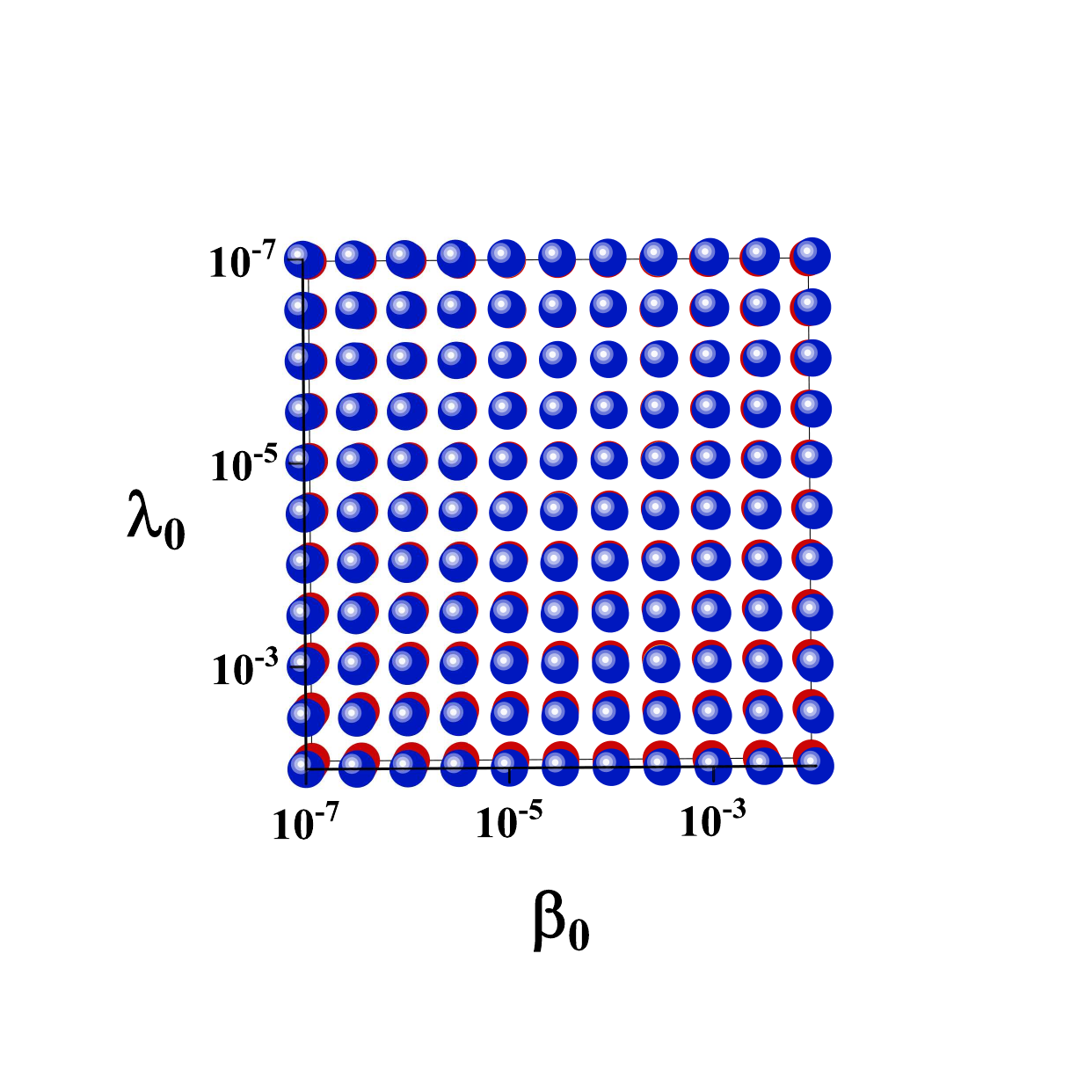}
}
\subfigure[]{
\includegraphics[width=1.5in]{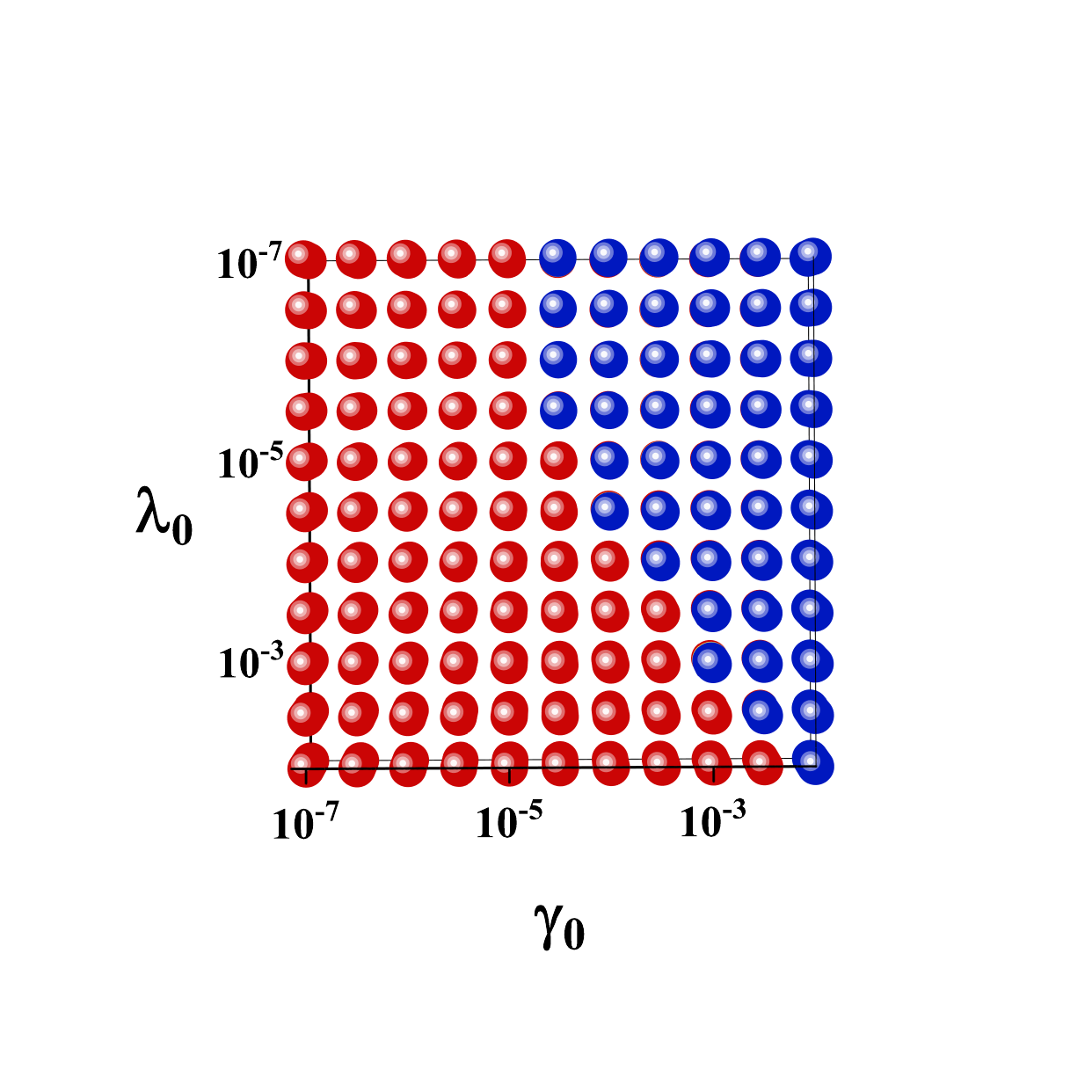}
}
\\
\subfigure[]{
\includegraphics[width=1.5in]{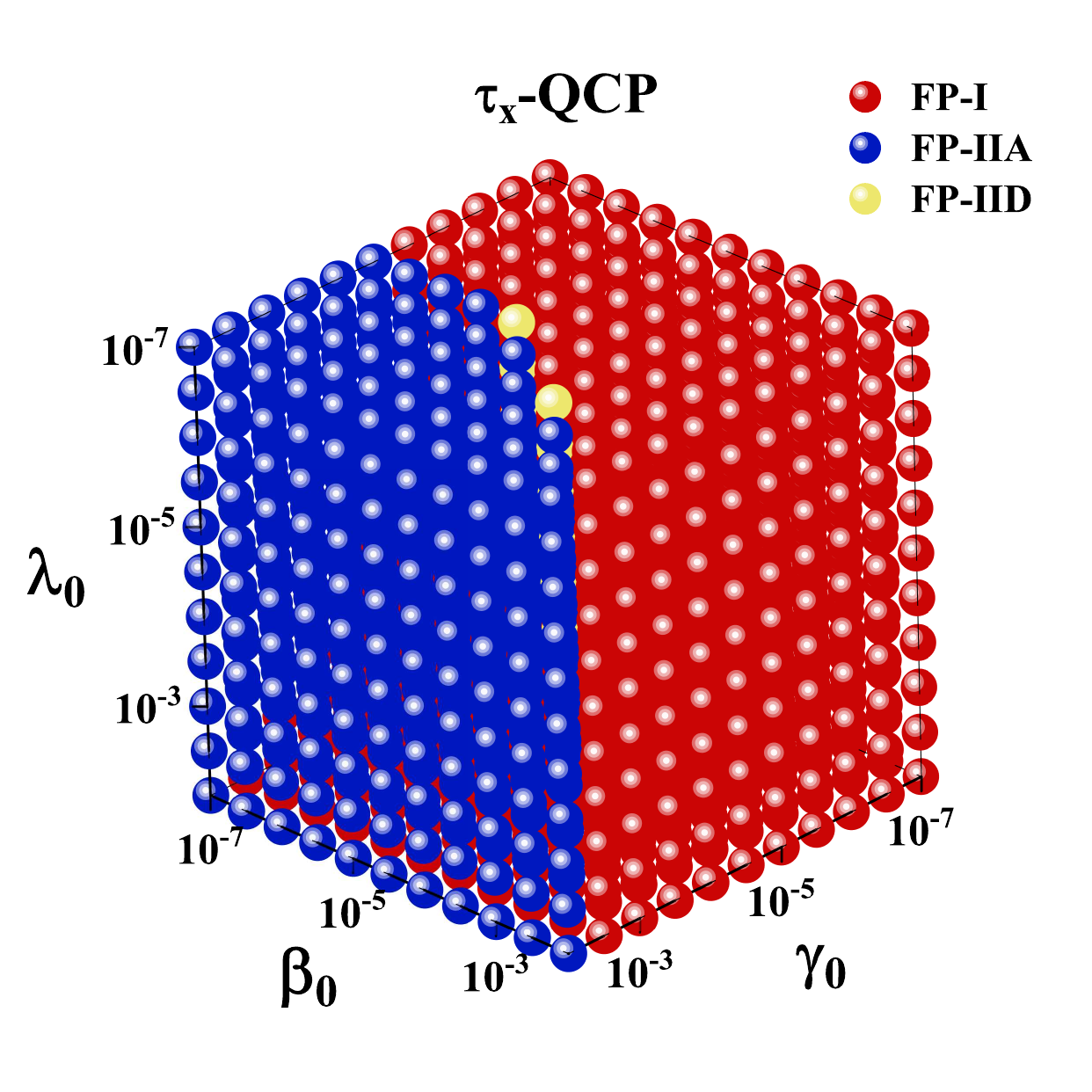}
}
\subfigure[]{
\includegraphics[width=1.5in]{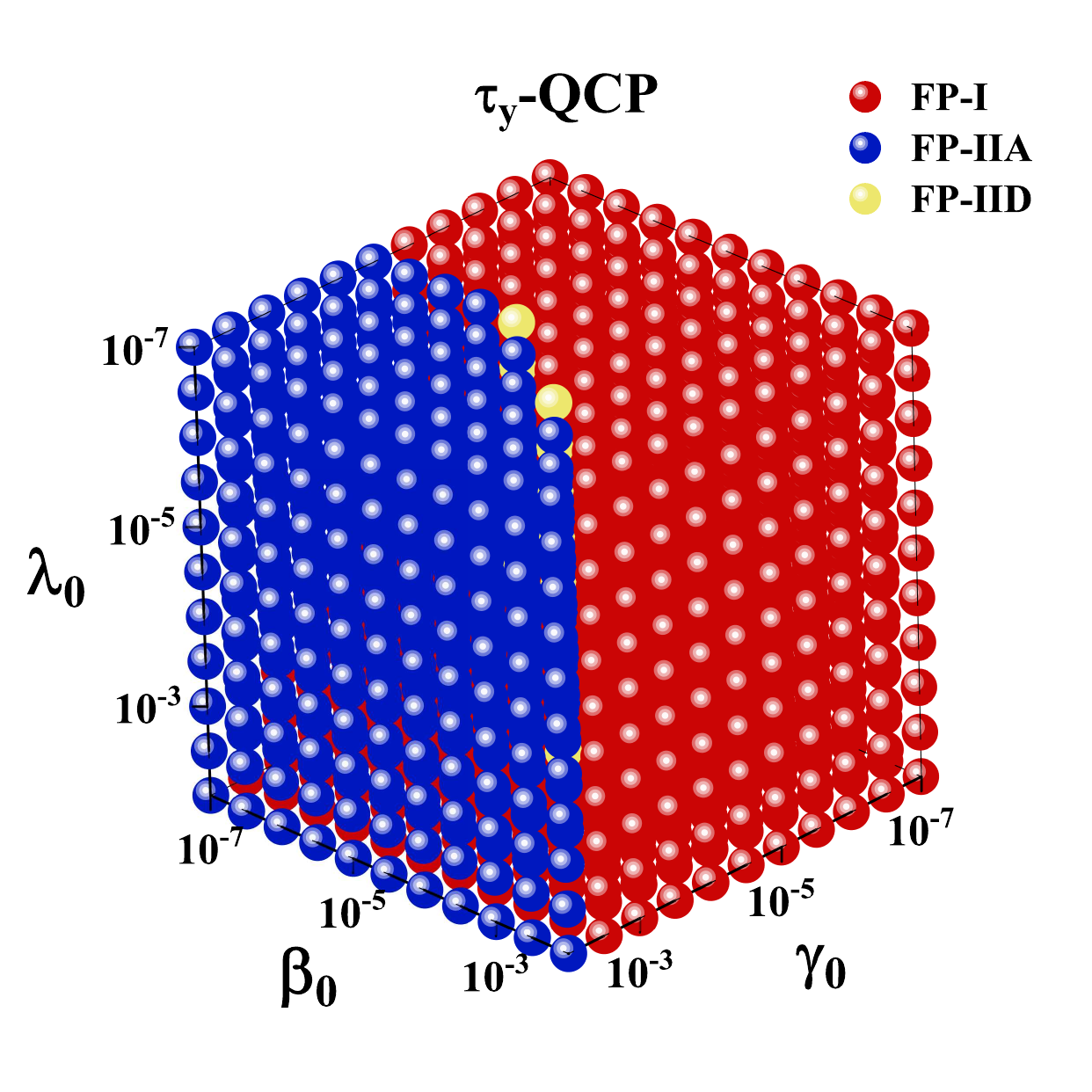}
}
\subfigure[]{
\includegraphics[width=1.5in]{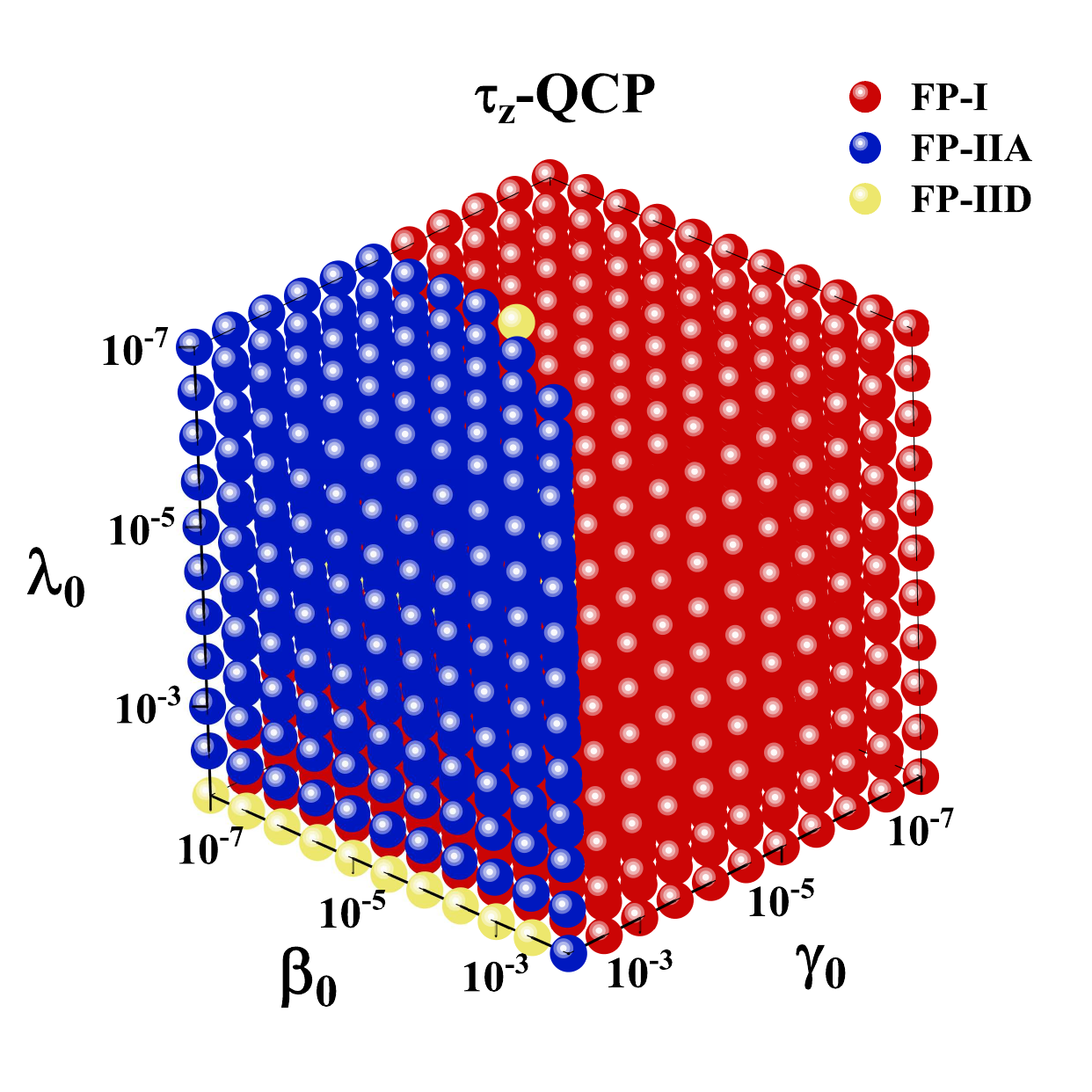}
}
\caption{(Color online) The dependence of FPs on three types of interaction parameters $\gamma$, $\beta$ and $\lambda$ for the Case-General as the system approaches: (a)-(c) the $\tau_{x}$-QCP with $v_{\Delta 0} / v_{F0} = 0.2$, (d) the $\tau_x$-QCP with $v_{\Delta 0} / v_{F0} = 0.05$, (e) the $\tau_y$-QCP with $v_{\Delta 0} / v_{F0} = 0.5$, and (f) the $\tau_z$-QCP with $v_{\Delta 0} / v_{F0} = 0.8$. Hereby, (b) and (c) correspond to the projections of (a) onto the $\beta_0-\lambda_0$ plane and the $\gamma_0-\lambda_0$ plane with a fixed $\gamma_0$ and $\beta_0$, respectively. These two panels indicate that the interaction parameters
$\gamma$ and $\beta$ play a much more significant role than $\lambda$ in determining the FPs.}
\label{Fig_4_3_1}
\end{figure*}

\begin{figure}
\centering
\subfigure[]{
\includegraphics[width=1.5in]{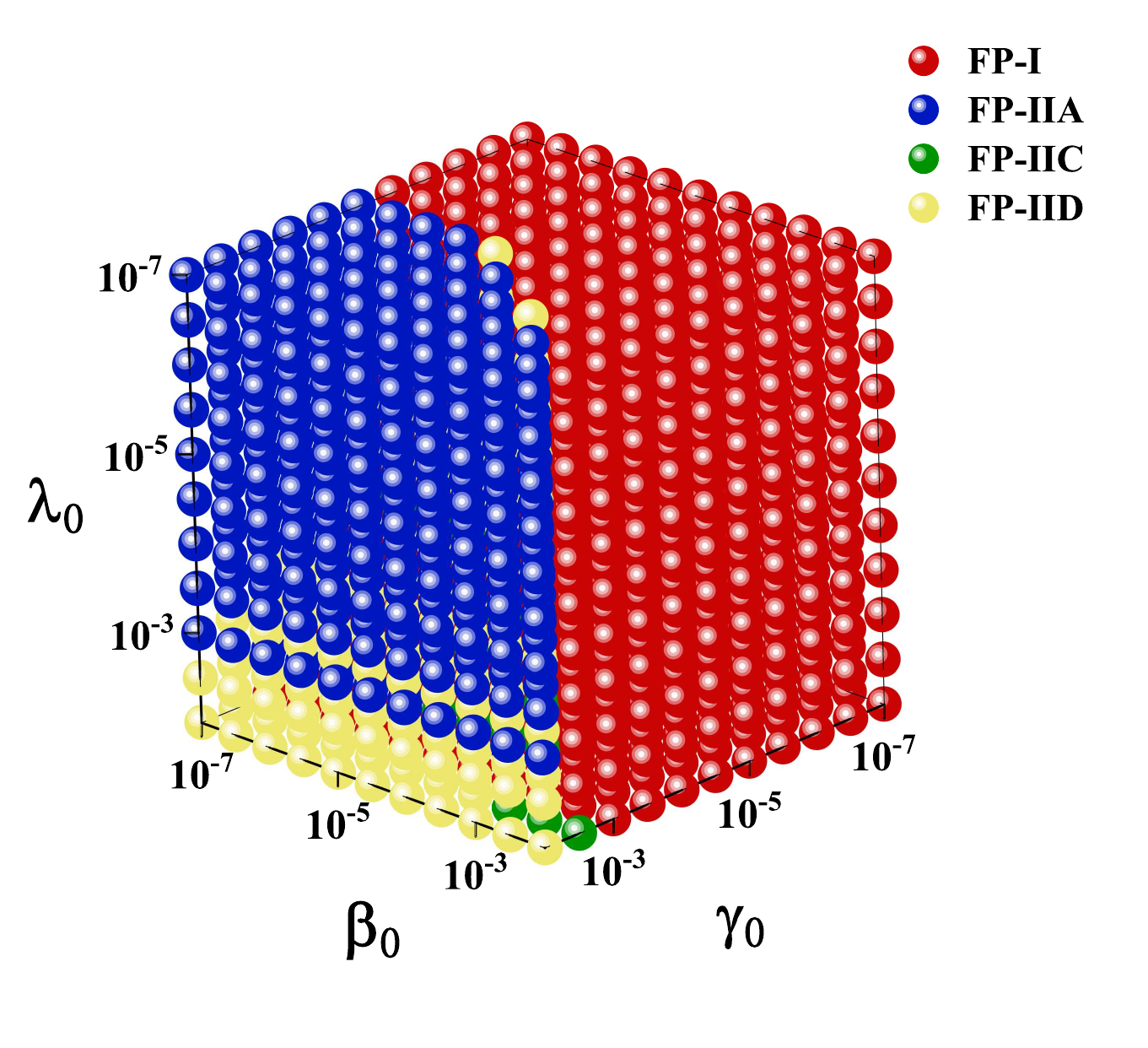}
}
\subfigure[]{
\includegraphics[width=1.5in]{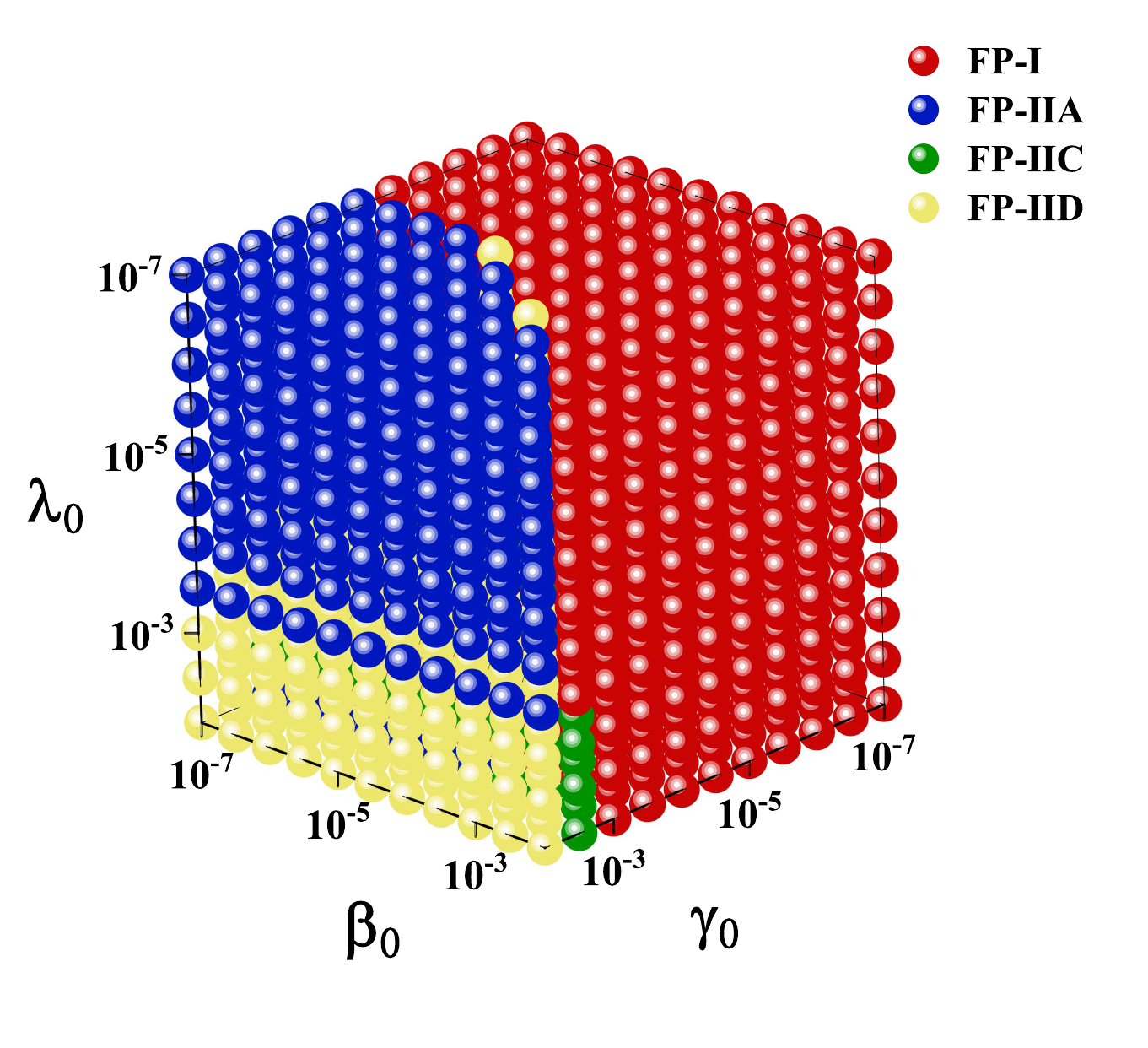}
}
\subfigure[]{
\includegraphics[width=1.5in]{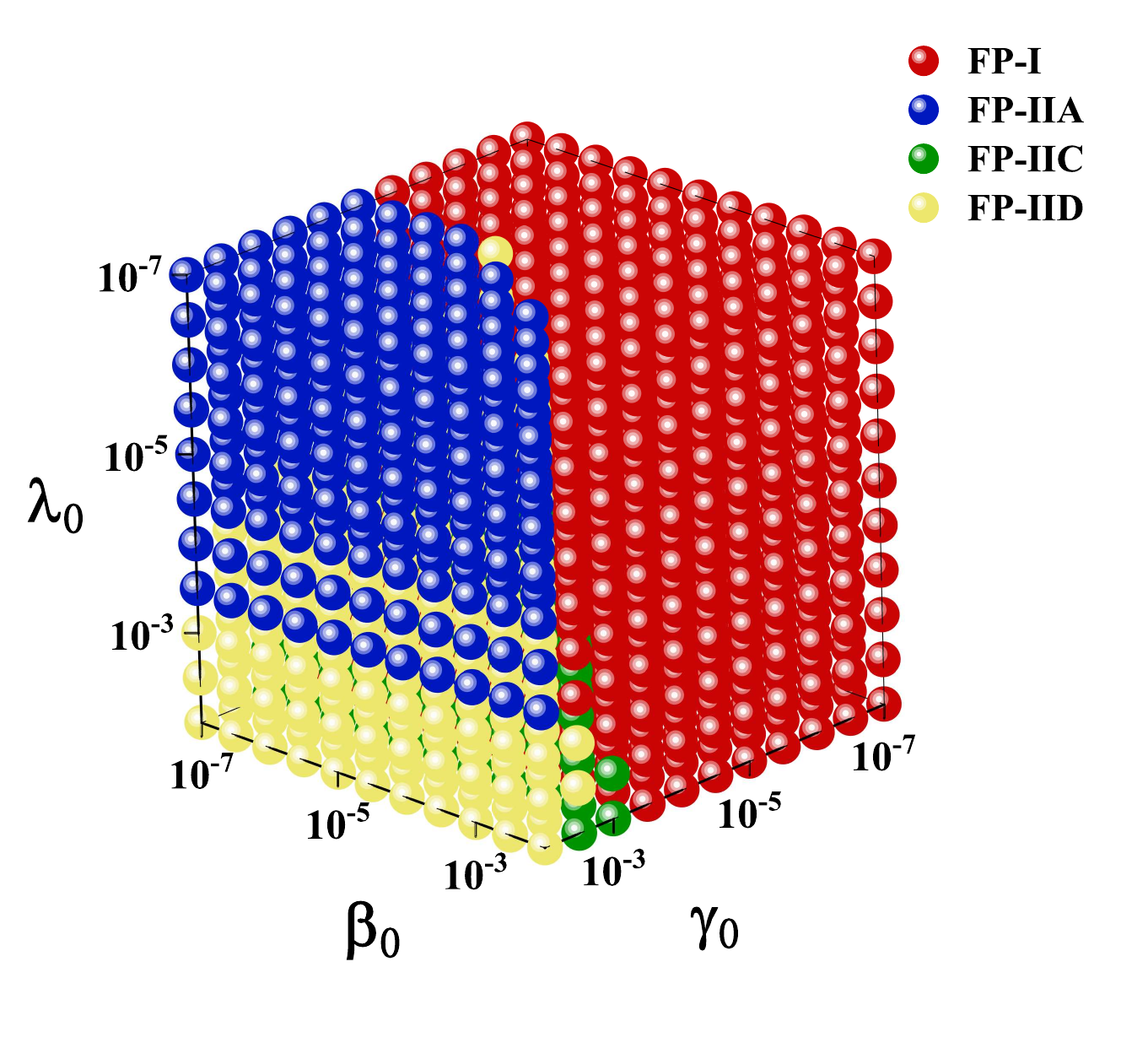}
}
\subfigure[]{
\includegraphics[width=1.5in]{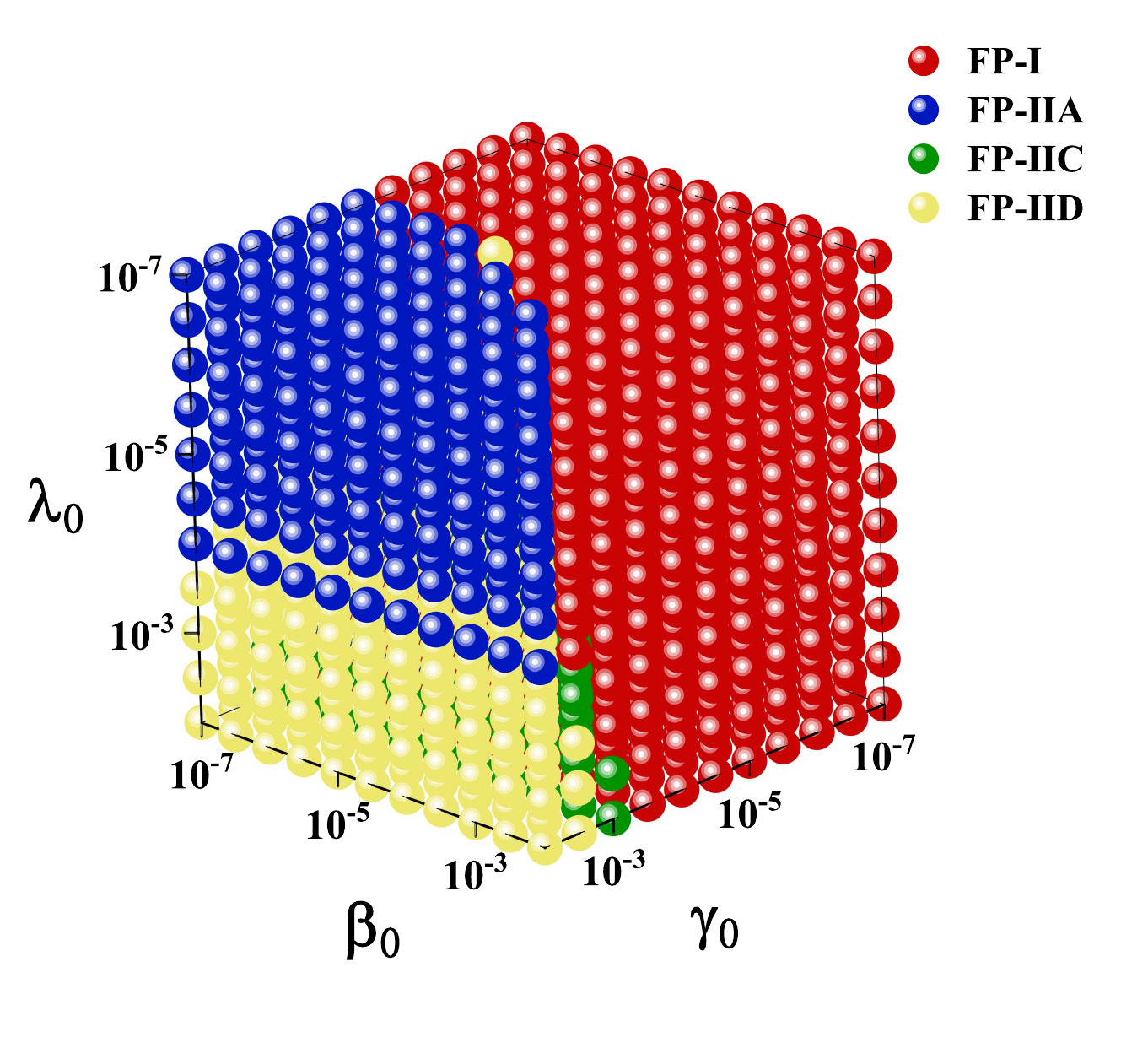}
}
\caption{(Color online) The dependence of FPs on three types of interaction parameters $\gamma$, $\beta$ and $\lambda$ for the Case-General as the system approaches the $\tau_0$-QCP for distinct initial values of fermion velocities:
(a) $v_{\Delta 0} / v_{F0} = 0.2$, (b) $v_{\Delta 0} / v_{F0} =0.4$, (c) $v_{\Delta 0} / v_{F0} =0.6$,
and (d) $v_{\Delta 0} / v_{F0} =0.8$.}
\label{Fig_4_3_2}
\end{figure}

Next, we move to study the impact of Case-General initial conditions on FPs.
In order to systematically examine the effects of the initial conditions of interaction parameters, we hereby
need to fix the velocities and select several representative values of $v_{\Delta0} / v_{F0}$.
By varying the initial values of $\gamma$, $\beta$ and $\lambda$ independently, careful numerical
calculations yield the central results summarized in Fig.~\ref{Fig_4_3_1} and Fig.~\ref{Fig_4_3_2}.

As for the $\tau_{x,y,z}$-QCP, Fig.~\ref{Fig_4_3_1} shows
that the basic distribution of FPs is similar.
Taking the $\tau_{x}$-QCP as an example, as shown in Fig.~\ref{Fig_4_3_1}(a)-(c), we realize that for the Case-General,
both $\gamma$ and $\beta$ play a dominate role in reshaping the FP, while $\lambda$ provides a much smaller contribution.
Concretely, Fig.~\ref{Fig_4_3_1}(a) illustrates that FP-I is dominant when all interaction parameters are less than $10^{-5}$.
This is also consistent with the basic tendencies presented in Sec.~\ref{Subsec_equal}.
However, by increasing $\gamma_0$ and choosing an appropriate $\beta_0$, FP-IIA can overcome FP-I and become
the leading FP. Furthermore, it is worth highlighting that the numerical analysis indicates that the critical
energy for FP-I is much lower than that for FP-II, which is crucial to determine the low-energy properties
as addressed in following section~\ref{Sec_T_c}.

In comparison, more interesting behavior is observed around the $\tau_0$-QCP, as depicted in Fig.~\ref{Fig_4_3_2}.
By fixing $v_{\Delta 0} / v_{F0}$ for several representative values, Fig.~\ref{Fig_4_3_2} shows that the system
can flow toward FP-I, FP-IIA, FP-IIC, and FP-IID, depending on the values of $\gamma_0$, $\beta_0$, and $\lambda_0$.
Specifically, FP-I is dominant for small values of $\gamma_0$. As $\gamma_0$ increases, either FP-IIA or
FP-IID has an opportunity to be the leading one when both $\lambda_0$ and $\beta_0$ are appropriately chosen.
In addition, FP-IID can appear within a very strict region.

\subsection{Impacts of disorder scatterings}\label{Subsec_impurity}

At last, we examine the effects of disorder scatterings on the fermion velocities and critical energy scales of FPs denoted by $l_c$. For convenience, the notations RCP, RM, and RGP-$x,z$ are employed to represent
the random chemical potential, the random mass, and the random gauge potential, respectively, addressed in Sec.~\ref{subsec_dis_eff}.

\begin{figure}
\centering
\subfigure[]{
\includegraphics[width=1.65in]{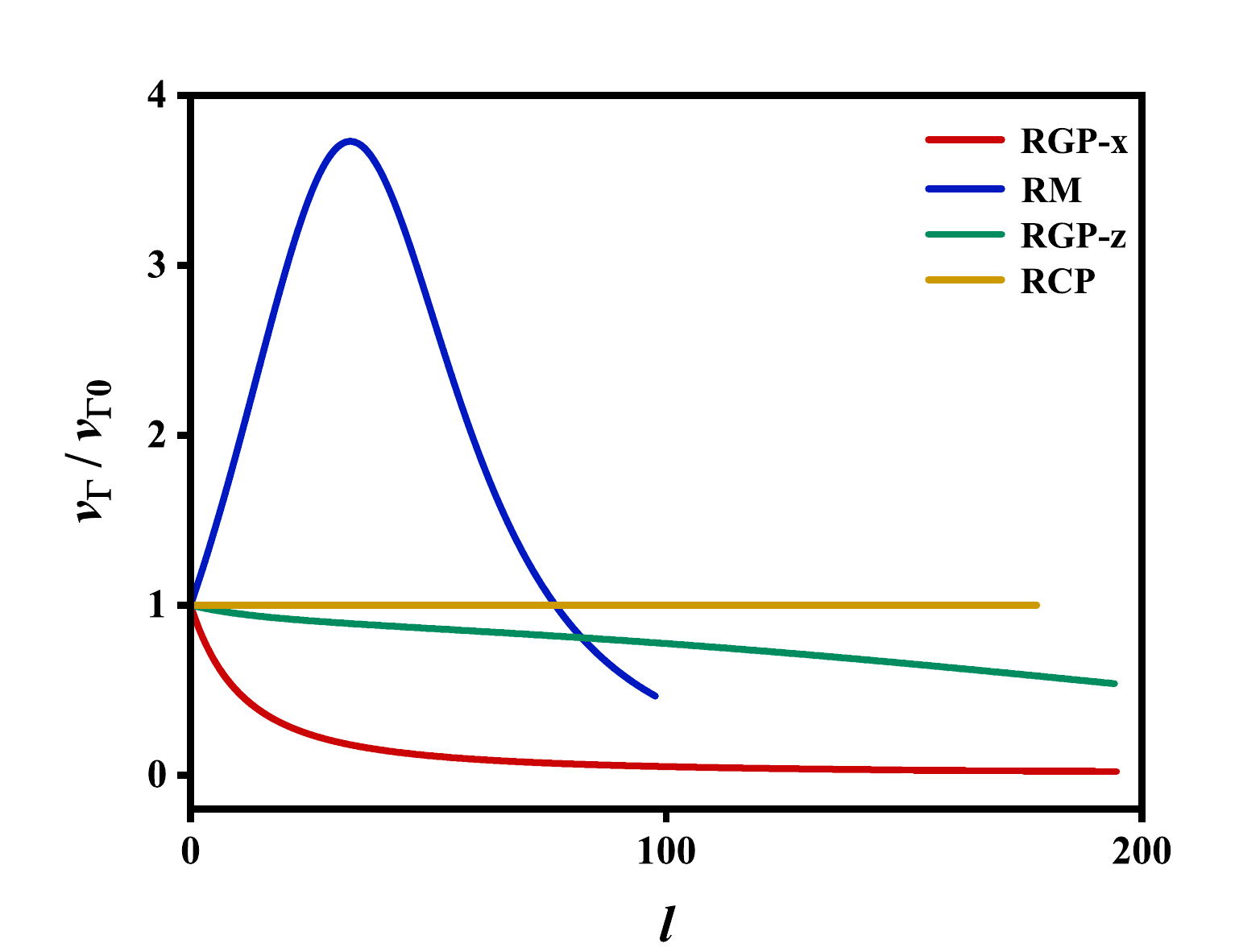}
}\hspace{-0.5cm}
\subfigure[]{
\includegraphics[width=1.65in]{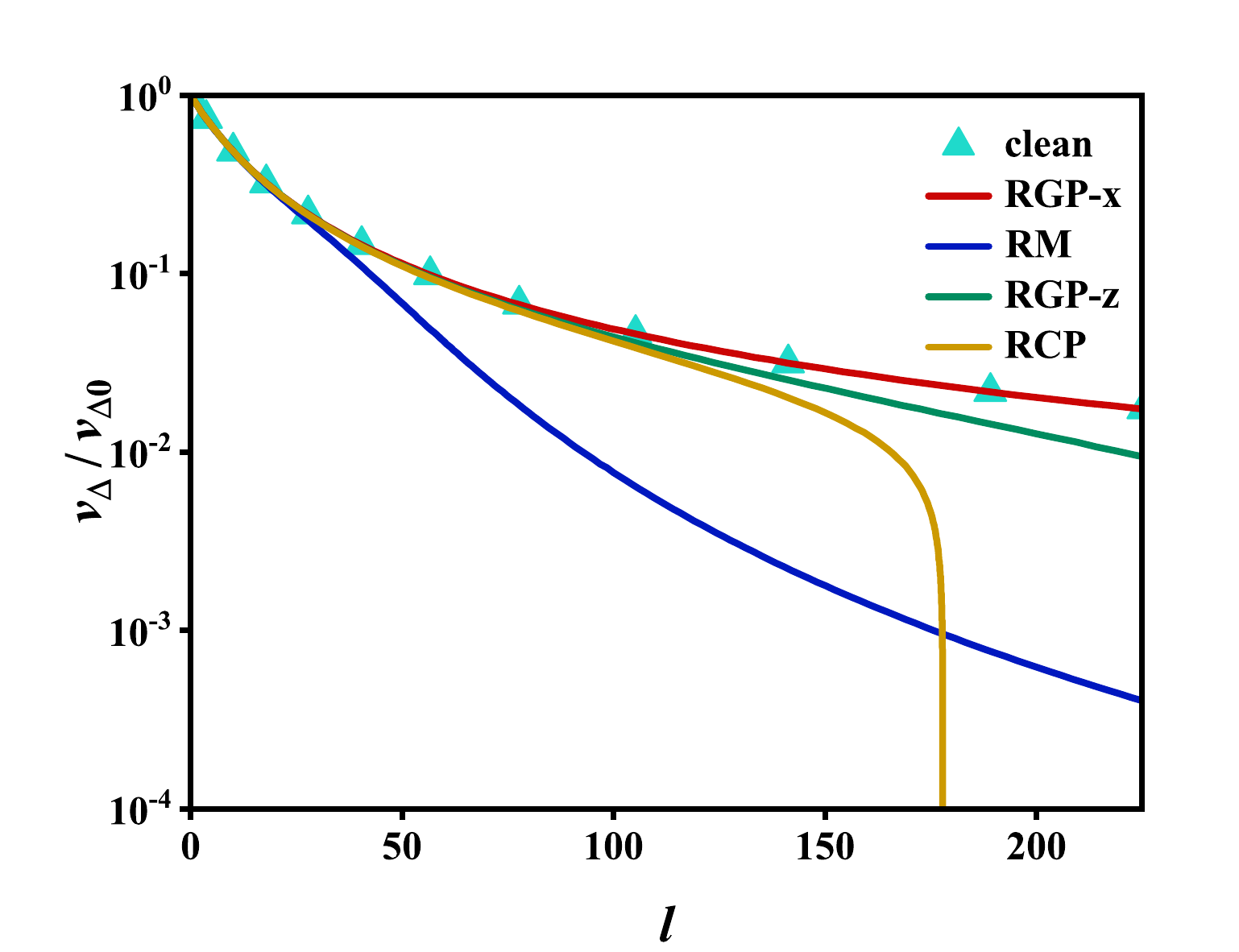}
}
\caption{(Color online) Energy-dependent evolutions near the $\tau_x$-QCP:
(a) distinct kinds of disorder strengths and (b) fermion velocities under the
influence of different sorts of disorders.}
\label{Fig_disorder}
\end{figure}

Numerical analysis shown in Fig.~\ref{Fig_disorder}(a) indicates that different kinds of disorders exhibit distinct
behavior as the energy scale decreases. From Fig.~\ref{Fig_disorder}(a), one can find that both random mass and random gauge potential
decrease and eventually vanish at the lowest-energy scale~\cite{Wang2011PRB,Wang2013PRB}.
In comparison, random chemical potential is marginal to the one-loop level.
Accordingly, they give rise to distinct impacts on the fermion velocities and critical energy scales.
Considering the fermion velocities, we find that around $\tau_{x}$-QCP shown in Fig.~\ref{Fig_disorder}(b)
they are heavily sensitive to the random chemical potential, which is able to vanish the
fermion velocities at a certain finite energy scale. However, both the random mass and random gauge potential
have negligible impacts on the fermion velocities, which gradually vanish at the lowest-energy limit.
Similar low-energy behavior is also observed around the $\tau_z$-QCP.
In sharp contrast, we find that around the $\tau_{y,0}$ QCPs, the fermion velocities are barely susceptible to
all kinds of disorders.

In addition, the disorder scatterings may influence the critical energy scales at which the FPs emerge.
The RG flow equations~(\ref{RG_v_F})-(\ref{RG_Gamma_z}) show that the disorder strength $v_\Gamma$ does not directly enter
the equations of interaction parameters, but it interacts with other parameters by modifying the fermion velocities.
For convenience, let us adopt $l_c$ to denote the critical energy scale, which is approximately $l_c\approx190$ and $l_c\in[3,5]$
for FP-I and FP-II, respectively. Although $v_\Gamma$ cannot modify the $l_c$ directly, it can influence the fermion velocities, potentially driving them to zero at a finite energy scale denoted by $l'_c$ when its initial value is appropriate.
In this sense, we need to stop the RG equations at $l'_c$ as fermion velocities become negative at $l>l'_c$.
Therefore, disorder scattering can affect the critical energy scale if $l'_c<l_c$.
Given that only the random chemical potential is marginal and could qualitatively contribute to the fermion velocities,
we hereby focus exclusively on this type of disorder as schematically illustrated in Fig.~\ref{Fig_schematic_l_c}.
Numerical analysis suggests that $l'_c>l_c$ around the $\tau_{y,0}$-QCP (Scenario A)
and $l''_c<l_c$ around the $\tau_{x,z}$-QCP (Scenario B). This implies that the random chemical potential can indeed reduce
the critical energy scale $l_c$ around the $\tau_{x,z}$-QCP.

To recapitulate, we have identified several FPs associated with distinct QCPs and analyzed the effects of disorder around these FPs. Based on these results, we are going to investigate the underlying
critical behavior in the vicinity of QCPs.

\section{Critical behavior induced by the QCPs}\label{Sec_T_c}

Due to the strong quantum fluctuations around the QCPs, all the interactions become coupled as
described by flow equations~(\ref{RG_v_F})-(\ref{RG_Gamma_z}).  These coupled interactions give rise to
several FPs for distinct kinds of QCPs as presented in Sec.~\ref{Sec_FP}, which are accompanied by the critical energy scales.
Armed with these, a natural question arises how do these FPs modify the critical behavior near the
QCPs in Fig.~\ref{fig1}. Since the critical temperature is an essential quantity for the superconductors, we hereby place our focus on
this very quantity and endeavor to provide a detailed analysis of its critical behavior in the vicinity of the QCPs.

\subsection{Renormalization-group-revised expressions of superfluid density and critical temperature}\label{Subsec_5_1}

Beneath the superconducting dome of a $d$-wave superconductor, the zero-temperature superfluid density ($\rho^s(0)$) exhibits
a linear relationship with the doping concentration ($x$), which can be expressed as~\cite{Orenstein1990PRB,Hardy1993PRL,Uemura1989PRL}
\begin{equation}
\rho^s(0) = \frac{x}{a^2},\label{Eq_rho^s_0}
\end{equation}
with $a$ being the lattice spacing constant. At a finite temperature, the normal nodal QPs can be excited from
the SC condensate, which provides an inverse feedback to the superfluid density.

At first, let us consider the limit case in which all the interactions and fluctuations are absent. We can obtain
the finite-temperature superfluid density ($\rho^s(T)$) as~\cite{Lee1997PRL}
\begin{equation}
\rho^s(T) = \rho^s(0) - \rho^n(T).\label{Eq_rho^s_T}
\end{equation}
where the normal QP density takes the form of
\begin{equation}
\rho^n(T) =m \frac{2\ln2}{\pi}\frac{v_F}{v_\Delta}T,\label{Eq_rho^n_T}
\end{equation}
with $m$ denoting the mass of nodal QPs. This therefore gives rise to the critical SC temperature at $\rho^s(T)=0$,
\begin{equation}
T_c = \frac{1}{2\ln2}\frac{v_\Delta}{v_F}\frac{x}{ma^2}.
\end{equation}
It is significant to emphasize that the fermion velocities thereby are some constants, which are equivalent to the
initial values $v_{\Delta0},v_{F0}$ mentioned in Sec.~\ref{Sec_FP}.

Next, we switch on the interactions and quantum fluctuations in the vicinity of the potential QCP shown in Fig.~\ref{fig1}.
As presented in Sec.~\ref{Sec_RG}, all the interaction parameters are closely coupled under the control of the RG Eqs.~(\ref{RG_v_F})-(\ref{RG_Gamma_z}). This accordingly renormalizes the fermion velocities and make
them energy-dependent from $v_{\Delta0},v_{F0}$ to $v_{\Delta}(k),v_{F}(k)$ that are dictated by the RG evolutions.
On the basis of these considerations, the normal QP density is renormalized and recast to~\cite{Lee1997PRL,Durst2000PRB,Liu2012PRB}
\begin{widetext}
\begin{equation}
\rho_R^n (T)=\frac{4m}{k_B T} \left[ \int_0^{\Lambda_c} \frac{d^2 \mathbf{k}}{(2\pi)^2}
\frac{v_F^2(\mathbf{k}) e^{\sqrt{v_F^2 (\mathbf{k}) k_x^2+v_\Delta^2(\mathbf{k}) k_y^2 }/k_B T}}
{\left(1+e^{\sqrt{v_F^2 (\mathbf{k}) k_x^2+v_\Delta^2 (\mathbf{k}) k_y^2 }/k_B T}\right)^2}
+\int_{\Lambda_c}^\infty \Bigl(v_{F,\Delta} (\mathbf{k})\rightarrow v_{F,\Delta} (l_c)\Bigr) \right],\label{Eq_rho_n_s}
\end{equation}
\end{widetext}
where $k_B$ denotes the Boltzmann constant. Here, $\Lambda_c$ represents the critical energy and is
associated with $l_c$ for a given FP. Several clarifications are necessary at this point. As aforementioned in Sec.~\ref{Subsec_eff_FP},
the RG equations are no longer valid beyond $l_c$. Therefore, we divide the whole energy regime into
two regions, namely $l\in[0,l_c)$ and $l\geq l_c$. In other words, for $l\in[0,l_c)$, the fermion velocities are energy-dependent due to the RG flow, which fully captures the effects of quantum fluctuations. In contrast, for $l\geq l_c$, we go beyond the mean-field approach, where the fermion velocities are treated as constants (i.e., $v_{\Delta0}$ and $v_{F0}$), but adopt the RG-renormalized mean-field method, in which these constants  $v_{\Delta0}$ and $v_{F0}$ are replaced by the $v_{\Delta}(l_c)$ and $v_{F}(l_c)$, which are modified by the quantum fluctuations at the energy scale $l_c$~\cite{Metzner2000PRL,Reiss2007PRB,Wang-EM2014PRB}. As a result, the presence of quantum fluctuations renormalizes the superfluid density~(\ref{Eq_rho^s_T}) to the following form,
\begin{equation}
\rho_R^s(T) = \rho^s(0) - \rho_R^n(T),\label{Eq_rho_R_s}
\end{equation}
from which the renormalized critical temperature ($T'_c$) can be obtained by taking $\rho_R^s(T'_c) = 0$.

\begin{figure}
\centering
\includegraphics[width=3.1in]{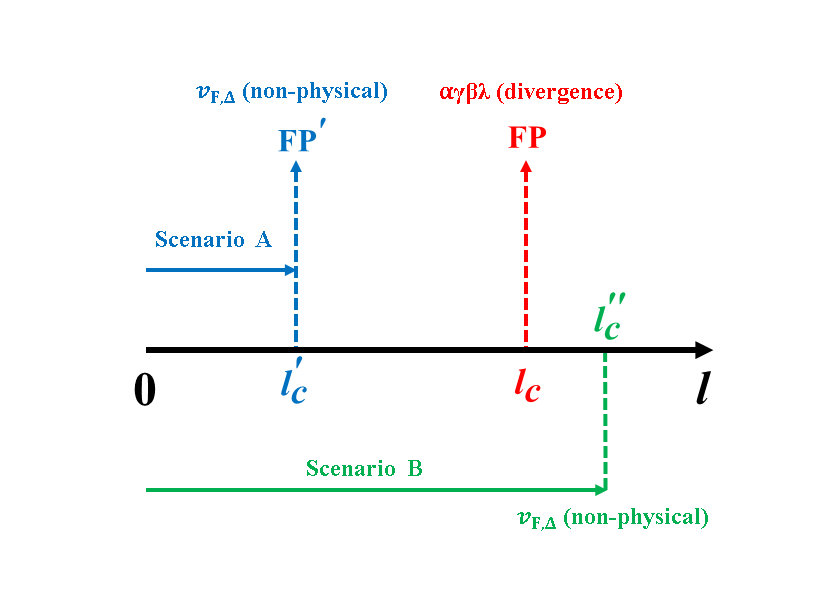} \vspace{-0.5cm}
\caption{(Color online) Schematic illustrations of distinct kinds of low-energy fates as approaching the QCPs.
The red line serves as the clean-limit case, in which the FP associated with divergence of interaction
parameters is determined at $l=l_c$. In the presence of the random chemical potential, there exist two potential
scenarios, namely Scenario A (blue) and Scenario B (green).}
\label{Fig_schematic_l_c}
\end{figure}

\begin{figure}
\hspace{-0.68cm}
\includegraphics[width=3in]{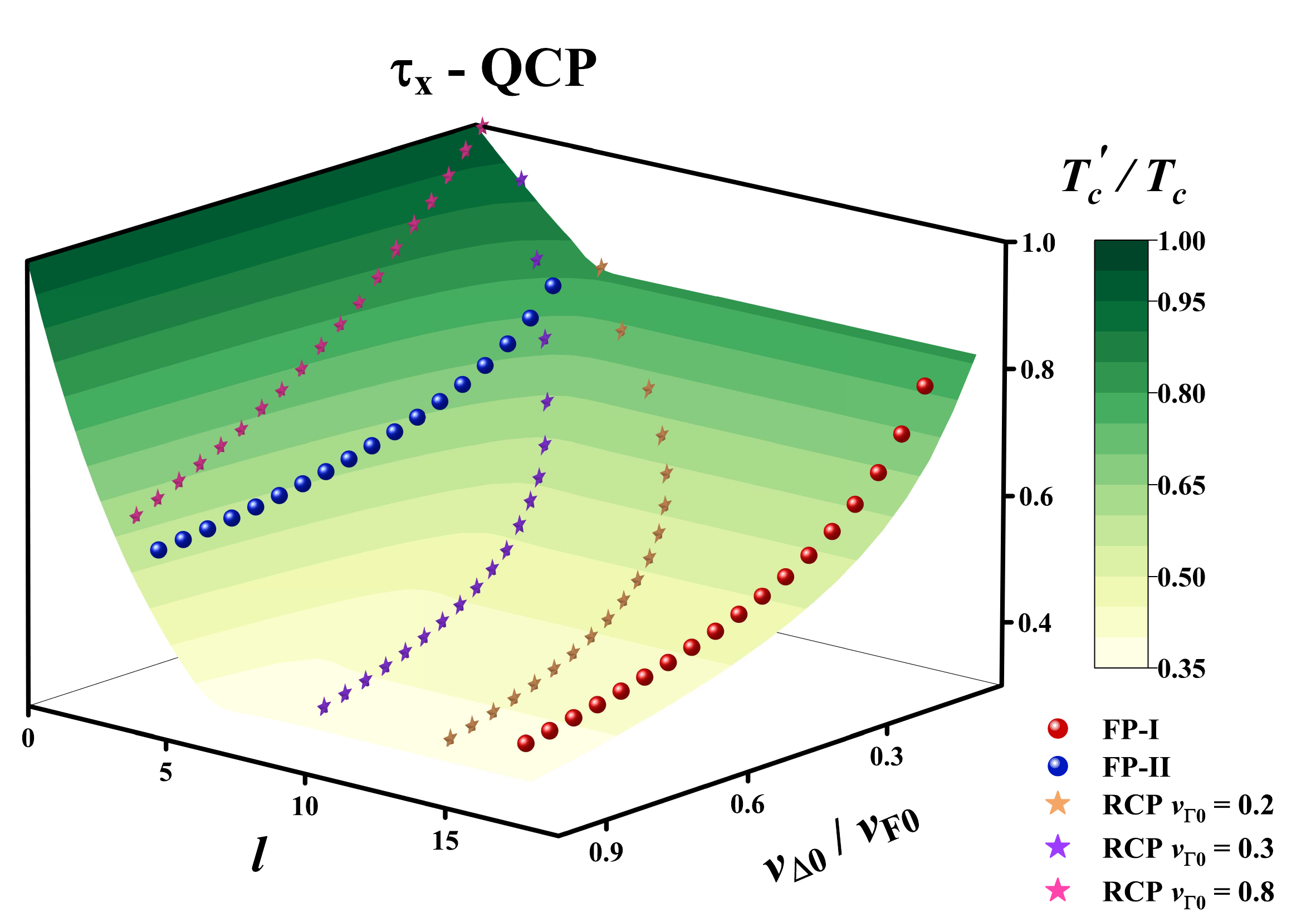}
\caption{(Color online) The tendencies of the renormalized critical temperature $T'_c$ relative to the initial critical temperature $T_c$ near distinct FPs as approaching the $\tau_x$-QCP. The value of $T'_c/T_c$ depends upon the energy scales ($l$) and the initial value of fermion velocities ($v_{\Delta 0}/v_{F0}$) in the presence of a random chemical potential.}
\label{Fig_x_Tc}
\end{figure}

\subsection{Numerical results for the renormalized critical temperature}\label{Subsec_T_c}

With all of the above on hand, we are now in a proper position to examine the behavior of critical temperature around different kinds of
QCPs. Reading from Eq.~(\ref{Eq_rho_n_s}) and Eq.~(\ref{Eq_rho_R_s}), we can infer that
there exist two significant items that are largely associated with the critical temperature: the
initial values of fermion velocities at $l=0$ and the critical energy scale denoted by
$l_c$ of the FP. In particular, learning from Sec.~\ref{Sec_FP} around the QCPs,
we find that the critical energy scale $l_c$ is closely related to the type
of FP.

\begin{figure}
\hspace{-0.68cm}
\subfigure[]{\includegraphics[width=3in]{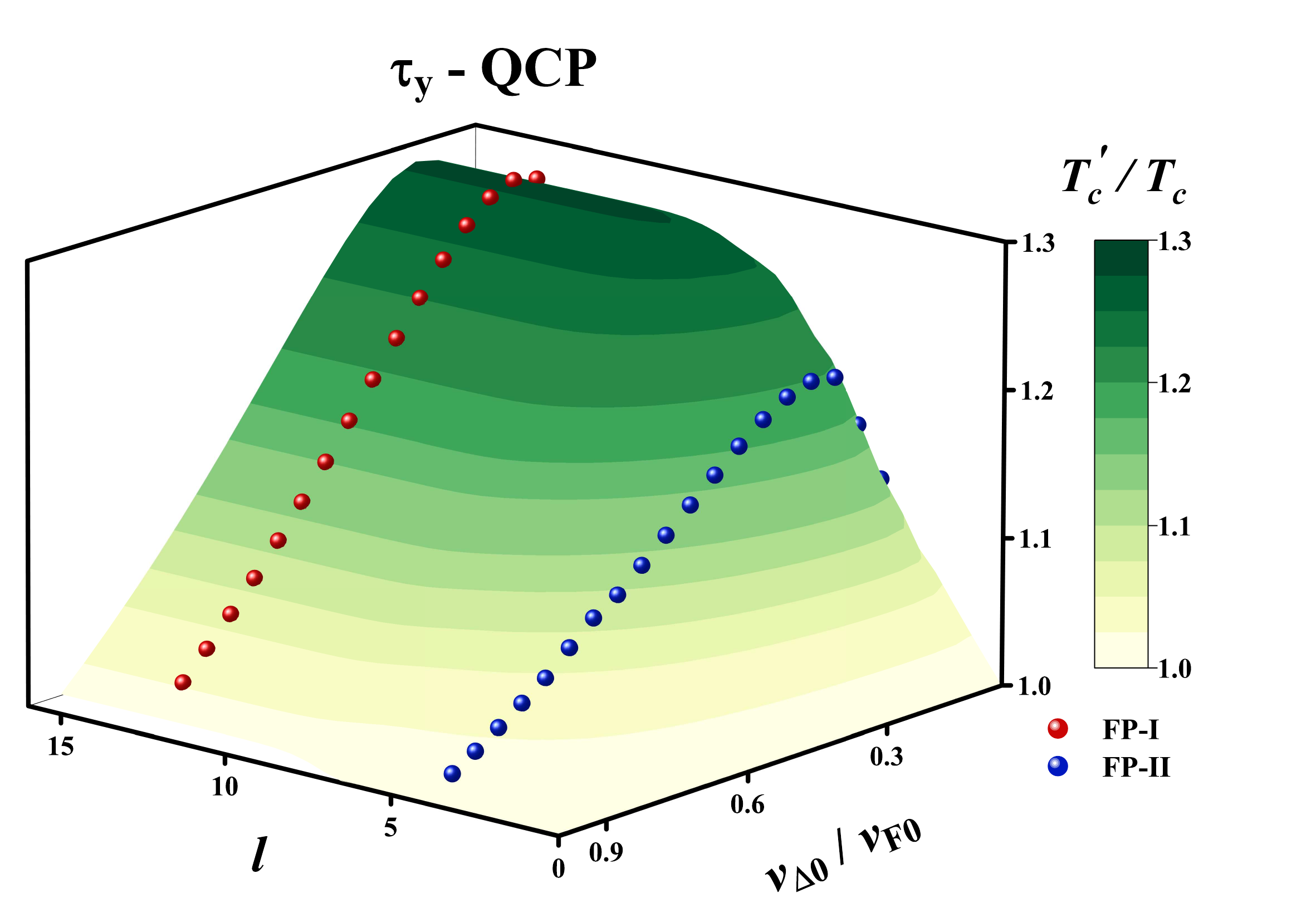}}\\
\subfigure[]{\includegraphics[width=3in]{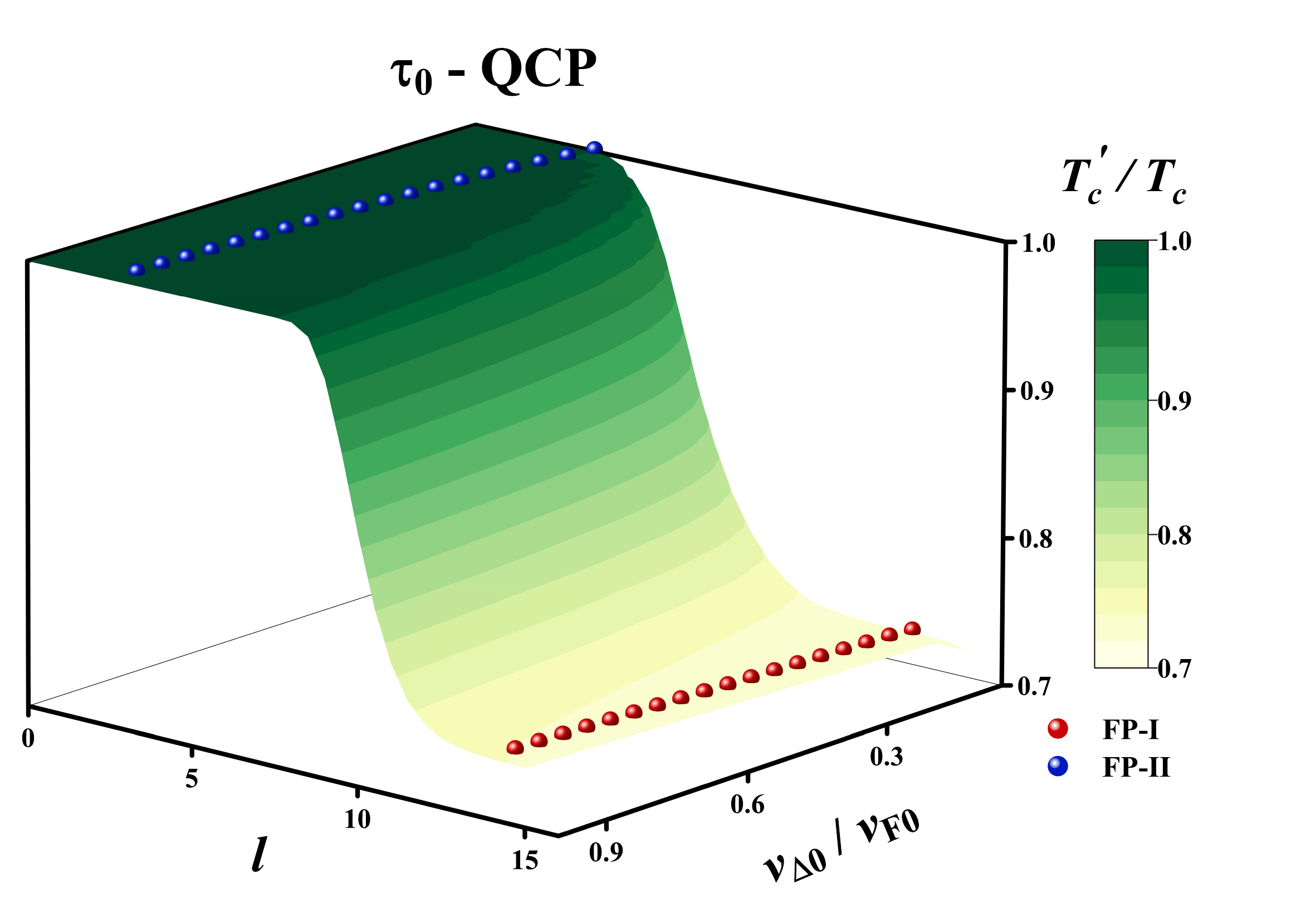}}
\caption{(Color online) The tendencies of the renormalized critical temperature $T'_c$ relative to the initial critical temperature $T_c$ near distinct FPs as approaching (a) the $\tau_y$-QCP and (b) the $\tau_0$-QCP. The value of $T'_c/T_c$ depends upon the energy scales ($l$) and the initial value of fermion velocities ($v_{\Delta 0}/v_{F0}$) in the presence of random chemical potential.}\label{Fig_y_Tc}
\end{figure}

After performing the detailed numerical analysis of the RG equations~(\ref{RG_v_F})-(\ref{RG_Gamma_z})
and the renormalized superfluid density~(\ref{Eq_rho_n_s})-(\ref{Eq_rho_R_s}), we examine the impacts of QCPs as shown
in Fig.~\ref{fig1} on the critical temperature. The results are presented in Fig.~\ref{Fig_x_Tc}-\ref{Fig_y_Tc}
for different kinds of FPs. These figures clearly illustrate the dependence of the renormalized critical temperature $T'_c/T_c$ on the initial ratio of fermion velocities $v_{\Delta0}/v_{F0}$ and the critical energy scale $l_c$ around distinct QCPs. Here, $T'_c$ and $T_c$ serve as the renormalized and noninteracting critical temperatures, respectively.

At the outset, we consider the $\tau_x$-QCP. One can find from Fig.~\ref{Fig_x_Tc} that the quantum fluctuations cause
a pronounced decrease in the critical temperature for both FP-I and FP-II, as discussed
in Sec.~\ref{Subsec_eff_FP}. Tuning up the $v_{\Delta0}/v_{F0}$ is prone to decreasing the critical temperature.
In addition, one can notice that $T'_c$ decreases a little more sharply as the system approaches FP-I compared to FP-II.
Furthermore, we study the effects of the random chemical potential which
is marginal and has an opportunity to provide nontrivial corrections as explained in Sec.~\ref{Subsec_impurity}. As shown in Fig.~\ref{Fig_schematic_l_c}, the occurrence of Scenario A or Scenario B depends on both the initial disorder strength $v_{\Gamma0}$ and the ratio $v_{\Delta0}/v_{F0}$. For FP-I, once $v_{\Gamma0}$ is suitable, the system always transitions to Scenario A, leading to a slight increase in $T'_c$ compared to the clean limit case. For FP-II, Scenario B is preferred under most conditions. However, Scenario A can also be realized when both $v_{\Delta0}/v_F$ and $v_{\Gamma0}$
are appropriately chosen, such as $v_{\Gamma_0}=0.3$ and $v_{\Delta0}/v_{F0}\textless0.1$. This accordingly indicates that the random chemical potential plays an important role around the $\tau_x$-QCP.

Subsequently, let us move to the $\tau_y$-QCP and $\tau_0$-QCP. We find that Scenario-A in Fig.~\ref{Fig_schematic_l_c}
triggered by disorder scatterings cannot be realized for these two QCPs. This implies that our analysis should focus on the
clean limit case. In sharp contrast to the $\tau_x$-QCP, near the $\tau_y$-QCP, Fig.~\ref{Fig_y_Tc}(a)
shows that the renormalized critical temperature $T'_c$ for both FP-I and FP-II increases
and would exhibit a peak at $x_c$ in Fig.~\ref{fig1} induced by the quantum fluctuations.
In particular, $T'_c$ is sensitive to the ratio $v_{\Delta0}/v_{F0}$, reaching its maximum at $v_{\Delta0}/v_{F0} \approx 0.25$
and gradually decreasing as $v_{\Delta0}/v_{F0}$ either increases or decreases. In addition, $T'_c$ for FP-I is slightly
higher than for FP-II. As we approach the $\tau_0$-QCP, the renormalized critical temperature becomes more sensitive to the types of
FPs, but less sensitive to the initial values of fermion velocities.
These are remarkably different from those around $\tau_{x}$-QCP and $\tau_y$-QCP.
Figure~\ref{Fig_y_Tc}(b) manifests that $T'_c$ is slightly smaller than the noninteracting
critical temperature $T_c$ in the absence of quantum fluctuations. However, when FP-I is accessed, $T'_c$ undergoes a significant dip, reaching a value of $T'_c/T_c = 0.72$.

At last, we offer several comments on the $\tau_z$-QCP.
By performing a similar numerical analysis as done for the $\tau_{x,y,0}$-QCP, we notice that
the renormalized temperature $T'_c$ around the $\tau_z$-QCP is heavily increased for all initial values of
fermion velocities. In particular, $T'_c/T_c$ exceeds $10$ at $v_{\Delta0}/v_{F0} = 0.1$.
Such a high critical temperature has not yet been
reported for the $d$-wave superconductors~\cite{Lee2006RMP,Fradkin2015RMP,Keimer2008Science,Orenstein2000Science,
Bozovic2016Nature,Phillips2022Science,Ramakrishnan2025}.  Under this circumstance,
although all four kinds of QCPs are allowed from a symmetry perspective,
we argue that the $\tau_z$-QCP can barely be realized from the physical perspective
and instead the other three QCPs own the opportunity to be the candidate QCP.

To be brief, the critical temperature of $d$-wave superconductors would be
considerably modified by the quantum fluctuations around the QCPs. Depending on the type of QCP, the critical temperature can either increase (around the $\tau_y$-QCP) or decrease (around the $\tau_{x,0}$-QCP). The exact value of $T'_c$ is associated with the
types of FPs and initial values of fermion velocities.

\section{Comparisons and discussions}

Before going further, let us briefly compare our study with prior research on quantum criticality
in $d$-wave superconductors~\cite{Vojta2000PRL,Kim-Kivelson2008PRB,Sachdev2008PRB,Xu2008PRB,Wang2013PRB,She2015PRB}
and highlight the key points, focusing on the low-energy effective theories and the key physical results
as well as implications for future studies.

At the outset, we consider the effective theory for the QPT. Quantum critical fluctuations near the QPT illustrated in Fig.~\ref{fig1} generally couple strongly with each other and with other degrees of freedom, giving rise to exotic physical behavior.

To explore the possible types of such a transition, Vojta \emph{et al.}~\cite{Vojta2000PRL} performed a complete group-theoretical analysis and identified seven distinct classes of QPTs, each corresponding to an $X$ state with a specific symmetry-breaking pattern. This classification provides the fundamental basis for developing a low-energy effective field theory to study critical behavior around the QCP. Such a theory involves constructing a low-energy effective model that incorporates nodal Dirac fermions and a bosonic order parameter field representing the $X$ state. To simplify the analysis, the previous studies usually neglect the fluctuations of superconducting order
parameter. In other words, when studying the quantum criticality of the competing order, the amplitude and phase fluctuations of superconductivity are considered negligible or ``frozen out'' at low energies, allowing one to focus purely on the quantum critical dynamics of the competing order parameter $\phi$~\cite{Vojta2000PRL,Kim-Kivelson2008PRB,Sachdev2008PRB,Xu2008PRB,Wang2013PRB,She2015PRB}.
In this sense, the effective action around the QCP includes contributions from nodal fermions $\Psi$, the $X$-state order parameter $\phi$, and the Yukawa coupling between them,
\begin{equation}
S = S_\Psi + S_\phi + S_{\Psi\phi}. \label{Eq_effective_action_0}
\end{equation}

In principle, near a QCP, both the $X$-state order parameter and the fluctuations of the
superconducting order parameter become significant, and their intimate competition may
fundamentally reshape the quantum critical physics~\cite{Wang2014PRD,Kleinert2003NPB}.
To fully capture the critical behavior near the QCP, we go beyond the standard approximation
that neglects superconducting fluctuations. Specifically, we explicitly introduce the quantum
fluctuations of the $d$-wave superconducting order parameter¡ªincluding its amplitude ($h$)
and phase ($\eta$) modes¡ªinto the low-energy effective theory. This yields our improved effective field
theory $S_{\mathrm{eff}}$~(\ref{Eq-S-eff}), which incorporates the fermions $\Psi$, the $X$-state order parameter $\phi$, the superconducting fluctuation fields ($h$ and $\eta$), and their mutual interactions. In addition, we consider three distinct types of impurities that affect the gapless quasiparticles and are inevitably present in real materials~\cite{Lee1985RMP,Mirlin2008RMP,Nersesyan1995NPB}.
This leads to the final effective theory $S^\star_{\mathrm{eff}}$~(\ref{Eq_S-phenomen})
with $S_{\mathrm{dis}}$~(\ref{Eq-S-dis}) denoting the impurity contribution. Accordingly, this model
advances the theoretical framework by treating superconducting fluctuations as essential dynamical
degrees of freedom and incorporating the interactions of fermions coupled to an $X$ state as well as
the competition between the $X$-state order parameter and the superconducting order parameter.

Next, we move to the critical behavior near the QCPs. For convenience, seven candidate states are divided
into four distinct classes, labeled as $\tau_{0,x,y,z}$-Type QCPs~\cite{Vojta2000PRL,Vojta2000PRB,Wang2013PRB}.

The fermion velocity ratio $v_{\Delta}/v_{F}$ has been identified as a pivotal quantity. It not only
characterizes the intrinsic anisotropy of nodal quasiparticles but also dictates the quantum critical behavior.
Its renormalization near a QCP directly governs low-energy observables, making it an indispensable component
of the theoretical framework for understanding such transitions. Utilizing the simplified model~(\ref{Eq_effective_action_0}),
Kim \emph{et al.}~\cite{Kim-Kivelson2008PRB} first examined the coupling between emergent nematic order and gapless
fermionic quasiparticles at this nematic QCP (i.e., the $\tau_x$-Type transition), demonstrating that such
a nematic-induced quantum phase transition is a stable continuous transition. To proceed, Huh and
Sachdev~\cite{Sachdev2008PRB} discovered a strongly anisotropic fixed point for the fermion velocities
($v_\Delta / v_F \rightarrow 0$) at the nematic QCP. Our earlier work~\cite{Wang2013PRB} investigated the $\tau_y$-Type
and $\tau_z$-Type transitions, obtaining fixed points $v_\Delta / v_F \rightarrow 1$ and $v_\Delta / v_F \rightarrow \infty$,
respectively. The physical consequences of $v_\Delta / v_F$ were also studied. It determines the wedge angle
of the spectral function and hence quasiparticle coherence~\cite{Kim-Kivelson2008PRB}, controls the scaling of the
local density of states~\cite{Xu2008PRB}, and provides logarithmic corrections to the penetration depth~\cite{She2015PRB}.

Starting from the effective theory $S^\star_{\mathrm{eff}}$~(\ref{Eq_S-phenomen}) that includes the
interplay among gapless fermions, distinct types of order parameter fluctuations, and disorder scatterings,
we systematically investigate how this interplay modifies the fixed-point structure for distinct types of phase transitions
and how it reshapes low-energy physical observables associated with experimental signatures, with particular focus on the superfluid density $\rho_s$ and the critical temperature $T_c$.

Given the importance of fermion velocity ratio, it is of particular necessity to
 establish a direct comparison with previous studies. By neglecting superconducting order parameter,
 our RG equations of interaction parameters~(\ref{RG_v_F})-(\ref{RG_Gamma_z})
can exactly reproduce the key tendencies of
$v_\Delta / v_F$ obtained from $S_0$~(\ref{Eq_effective_action_0}) and thus the consequent critical properties associated with
the fermion velocities~\cite{Kim-Kivelson2008PRB,Sachdev2008PRB,Xu2008PRB}.

Going beyond the fermion velocities, we discover several new emergent FPs from
$S^\star_{\mathrm{eff}}$~(\ref{Eq_S-phenomen}). Specifically, we identify emergent Type-I and Type-II FPs,
with the latter comprising five distinct subclasses labeled FP-IIA through FP-IIE. As illustrated in Figs.~\ref{Fig_4_2_2_2D}-\ref{Fig_4_3_1}, these FPs are distributed across different QCPs depending on the
initial conditions of the interaction parameters. In addition, we briefly discuss the effects of disorder on these FPs presented in
Sec.~\ref{Subsec_impurity}.

In principle, these emergent FPs govern the critical behavior of physical observables, most notably causing a
significant renormalization of the superconducting critical temperature $T_c$ near the QCP,
as schematically depicted in Fig.~\ref{fig1}.The $\tau_{x,y,0}$-QCPs exhibit distinctive $T_c$ renormalization trends with experimentally distinguishable features. For instance, at the $\tau_y$-QCP, the coupling between different critical fluctuations leads to a significant enhancement of $T_c$, as shown in Fig.~\ref{Fig_y_Tc}(a) (the corresponding results for $\tau_{x}$-QCP
and $\tau_{0}$-QCP are presented in Fig.~\ref{Fig_x_Tc}
and Fig.~\ref{Fig_y_Tc}(b), respectively). In sharp contrast, our calculations for the $\tau_z$-QCP yield a renormalized
$T_c$ that lies beyond the experimentally plausible regime for typical $d$-wave superconductors.
It indicates that the $\tau_z$-QCP is unlikely a viable candidate for the QCPs in Fig.~\ref{fig1}
and hence cannot be experimentally realized in real materials.

Finally, we present several comments on the implications for future studies.
As a theoretical study, our primary aim is to explore
and uncover the quantum critical behavior dictated by the potential multi-order-parameter
quantum criticality in $d$-wave superconductors. Compared to the complicated realistic systems that
contain numerous concurrent physical processes, our theoretical models are typically built on
simplified Hamiltonians consisting of a limited set of related parameters. In consequence,
a perfect quantitative match between theory and experiment remains a challenging goal.
While direct quantitative comparison with specific materials is beyond our scope,
these findings would offer underlying guidance for further studies.
In particular, the distinct $T_c$ signatures associated with different QCPs provide
a diagnostic fingerprint for identifying the symmetry of fluctuations in real systems.
In addition, avoiding $\tau_x$-Type criticality protects $T_c$ from suppression,
whereas the predicted $T_c$ enhancement near the $\tau_y$-QCP suggests a concrete strategy
for quantum-criticality-enhanced superconductivity.


To recapitulate, this work extends the quantum critical framework for $d$-wave superconductors by incorporating superconducting fluctuations. Our multi-field effective theory reveals that the competition between superconducting and $X$-state fluctuations
generates new fixed points and alters low-energy physics. We anticipate that these results will advance the understanding of quantum criticality in $d$-wave cuprate superconductors and offer a conceptual guide for future materials research.


\section{Summary}\label{Sec_summary}

In summary, we systematically investigate a quantum phase transition (QPT) beneath the superconducting dome of a $d$-wave cuprate superconductor. This transition evolves from a pure $d$-wave superconducting state to a mixed state characterized by $d$-wave superconductivity coexisting with an additional order parameter $X$, as illustrated in Fig.~\ref{fig1}. For completeness, we examine four potential quantum critical points (QCPs) associated with this QPT: $\tau_{0}$-Type, $\tau_{x}$-Type, $\tau_{y}$-Type, and $\tau_{z}$-Type QCPs~\cite{Vojta2000PRL,Vojta2000PRB,Wang2013PRB}. To equally account for distinct degrees of freedom near these QCPs, we employ the RG approach~\cite{Shankar1994RMP,Wilson1975RMP,Polchinski1992}, which allows us to derive the coupled, energy-dependent flow equations for all relevant interaction parameters within our effective theory. After analyzing these equations, several critical phenomena
near the QCPs are induced due to strong quantum fluctuations.

We begin by performing a numerical
analysis of the coupled RG flow equations~(\ref{RG_v_F})-(\ref{RG_Gamma_z}). Our results indicate that the overall behavior of
the interaction parameters is governed by a set of fixed points (FPs) located at $l = l_c$,
which are closely tied to the low-energy critical behavior of the system~\cite{Maiti2010PRB,Vojta2003RPP,
Roy2018PRX,Wang2017QBCP,Vafek2012PRB,Vafek2014PRB,
Wang2020PRB,Chubukov2012ARCMP,Chubukov2016PRX,Nandkishore2012NP,Wang2020NPB}.
At the clean limit, varying the initial conditions leads to the emergence of two distinct types of FPs.
FP-I is characterized by the divergence of the quadratic parameter, which remains largely unaffected by
other interaction parameters. In contrast, FP-II is dominated by the cubic and quartic interaction parameters.
In addition, since the specific values of FP-II are highly sensitive to the initial conditions, as illustrated
in Fig.~\ref{Fig_4_2_1}, it further divides into five subclasses: FP-IIA, FP-IIB, FP-IIC, FP-IID,
and FP-IIE. To proceed, we introduce three types of disorder and examine the effects of fermion-disorder coupling
(\ref{Eq-S-dis}) on the stability of these FPs. Numerical analysis demonstrates that both random mass
and random gauge potential provide negligible impacts on FPs for all QCPs. However, the random chemical potential
significantly influences either the fermion velocities or the critical energy scales. Specifically, it tends to
reduce the critical energy scale $l_c$ near the $\tau_{x,z}$ QCPs, as shown in Fig.~\ref{Fig_schematic_l_c}.

Next, we carefully investigate the impacts of strong quantum fluctuations on the critical
temperature as the system approaches the FPs for all potential QCPs in Fig.~\ref{fig1}.
For the $\tau_x$-QCP, the critical temperature is significantly suppressed for both FP-I and FP-II,
as depicted in Fig.~\ref{Fig_x_Tc}. In particular, the critical temperature gradually decreases as
$v_{\Delta0}/v_{F0}$ increases and it decreases a little more sharply when approaching FP-I
compared to FP-II. In addition, the random chemical potential can lead to either Scenario A or
Scenario B in Fig.~\ref{Fig_schematic_l_c} as long as the initial conditions are appropriately tuned
around the $\tau_x$-QCP. In particular, the critical temperature decreases gradually with increasing
$v_{\Delta0}/v_{F0}$, exhibiting a slightly sharper decline near FP-I than near FP-II. With respect to
the $\tau_y$-QCP and $\tau_0$-QCP, the critical temperature, as depicted in Fig.~\ref{Fig_y_Tc}, is
primarily determined by the ratio $v_{\Delta0} / v_{F0}$ and the critical energy scale $l_c$, while remaining
largely insensitive to disorder scattering. This is because Scenario A in Fig.~\ref{Fig_schematic_l_c} cannot
be realized in these cases. Notably, for the $\tau_y$-QCP, the critical temperature increases and exhibits a
strong dependence on the fermion velocities, reaching a distinct peak at $v_{\Delta 0}/v_{F0}\sim 0.25$.
In comparison, for the $\tau_0$-QCP, the critical temperature becomes more sensitive to the type of FP
but less sensitive to the initial values of the fermion velocities. In this sense, the critical temperature
is slightly lower and undergoes a significant dip for FP-I and FP-II, respectively.

Considering the $\tau_z$-QCP, direct numerical analysis indicates that
the critical temperature can be significantly enhanced for all initial values of the fermion velocities.
However, this enhancement lies well beyond the range of critical temperatures $T_c$ typically observed in $d$-wave superconductors~\cite{Lee2006RMP,Fradkin2015RMP,Keimer2008Science,Orenstein2000Science,Bozovic2016Nature,
Phillips2022Science,Ramakrishnan2025}. Consequently,
we infer that such a QCP is unlikely to be physically realizable. This analysis of critical behavior provides
a strategy to exclude the $\tau_z$-QCP, as illustrated in Fig.~\ref{fig1}, while supporting the potential
existence of the other three QCPs. We hope that these findings will contribute to clarifying the critical
behavior of $d$-wave superconductors and improve our understanding of these materials and related systems in the future.

\section*{ACKNOWLEDGEMENTS}

Q.Q.Y and Y.S.F. thank Wen Liu and Wen-Hao Bian for the helpful discussions.
J.W. is supported by Tianjin Natural Science Foundation Project (25JCYBJC01640)
and was partially supported by the National Natural
Science Foundation of China under Grant No. 11504360.

\section*{DATA AVAILABILITY STATEMENT}

The data that support the findings of this study are available from the corresponding author upon reasonable request.

\appendix
		
\section{One-loop corrections}\label{Appendix_one-loop-corrections}

\begin{figure}[H]
\centering
\includegraphics[width=3.3in]{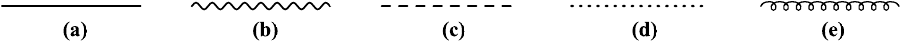}
\hspace{1.5cm}
\vspace{-0.1cm}
\caption{Free propagators for distinct kinds of fields: (a) $\Psi$, (b) $\phi$, (c) $h$, (d) $\eta$, and (e) $A$.}\label{Fig_free_propagators}
\end{figure}

\begin{figure}[H]
\centering
\includegraphics[width=3in]{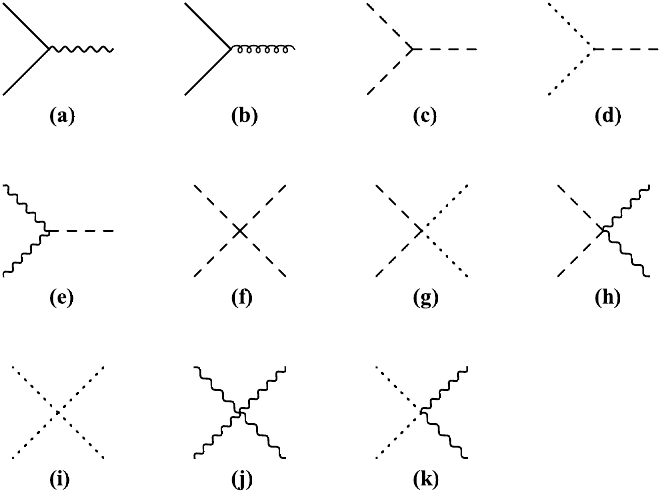}
\hspace{1cm}
\vspace{-0.1cm}
\caption{Tree-level vertices: (a) $\Psi^2\phi$; (b) $\Psi^2\Gamma$; (c) $h^3$;
(d) $\eta^2 h$; (e) $\phi^2 h$; (f) $h^4$; (g) $h^2\eta^2$; (h) $h^2\phi^2$;
(i) $\eta^4$; (j) $\phi^4$; (k) $\eta^2\phi^2$.}\label{Fig_free_vertex}
\end{figure}

On the basis of the total phenomenological effective theory
around the QCP~(\ref{Eq_S-phenomen}), we obtain the tree level and one-loop Feynman diagrams as shown in
Figs.~\ref{Fig_free_propagators}-\ref{Fig_1L_lambda_eta_h}. After performing long but straightforward algebraic calculations~\cite{Shankar1994RMP,Kim-Kivelson2008PRB,Sachdev2008PRB,
She2010PRB,She2015PRB,Wang2011PRB,Wang2013PRB,Wang2014PRD,Wang2015PRB,Wang2017PRB,Wang2017QBCP,
Vafek2012PRB,Vafek2014PRB,Wang2022SST}, we derive
all the one-loop corrections as follows,

\begin{figure}[H]
\centering
\includegraphics[width=3.5in]{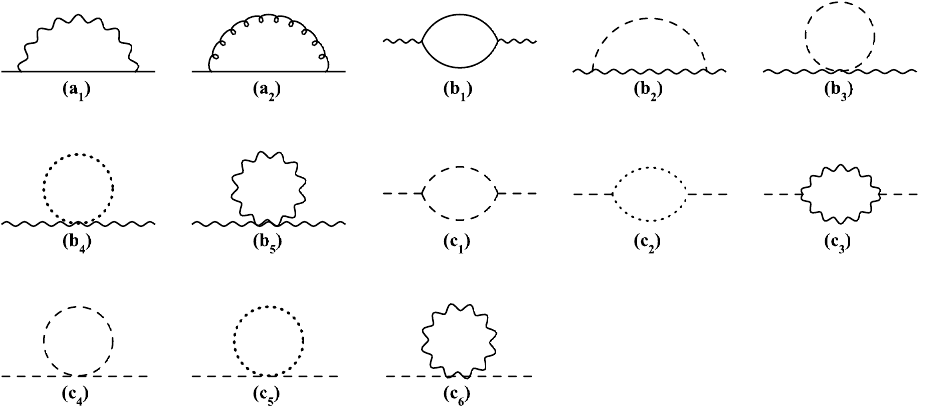}\hspace{1.5cm}
\vspace{-0.1cm}
\caption{One-loop corrections to the propagators of fields:
 ($\rm{a_1}$)-($\rm{a_2}$) for $\Psi$, ($\rm{b_1}$)-($\rm{b_5}$) for $\phi$, and ($\rm{c_1}$)-($\rm{c_6}$) for $h$
 due to the ordering competition.}\label{Fig_1L_propagators}
\end{figure}

\begin{figure}[H]
\centering
\includegraphics[width=3.3in]{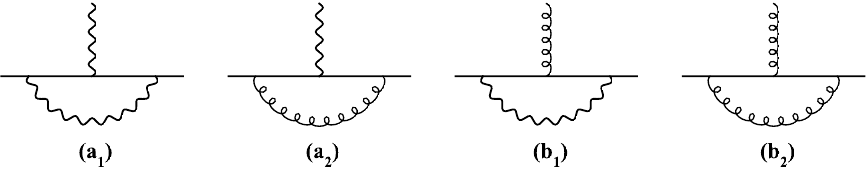}\hspace{1.5cm}
\vspace{-0.1cm}
\caption{One-loop corrections to the fermion-disorder coupling for($\rm{a_1}$)-($\rm{a_2}$) and
the Yukawa coupling between the QPs and $X$-state order parameter for($\rm{b_1}$)-($\rm{b_2}$).}\label{Fig_1L_disorder}
\end{figure}

\begin{widetext}

\begin{figure}[H]
\centering
\includegraphics[width=6in]{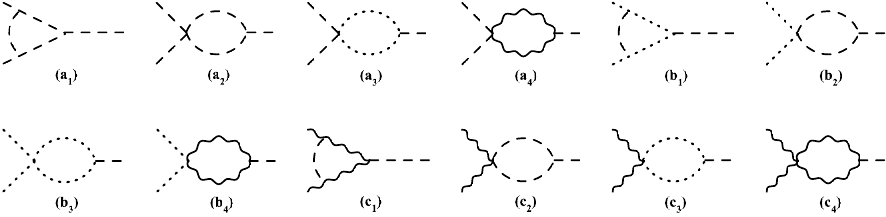}\hspace{1.5cm}
\vspace{-0.1cm}
\caption{One-loop corrections to $\gamma_h$ ($\rm{a_1}$)-($\rm{a_4}$),
$\gamma_{\eta h}$ ($\rm{b_1}$)-($\rm{b_4}$), and $\gamma_{\phi h}$ ($\rm{c_1}$)-($\rm{c_4}$)
due to the ordering competition.}\label{Fig_1L_propagators}
\end{figure}

\begin{figure}[H]
\centering
\includegraphics[width=5.5in]{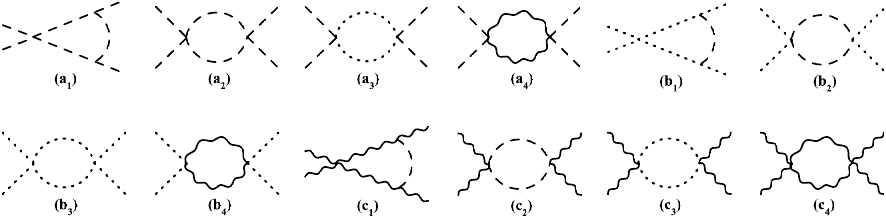}\hspace{1.5cm}
\vspace{-0.1cm}
\caption{One-loop corrections to $\beta_\phi$ ($\rm{a_1}$)-($\rm{a_4}$),
$\beta_\eta$ ($\rm{b_1}$)-($\rm{b_4}$), and $\beta_\phi$ ($\rm{c_1}$)-($\rm{c_4}$)
due to the ordering competition.}\label{Fig_1L_beta_phi}
\end{figure}

\begin{figure}[H]
\centering
\includegraphics[width=5.5in]{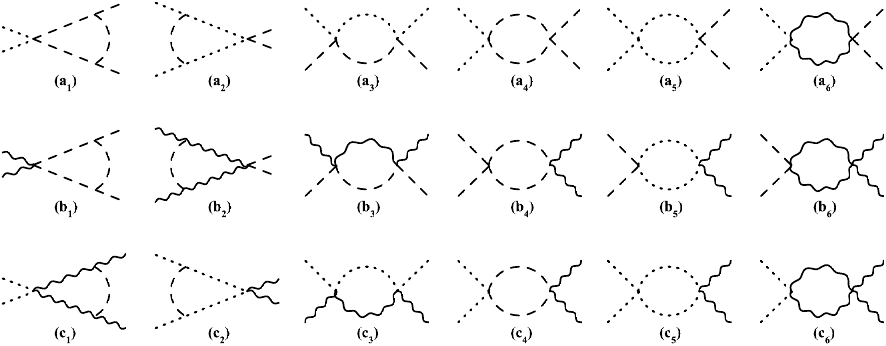}\hspace{1.5cm}
\vspace{-0.1cm}
\caption{One-loop corrections to $\lambda_{\eta h}$ ($\rm{a_1}$)-($\rm{a_6}$),
$\lambda_{h\phi}$ ($\rm{b_1}$)-($\rm{b_6}$), and $\lambda_{\eta\phi}$ ($\rm{c_1}$)-($\rm{c_6}$)
due to the ordering competition.}\label{Fig_1L_lambda_eta_h}
\end{figure}

		\begin{eqnarray}
		\Delta S^\Psi
         &=&-2\lambda_0^2(\mathcal{A}^i_1(-i\omega)+\mathcal{A}^i_2v_Fk_x\tau^z+\mathcal{A}^i_3v_\Delta k_y\tau^x)l-2C_gi\omega l,\\
           \Delta S^{\alpha_h}
            &=&\frac1{2\pi^2}\Big[-3\gamma_h^2-\gamma_{\eta h}^2
            -\frac{64}\pi\gamma_{\phi h}^2 v_{F}^2 v_\Delta^2\mathcal{F}_2^i(\theta,\varphi)
            +\frac{3}{4}\beta_{h}(1-2\alpha_{h})+\frac{1}{2}\lambda_{\eta h}\nonumber\\
            &&+\frac{2v_{F} v_\Delta}\pi\lambda_{h\phi}\mathcal{F}_1^i(\theta,\varphi)\bigg]l,\\
		\Delta S^{v_\Gamma}
			&=&4v_\Gamma\left[ (\mathcal{A}^i_1-3C_g)I+(\mathcal{A}^i_1+\mathcal{A}^i_2-\mathcal{A}^i_3+3C_g)\tau^y
			+\mathcal{A}^i_3\tau^x+\mathcal{A}^i_2\tau^z\right] l,\\
		\Delta S^{\gamma_h}
			&=&\frac{1}{4\pi^2}\bigg[
			-8\gamma_{h}^3(1-6\alpha_h)+\beta_h\gamma_{h}(1-4\alpha_h)-\lambda_{\eta h}\gamma_{\eta h}
            -\frac{\lambda_{h\phi}\gamma_{\phi h}}{2\pi}(16v_Fv_\Delta)^2\mathcal{F}^i_2(\theta,\varphi)\bigg] l,\\
        \Delta S^{\gamma_{\eta h}}
			&=&\frac{1}{4\pi^2}\bigg[ -2\gamma_{\eta h}^3(1-2\alpha_h)+\beta_{\eta}\gamma_{\eta h}
            -2\lambda_{\eta h}\gamma_{h}(1-4\alpha_h)-\frac{\lambda_{\eta\phi}\gamma_{\phi h}}
			{2\pi}(16v_Fv_\Delta)^2\mathcal{F}^i_2(\theta,\varphi)\bigg] l,\\
        \Delta S^{\gamma_{\phi h}}
			&=&\frac{1}{2\pi^2}\bigg[-\frac{\gamma_{\phi h}^3}{4\pi}(1-2\alpha_h)(16v_Fv_\Delta)^2\mathcal{F}^i_2(\theta,\varphi)
			-\lambda_{h\phi}\gamma_{h}(1-4\alpha_h)-\lambda_{\eta\phi}\gamma_{\eta h}\nonumber\\
            &&+\frac{\beta_{\phi}\gamma_{\phi h}}{8\pi}(16v_Fv_\Delta)^2\mathcal{F}^i_2(\theta,\varphi)\bigg] l,\\
		\Delta S^{\beta_h}
			&=&\frac{1}{4\pi^2}\bigg[
			-4\beta_h\gamma_{h}^2(1-6\alpha_h)+\frac{\beta_h^2}{2}(1-4\alpha_h)-2\lambda_{\eta h}^2-\frac{\lambda_{h\phi}^2}
			{2\pi}(16v_Fv_\Delta)^2\mathcal{F}^i_2(\theta,\varphi)\bigg] l,\\
		\Delta S^{\beta_\eta}
			&=&\frac{1}{4\pi^2}\bigg[
			-\beta_\eta\gamma_{\eta h}^2(1-2\alpha_h)+\frac{1}{2}\beta_\eta^2-2\lambda_{\eta h}^2(1-4\alpha_h)-\frac{\lambda_{h\phi}^2}
			{2\pi}(16v_Fv_\Delta)^2\mathcal{F}^i_2(\theta,\varphi)\bigg] l,\\
        \Delta S^{\beta_\phi}
			&=&\frac{1}{4\pi^2}\bigg[ -\frac{\beta_\phi\gamma_{\phi h}^2}{4\pi}(1-2\alpha_h)(16v_Fv_\Delta)^2\mathcal{F}^i_2(\theta,\varphi)
			-2\lambda_{\phi h}^2(1-4\alpha_h)\nonumber\\
            &&-2\lambda_{\eta\phi}^2+\frac{\beta_\phi^2}{8\pi}(16v_Fv_\Delta)^2\mathcal{F}^i_2(\theta,\varphi)\bigg]l\\
		\Delta S^{\lambda_{\eta h}}
			&=&\frac{1}{2\pi^2}\bigg[
			-4\lambda_{\eta h}\gamma_{h}^2(1-6\alpha_h)+\frac{\lambda_{\eta h}^2}{2}(1-2\alpha_h)
            +\frac{\beta_{h}\lambda_{\eta h}}{2}(1-4\alpha_h)+\frac{\beta_{h}\lambda_{\eta h}}{2}\nonumber\\
			&&-4\lambda_{\eta h}\gamma_{\eta h}^2(1-2\alpha_h)
            -\frac{\lambda_{\eta\phi}\lambda_{h\phi}}{4\pi}(16v_Fv_\Delta)^2\mathcal{F}^i_2(\theta,\varphi)\bigg] l,\\
        \Delta S^{\lambda_{h\phi}}
            &=&\frac{1}{2\pi^{2}}\bigg[-4\lambda_{h\phi}\gamma_{h}^{2}(1-6\alpha_{h})
            +\frac{\lambda_{h\phi}^{2}}{8\pi}(1-2\alpha_h)(16v_Fv_{\Delta})\mathcal{F}_{1}^{i}(\theta,\varphi)
            +\frac{\beta_\phi\lambda_{h\phi}}{8\pi}(16v_Fv_\Delta)^2\mathcal{F}_2^i(\theta,\varphi)-\lambda_{\eta h}\lambda_{\eta\phi}\nonumber\\
            &&+\frac{\beta_h\lambda_{h\phi}}2(1-4\alpha_h)-\frac{1}{4\pi}\lambda_{h\phi}\gamma_{\phi h}^2(1-2\alpha_h)
            (16v_Fv_\Delta)^2\mathcal{F}_2^i(\theta,\varphi)\bigg]l,\\
		\Delta S^{\lambda_{\eta\phi}}
			&=&\frac{1}{4\pi^2}\bigg[
			-2\lambda_{\eta\phi}\gamma_{\eta h}^2(1-2\alpha_h)
			-\frac{\lambda_{\eta\phi}\gamma_{\phi h}^2}{2\pi}(1-2\alpha_h)(16v_Fv_\Delta)^2\mathcal{F}^i_2(\theta,\varphi)
            -2\lambda_{\eta\phi}\lambda_{h\phi}(1-4\alpha_h)+\beta_{\eta}\lambda_{\eta\phi}\nonumber\\
			&&+\frac{\beta_\phi\lambda_{\eta\phi}}{4\pi}(16v_Fv_\Delta)^2\mathcal{F}^i_2(\theta,\varphi)
			+\frac{\lambda_{\eta\phi}^2}{4\pi}(16v_Fv_\Delta)\mathcal{F}^i_1(\theta,\varphi)\bigg] l.
		\end{eqnarray}
Hereby, the superscripts $i=\tau_{x,y,z,0}$ of the related coefficients $\mathcal{A}^i_{1,2,3}$ and $\mathcal{F}^i_{1,2}$
are adopted to specify type-$\tau_{x,y,z,0}$ QCPs. Specifically, they are denominated as
		\begin{eqnarray}
			\mathcal{A}^{\tau_x}_1&\equiv&\frac{2(v_\Delta/v_F)}{N_f\pi^3}\int_{-\infty}^{\infty}dx\int_{0}^{2\pi}d\theta
			     \frac{x^2-\cos^2\theta-(v_\Delta/v_F)^2\sin^2\theta}{[x^2+\cos^2\theta+(v_\Delta/v_F)^2\sin^2\theta]^2}\mathcal{G}_x(x,\theta),\\
			\mathcal{A}^{\tau_x}_2&\equiv&\frac{2(v_\Delta/v_F)}{N_f\pi^3}\int_{-\infty}^{\infty}dx\int_{0}^{2\pi}d\theta
			     \frac{-x^2+\cos^2\theta-(v_\Delta/v_F)^2\sin^2\theta}{[x^2+\cos^2\theta+(v_\Delta/v_F)^2\sin^2\theta]^2}\mathcal{G}_x(x,\theta),\\
			\mathcal{A}^{\tau_x}_3&\equiv&\frac{2(v_\Delta/v_F)}{N_f\pi^3}\int_{-\infty}^{\infty}dx\int_{0}^{2\pi}d\theta
			     \frac{x^2+\cos^2\theta-(v_\Delta/v_F)^2\sin^2\theta}{[x^2+\cos^2\theta+(v_\Delta/v_F)^2\sin^2\theta]^2}\mathcal{G}_x(x,\theta),\\
			\mathcal{R}^{\tau_x}&\equiv&\frac{2(v_\Delta/v_F)}{N_f\pi^3}\int^\infty_{-\infty} dx
			     \int_{0}^{2\pi}d\theta\frac{-x^2-\cos^2\theta+(v_\Delta/v_F)^2\sin\theta^2}
			     {[x^2+\cos^2\theta+(v_\Delta/v_F)^2\sin^2\theta]^2}\mathcal{G}_x(x,\theta),\\
            \mathcal{G}_x^{-1}(x,\theta)&\equiv&\frac{x^2+\cos^2\theta}{\sqrt{x^2+\cos^2\theta
			     +(v_\Delta/v_F)^2\sin^2\theta}}+\frac{x^2+\sin^2\theta}{\sqrt{x^2+\sin^2\theta+(v_\Delta/v_F)^2\cos^2\theta}},\\
			\mathcal{F}^{\tau_x}(\theta,\varphi)&\equiv&\frac{\cos^2\theta+v_F^2\cos^2\varphi\sin^2\theta}{\sqrt{\cos^2\theta
                +v_F^2\cos^2\varphi\sin^2\theta+v_\Delta^2\sin^2\varphi\sin^2\theta}}+\frac{\cos^2\theta+v_F^2\sin^2\varphi\sin^2\theta}
                {\sqrt{\cos^2\theta+v_F^2\sin^2\varphi\sin^2\theta+v_\Delta^2\cos^2\varphi\sin^2\theta}},\\
            \mathcal{F}^{\tau_x}_1(\theta,\varphi)&\equiv&\int_{0}^{\pi}d\theta
                \int_{0}^{2\pi}d\varphi\frac{\sin\theta}{\mathcal{F}^{\tau_x}(\theta,\varphi)},\\
            \mathcal{F}^{\tau_x}_2(\theta,\varphi)&\equiv&\int_{0}^{\pi}d\theta
                \int_{0}^{2\pi}d\varphi\frac{\sin\theta}{[\mathcal{F}^{\tau_x}(\theta,\varphi)]^2}.
		\end{eqnarray}
For the type $\tau_x$-QCP,
		\begin{eqnarray}
			\mathcal{A}^{\tau_y}_1&\equiv&\frac{2(v_\Delta/v_F)}{N_f\pi^3}
	            \int_{-\infty}^{\infty}dx\int_{0}^{2\pi}d\theta\frac{x^2-\cos^2\theta-(v_\Delta/v_F)^2\sin^2\theta}
	            {[x^2+\cos^2\theta+(v_\Delta/v_F)^2\sin^2\theta]^2}\mathcal{G}_y(x,\theta),\\	
            \mathcal{A}^{\tau_y}_2&\equiv&\frac{2(v_\Delta/v_F)}{N_f\pi^3}
		        \int_{-\infty}^{\infty}dx\int_{0}^{2\pi}d\theta\frac{-x^2+\cos^2\theta-(v_\Delta/v_F)^2\sin^2\theta}
		        {[x^2+\cos^2\theta+(v_\Delta/v_F)^2\sin^2\theta]^2}\mathcal{G}_y(x,\theta),\\
	        \mathcal{A}^{\tau_y}_3&\equiv&\frac{2(v_\Delta/v_F)}{N_f\pi^3}
	             \int_{-\infty}^{\infty}dx\int_{0}^{2\pi}d\theta\frac{-x^2-\cos^2\theta+(v_\Delta/v_F)^2\sin^2\theta}
		         {[x^2+\cos^2\theta+(v_\Delta/v_F)^2\sin^2\theta]^2}\mathcal{G}_y(x,\theta),\\
            \mathcal{R}^{\tau_y}&\equiv&\frac{2(v_\Delta/v_F)}{N_f\pi^3}\int^\infty_{-\infty} dx
		         \int_{0}^{2\pi}d\theta\frac{-x^2-\cos^2\theta+(v_\Delta/v_F)^2\sin\theta^2}
		         {[x^2+\cos^2\theta+(v_\Delta/v_F)^2\sin^2\theta]^2}\mathcal{G}_y(x,\theta),\\
	        \mathcal{G}_y^{-1}(x,\theta)
		         &\equiv&\sqrt{x^2+\cos^2\theta+(v_\Delta/v_F)^2\sin^2\theta}+\sqrt{x^2+\sin^2\theta+(v_\Delta/v_F)^2\cos^2\theta},\\
	        \mathcal{F}^{\tau_y}(\theta,\varphi)&\equiv&\sqrt{\cos^2\theta+v_F^2\cos^2\varphi\sin^2\theta+v_\Delta^2\sin^2\varphi\sin^2\theta}
                +\sqrt{\cos^2\theta+v_F^2\sin^2\varphi\sin^2\theta+v_\Delta^2\cos^2\varphi\sin^2\theta},\\
            \mathcal{F}^{\tau_y}_1(\theta,\varphi)&\equiv&\int_{0}^{\pi}d\theta
                \int_{0}^{2\pi}d\varphi\frac{\sin\theta}{\mathcal{F}^{\tau_y}(\theta,\varphi)},\\
            \mathcal{F}^{\tau_y}_2(\theta,\varphi)&\equiv&\int_{0}^{\pi}d\theta
                \int_{0}^{2\pi}d\varphi\frac{\sin\theta}{[\mathcal{F}^{\tau_y}(\theta,\varphi)]^2}.
		\end{eqnarray}
For the type $\tau_y$-QCP,
        \begin{eqnarray}
	       \mathcal{A}^{\tau_z}_1&\equiv&\frac{2(v_F/v_\Delta)}{N_f\pi^3}
	           \int_{-\infty}^{\infty}dx\int_{0}^{2\pi}d\theta\frac{x^2-\cos^2\theta-(v_\Delta/v_F)^2\sin^2\theta}
	           {[x^2+\cos^2\theta+(v_\Delta/v_F)^2\sin^2\theta]^2}\mathcal{G}_z(x,\theta),\\	
            \mathcal{A}^{\tau_z}_2&\equiv&\frac{2(v_F/v_\Delta)}{N_f\pi^3}
		        \int_{-\infty}^{\infty}dx\int_{0}^{2\pi}d\theta\frac{x^2-\cos^2\theta+(v_\Delta/v_F)^2\sin^2\theta}
	    	    {[x^2+\cos^2\theta+(v_\Delta/v_F)^2\sin^2\theta]^2}\mathcal{G}_z(x,\theta),\\
	        \mathcal{A}^{\tau_z}_3&\equiv&\frac{2(v_F/v_\Delta)}{N_f\pi^3}
	            \int_{-\infty}^{\infty}dx\int_{0}^{2\pi}d\theta\frac{-x^2-\cos^2\theta+(v_\Delta/v_F)^2\sin^2\theta}
		        {[x^2+\cos^2\theta+(v_\Delta/v_F)^2\sin^2\theta]^2}\mathcal{G}_z(x,\theta),\\
            \mathcal{R}^{\tau_z}&\equiv&\frac{2(v_\Delta/v_F)}{N_f\pi^3}\int^\infty_{-\infty} dx
		        \int_{0}^{2\pi}d\theta\frac{-x^2-\cos^2\theta+(v_\Delta/v_F)^2\sin\theta^2}
		        {[x^2+\cos^2\theta+(v_\Delta/v_F)^2\sin^2\theta]^2}\mathcal{G}_z(x,\theta),\\
	        \mathcal{G}_z^{-1}(x,\theta)
		        &\equiv&\frac{x^2+(v_\Delta/v_F)^2\sin^2\theta}{\sqrt{x^2+\cos^2\theta+(v_\Delta/v_F)^2\sin^2\theta}}
                +\frac{x^2+(v_\Delta/v_F)^2\cos^2\theta}{\sqrt{x^2+\sin^2\theta+(v_\Delta/v_F)^2\cos^2\theta}},\\
	        \mathcal{F}^{\tau_z}(\theta,\varphi)&\equiv&\frac{\cos^2\theta+v_\Delta^2\sin^2\varphi\sin^2\theta}
                {\sqrt{\cos^2\theta+v_F^2\cos^2\varphi\sin^2\theta+v_\Delta^2\sin^2\varphi\sin^2\theta}}+\frac{\cos^2\theta
                +v_\Delta^2\cos^2\varphi\sin^2\theta}{\sqrt{\cos^2\theta+v_F^2\sin^2\varphi\sin^2\theta+v_\Delta^2\cos^2\varphi\sin^2\theta}},\\
            \mathcal{F}^{\tau_z}_1(\theta,\varphi)&\equiv&\int_{0}^{\pi}d\theta
                \int_{0}^{2\pi}d\varphi\frac{\sin\theta}{\mathcal{F}^{\tau_z}(\theta,\varphi)},\\
            \mathcal{F}^{\tau_z}_2(\theta,\varphi)&\equiv&\int_{0}^{\pi}d\theta
                \int_{0}^{2\pi}d\varphi\frac{\sin\theta}{[\mathcal{F}^{\tau_z}(\theta,\varphi)]^2}.
        \end{eqnarray}
And for the type $\tau_0$-QCP,
\begin{eqnarray}
	\mathcal{A}^{\tau_0^A}_1&\equiv&\frac{2(v_\Delta/v_F)}{N_f\pi^3}
	   \int_{-\infty}^{\infty}dx\int_{0}^{2\pi}d\theta\frac{x^2-\cos^2\theta-(v_\Delta/v_F)^2\sin^2\theta}
	   {[x^2+\cos^2\theta+(v_\Delta/v_F)^2\sin^2\theta]^2}\mathcal{G}_0^A(x,\theta),\\
    \mathcal{A}^{\tau_0^A}_2&\equiv&\frac{2(v_\Delta/v_F)}{N_f\pi^3}
	   \int_{-\infty}^{\infty}dx\int_{0}^{2\pi}d\theta\frac{x^2-\cos^2\theta+(v_\Delta/v_F)^2\sin^2\theta}
	   {[x^2+\cos^2\theta+(v_\Delta/v_F)^2\sin^2\theta]^2}\mathcal{G}_0^A(x,\theta),\\
    \mathcal{A}^{\tau_0^A}_3&\equiv&\frac{2(v_\Delta/v_F)}{N_f\pi^3}
	   \int_{-\infty}^{\infty}dx\int_{0}^{2\pi}d\theta\frac{x^2+\cos^2\theta-(v_\Delta/v_F)^2\sin^2\theta}
	   {[x^2+\cos^2\theta+(v_\Delta/v_F)^2\sin^2\theta]^2}\mathcal{G}_0^A(x,\theta),\\
	\mathcal{R}^{\tau_0^A}&\equiv&\frac{2(v_\Delta/v_F)}{N_f\pi^3}\int^\infty_{-\infty} dx
		\int_{0}^{2\pi}d\theta\frac{-x^2-\cos^2\theta+(v_\Delta/v_F)^2\sin\theta^2}
		{[x^2+\cos^2\theta+(v_\Delta/v_F)^2\sin^2\theta]^2}\mathcal{G}_0^{A}(x,\theta),\\
    \mathcal{G}_0^{A-1}(x,\theta)
		&\equiv&-\sqrt{x^2+\cos^2\theta+(v_\Delta/v_F)^2\sin^2\theta},\\
	\mathcal{F}^{\tau_0^A}(\theta,\varphi)&\equiv&\sqrt{\cos^2\theta+v_F^2\cos^2\varphi\sin^2\theta+v_\Delta^2\sin^2\varphi\sin^2\theta},\\
    \mathcal{F}^{\tau_0^A}_1(\theta,\varphi)&\equiv&\int_{0}^{\pi}d\theta
    \int_{0}^{2\pi}d\varphi\frac{\sin\theta}{\mathcal{F}^{\tau_0^A}(\theta,\varphi)},\\
    \mathcal{F}^{\tau_0^A}_2(\theta,\varphi)&\equiv&\int_{0}^{\pi}d\theta
    \int_{0}^{2\pi}d\varphi\frac{\sin\theta}{[\mathcal{F}^{\tau_0^A}(\theta,\varphi)]^2},
\end{eqnarray}
together with
\begin{eqnarray}
	\mathcal{A}^{\tau_0^B}_1&\equiv&\frac{2(v_\Delta/v_F)}{N_f\pi^3}
	   \int_{-\infty}^{\infty}dx\int_{0}^{2\pi}d\theta\frac{x^2-\sin^2\theta-(v_\Delta/v_F)^2\cos^2\theta}
	   {[x^2+\sin^2\theta+(v_\Delta/v_F)^2\cos^2\theta]^2}\mathcal{G}_0^B(x,\theta),\\
    \mathcal{A}^{\tau_0^B}_2&\equiv&\frac{2(v_\Delta/v_F)}{N_f\pi^3}
	   \int_{-\infty}^{\infty}dx\int_{0}^{2\pi}d\theta\frac{x^2-\sin^2\theta+(v_\Delta/v_F)^2\cos^2\theta}
	   {[x^2+\sin^2\theta+(v_\Delta/v_F)^2\cos^2\theta]^2}\mathcal{G}_0^B(x,\theta),\\
    \mathcal{A}^{\tau_0^B}_3&\equiv&\frac{2(v_\Delta/v_F)}{N_f\pi^3}
	   \int_{-\infty}^{\infty}dx\int_{0}^{2\pi}d\theta\frac{x^2+\sin^2\theta-(v_\Delta/v_F)^2\cos^2\theta}
	   {[x^2+\sin^2\theta+(v_\Delta/v_F)^2\cos^2\theta]^2}\mathcal{G}_0^B(x,\theta),\\
	\mathcal{R}^{\tau_0^B}&\equiv&\frac{2(v_\Delta/v_F)}{N_f\pi^3}\int^\infty_{-\infty} dx
		\int_{0}^{2\pi}d\theta\frac{-x^2-\cos^2\theta+(v_\Delta/v_F)^2\sin\theta^2}
		{[x^2+\cos^2\theta+(v_\Delta/v_F)^2\sin^2\theta]^2}\mathcal{G}_0^{B}(x,\theta),\\
    \mathcal{G}_0^{B-1}(x,\theta)
		&\equiv&-\sqrt{x^2+\sin^2\theta+(v_\Delta/v_F)^2\cos^2\theta},\\
	\mathcal{F}^{\tau_0^B}(\theta,\varphi)&\equiv&\sqrt{\cos^2\theta+v_F^2\sin^2\varphi\sin^2\theta+v_\Delta^2\cos^2\varphi\sin^2\theta},\\
    \mathcal{F}^{\tau_0^B}_1(\theta,\varphi)&\equiv&\int_{0}^{\pi}d\theta
    \int_{0}^{2\pi}d\varphi\frac{\sin\theta}{\mathcal{F}^{\tau_0^B}(\theta,\varphi)},\\
    \mathcal{F}^{\tau_0^B}_2(\theta,\varphi)&\equiv&\int_{0}^{\pi}d\theta
    \int_{0}^{2\pi}d\varphi\frac{\sin\theta}{[\mathcal{F}^{\tau_0^B}(\theta,\varphi)]^2}.
\end{eqnarray}

\end{widetext}



\end{document}